\documentclass[11pt]{cernrep}   
\usepackage{graphicx}
\usepackage{here}
\usepackage{latexsym}
\usepackage{cite}

\renewcommand{\theequation}{\arabic{section}.\arabic{equation}}
\renewcommand{\thesection}{\arabic{section}}
\renewcommand{\thesubsection}{\arabic{section}.\arabic{subsection}}

\usepackage{amssymb}
\def\Journal#1#2#3#4{#1 {\bf #2} (#3) #4}
\def\refjl#1#2#3#4#5#6{\bibitem{#1} #2, \Journal{#3}{#4}{#5}{#6}.}
\def\refbk#1#2#3#4{\bibitem{#1} #2, #3, #4.}

\def\PRL{Phys. Rev. Lett.}

\def\NP{Nucl. Phys.}

\def\PL{Phys. Lett.}
\def\PR{Phys. Rev.}

\def\ZP{Z. Phys.}

\newcommand{\eqn}[1]{(\ref{#1})}
\newcommand{\be}{\begin{equation}}
\newcommand{\ee}{\end{equation}}
\newcommand{\no}{\nonumber}
\newcommand{\bel}[1]{\be\label{#1}}
\newcommand{\ba}{\begin{array}{c}}
\newcommand{\bat}{\begin{array}{cc}}
\newcommand{\bath}{\begin{array}{ccc}}
\newcommand{\ea}{\end{array}}
\newcommand{\beqn}{\begin{eqnarray}}
\newcommand{\eeqn}{\end{eqnarray}}

\newcommand{\bi}{\begin{itemize}}
\newcommand{\ei}{\end{itemize}}

\newcommand{\rms}{\rm\scriptstyle}

\newcommand{\toG}{\stackrel{G}{\,\longrightarrow\,}}

\newcommand{\ssb}{\stackrel{\mbox{\rm\scriptsize SSB}}{\longrightarrow}}
\newcommand{\toU}{\stackrel{\mbox{\rm\scriptsize U(1)}}{\longrightarrow}}

\newcommand{\toTheta}{\stackrel{\theta^i=0}{\longrightarrow}}

\def\gap{\;\lower3pt\hbox{$\buildrel > \over \sim$}\;}
\def\lap{\;\lower3pt\hbox{$\buildrel < \over \sim$}\;}
\newcommand{\cL}{{\cal L}}

\newcommand{\cM}{{\cal M}}
\newcommand{\cO}{{\cal O}}
\newcommand{\cP}{{\cal P}}
\newcommand{\cQ}{{\cal Q}}

\newcommand{\cA}{{\cal A}}
\newcommand{\cI}{{\cal I}}

\newcommand{\cJ}{{\cal J}}
\newcommand{\cC}{{\cal C}}
\newcommand{\cCP}{{\cal CP}}
\newcommand{\cT}{{\cal T}}
\newcommand{\cCPT}{{\cal CPT}}
\def\bV{\mathbf{V}}
\def\bM{\mathbf{M}}
\def\bU{\mathbf{U}}
\def\bS{\mathbf{S}}
\def\bH{\mathbf{H}}
\def\bmd{\mathbf{d}}
\def\bmu{\mathbf{u}}
\def\bml{\mathbf{l}}
\def\bmf{\mathbf{f}}

\newcommand{\e}{\mbox{\rm e}}
\def\CR{\nonumber \\ }
\def\gp{g\hskip .7pt\raisebox{-1.5pt}{$'$}}
\def\dop{d\hskip .6pt'}
\def\up{u\hskip .6pt'}
\def\lp{l\hskip .2pt'}
\def\nup{\nu\hskip .7pt'}
\def\sp{s\hskip .7pt'}
\def\bp{b\hskip .6pt'}
\def\sweff{\sin^2{\theta\hskip .2pt^{\rms lept}_{\rms eff}}}
\newcommand{\Br}{\mathrm{Br}}

\begin{document}


\title{The Standard Model of Electroweak Interactions}
\author{A. Pich}
\institute{ 
IFIC, University of Val\`encia -- CSIC,\\
Val\`encia, Spain}
\maketitle
\begin{abstract}
Gauge invariance is a powerful tool to
determine the dynamics of the electroweak and strong forces.
The particle content, structure and symmetries of the
Standard Model Lagrangian are discussed.
Special emphasis is given to the many phenomenological tests
which have established this theoretical framework
as the Standard Theory of electroweak interactions.
\end{abstract}



\section{Introduction}

The Standard Model (SM) is a gauge theory, based on the symmetry
group $SU(3)_C \otimes SU(2)_L \otimes U(1)_Y$, which describes
strong, weak and electromagnetic interactions, via the exchange of
the corresponding spin-1 gauge fields: eight massless gluons and one
massless photon, respectively, for the strong and electromagnetic
interactions, and three massive bosons, $W^\pm$ and $Z$, for the
weak interaction. The fermionic matter content is given by the known
leptons and quarks, which are organized in a three-fold family
structure:
\bel{eq:families} \left[\bat \nu_e & u \\  e^- & \dop \ea \right]
\quad , \quad \left[\bat \nu_\mu & c \\  \mu^- & \sp \ea \right]
\quad , \quad \left[\bat \nu_\tau & t \\  \tau^- & \bp \ea \right]
\quad , \ee
where (each quark appears in three different colours)
\bel{eq:structure}
\left[\bat \nu^{}_l & q^{}_u \\  l^- & q^{}_d \ea \right] \quad\equiv\quad
\left(\ba \nu^{}_l \\ l^- \ea \right)_L\; , \;\,
\left(\ba q^{}_u \\ q^{}_d \ea \right)_L\; , \;\, l^-_R\; ,
\;\, q^{\phantom{j}}_{uR}\; , \;\,
q^{\phantom{j}}_{dR}\; ,
\ee
plus the corresponding antiparticles. Thus, the left-handed fields
are $SU(2)_L$ doublets, while their right-handed partners transform
as $SU(2)_L$ singlets. The three fermionic families in
Eq.~\eqn{eq:families} appear to have identical properties (gauge
interactions); they differ only by their mass and their flavour
quantum number.

The gauge symmetry is broken by the vacuum,
which triggers the Spontaneous Symmetry Breaking (SSB)
of the electroweak group to the electromagnetic subgroup:
\bel{eq:ssb}
SU(3)_C \otimes SU(2)_L \otimes U(1)_Y \quad \ssb\quad
SU(3)_C \otimes U(1)_{\mathrm{QED}} \, .
\ee
The SSB mechanism generates the masses of the weak gauge bosons, and
gives rise to the appearance of a physical scalar particle in the
model, the so-called Higgs. The fermion masses and mixings are also
generated through the SSB.

The SM constitutes one of the most successful achievements
in modern physics. It provides a very elegant theoretical
framework, which is able to describe the
known experimental facts in particle physics with high precision.
These lectures \cite{SanFeliu} provide an introduction to the
electroweak sector of the SM, i.e., the $SU(2)_L \otimes U(1)_Y$
part \cite{GL:61,WE:67,SA:69,GIM:70}. The strong $SU(3)_C$ piece is
discussed in more detail in Ref.~\cite{PI:00}. The power of the
gauge principle is shown in Section~2, where the simpler Lagrangians
of quantum electrodynamics and quantum chromodynamics are derived.
The electroweak theoretical framework is presented in Sections~3 and
4, which discuss, respectively, the gauge structure and the SSB
mechanism. Section~5 summarizes the present phenomenological status
and shows the main precision tests performed at the $Z$ peak. The
flavour structure is discussed in Section~6, where knowledge of the
quark mixing angles is briefly reviewed and the importance of $\cCP$
violation tests is emphasized. Finally, a few comments on open
questions, to be investigated at future facilities, are given in the
summary.

Some useful but more technical information has been collected in
several appendices: a minimal amount of quantum field theory
concepts are given in Appendix~A; Appendix~B summarizes the most
important algebraic properties of $SU(N)$ matrices; and a short
discussion on gauge anomalies is presented in Appendix~C.


\setcounter{equation}{0}
\section{Gauge Invariance}
\label{sec:gauge}
\subsection{Quantum electrodynamics}
\label{sec:qed}

Let us consider the Lagrangian describing a free Dirac fermion:
\bel{eq:l_free}
\cL_0\, =\, i \,\overline{\psi}(x)\gamma^\mu\partial_\mu\psi(x)
\, - \, m\, \overline{\psi}(x)\psi(x) \, .
\ee
$\cL_0$ is invariant under {\em global}\ $U(1)$ transformations
\bel{eq:global}
\psi(x) \quad\toU\quad \psi'(x)\,\equiv\,\exp{\{i Q \theta\}}\,\psi(x) \, ,
\ee
where $Q\theta$ is an arbitrary real constant. The phase of
$\psi(x)$ is then a pure convention-dependent quantity without
physical meaning.
However, the free Lagrangian is no longer invariant if one allows
the phase transformation to depend on the space-time coordinate,
i.e., under {\em local} phase redefinitions $\theta=\theta(x)$,
because
\bel{eq:local}
\partial_\mu\psi(x) \quad\toU\quad \exp{\{i Q \theta\}}\;
\left(\partial_\mu + i Q \,\partial_\mu\theta\right)\,
\psi(x) \, .
\ee
Thus, once a given phase convention has been adopted at the
reference point $x_0$, the same convention must be taken at all
space-time points. This looks very unnatural.

The `gauge principle' is the requirement that the $U(1)$ phase
invariance should hold {\em locally}. This is only possible if one
adds an extra piece to the Lagrangian, transforming in such a
way as to cancel the $\partial_\mu\theta$ term in
Eq.~\eqn{eq:local}. The needed modification is completely fixed by
the transformation \eqn{eq:local}: one introduces a new spin-1
(since $\partial_\mu\theta$  has a Lorentz index) field $A_\mu(x)$,
transforming as
\bel{eq:a_transf}
 A_\mu(x)\quad\toU\quad A_\mu'(x)\,\equiv\,
 A_\mu(x) - {1\over e}\, \partial_\mu\theta\, ,
\ee
and defines the covariant derivative
\bel{eq:d_covariant}
D_\mu\psi(x)\,\equiv\,\left[\partial_\mu+ieQA_\mu(x)\right]
\,\psi(x)\, ,
\ee
which has the required property of transforming like the field itself:
\bel{eq:d_transf}
D_\mu\psi(x)\quad\toU\quad\left(D_\mu\psi\right)'(x)\,\equiv\,
\exp{\{i Q \theta\}}\,D_\mu\psi(x)\,.
\ee
The Lagrangian
\bel{eq:l_new}
\cL\,\equiv\,
i \,\overline{\psi}(x)\gamma^\mu D_\mu\psi(x)
\, - \, m\, \overline{\psi}(x)\psi(x)
\, =\, \cL_0\, -\, e Q A_\mu(x)\, \overline{\psi}(x)\gamma^\mu\psi(x)
\ee
is then invariant under local $U(1)$ transformations.

The gauge principle has generated an interaction between the Dirac
spinor and the gauge field $A_\mu$, which is nothing else than the
familiar vertex of Quantum Electrodynamics (QED). Note that the
corresponding electromagnetic charge $Q$ is completely arbitrary. If
one wants $A_\mu$ to be a true propagating field, one needs to add a
gauge-invariant kinetic term
\bel{eq:l_kinetic}
\cL_{\rms Kin}\,\equiv\, -{1\over 4}\, F_{\mu\nu}(x)\, F^{\mu\nu}(x)\,,
\ee
where $F_{\mu\nu}\,\equiv\, \partial_\mu A_\nu -\partial_\nu A_\mu$
is the usual electromagnetic field strength. A possible mass term
for the gauge field, $\cL_m = {1\over 2}m^2A^\mu A_\mu$, is
forbidden because it would violate gauge invariance; therefore, the
photon field is predicted to be massless. Experimentally, we know
that $m_\gamma < 6\cdot 10^{-17}$~eV \cite{PDG}.

The total Lagrangian in Eqs. \eqn{eq:l_new} and \eqn{eq:l_kinetic}
gives rise to the well-known Maxwell equations:
\bel{eq:Maxwell_QED}
\partial_\mu F^{\mu\nu}\, =\, J^\nu\,\equiv\,
  e Q\, \overline{\psi}\gamma^\nu\psi\, ,
\ee
where $J^\nu$ is the fermion electromagnetic current.
From a simple gauge-symmetry requirement, we have deduced the right
QED Lagrangian, which leads to a very  successful quantum field
theory.

\subsubsection{Lepton anomalous magnetic moments}
\label{sec:g-2}

\begin{figure}[tbh]
\begin{center}
\includegraphics[width=9.5cm]{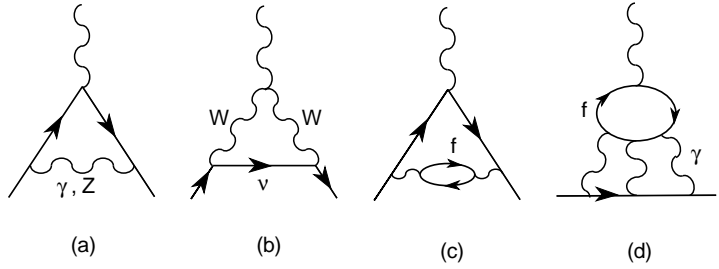}
\caption{Feynman diagrams contributing to the lepton anomalous magnetic moment.}
\label{fig:AnMagMom}
\end{center}
\end{figure}

The most stringent QED test comes from the high-precision
measurements of the $e$ \cite{OHUG:06} and $\mu$ \cite{E821:06}
anomalous magnetic moments \ $a_l\equiv (g^\gamma_l-2)/2\, $,
where \ $\vec\mu_l\equiv g^\gamma_l \,(e/2m_l)\, \vec{S}_l$:
\bel{eq:a_mu}
 a_e = (1\; 159\; 652\; 180.85\pm 0.76) \,\cdot\, 10^{-12}\, ,
 \qquad\qquad
 a_\mu = (11\; 659\; 208.0\pm 6.3) \,\cdot\, 10^{-10}\, .
\ee

To a measurable level, $a_e$ arises entirely from virtual electrons
and photons; these contributions are fully known to $O(\alpha^4)$
and some $O(\alpha^5)$ corrections have been already computed
\cite{DM:04,MRR:07,KN:06,AHKN:06,KA:06}. The impressive agreement
achieved between theory and experiment has promoted QED to the level
of the best theory ever built to describe Nature. The theoretical
error is dominated by the uncertainty in the input value of the QED
coupling $\alpha\equiv e^2/(4\pi)$. Turning things around, $a_e$
provides the most accurate determination of the fine structure
constant \cite{GHKNO:06}:
\be
\alpha^{-1} = 137.035\; 999\; 710 \,\pm\, 0.000\; 000\; 096\, .\ee

The anomalous magnetic moment of the muon is sensitive to small
corrections from virtual heavier states; compared to $a_e$, they
scale with the mass ratio $m_\mu^2/m_e^2$. Electroweak effects from
virtual $W^\pm$ and $Z$ bosons amount to a contribution of $(15.4
\pm 0.2)\cdot 10^{-10}$ \cite{DM:04,MRR:07}, which is larger than
the present experimental precision. Thus $a_\mu$ allows one to test
the entire SM. The main theoretical uncertainty comes from strong
interactions. Since quarks have electric charge, virtual
quark-antiquark pairs induce {\it hadronic vacuum polarization}
corrections to the photon propagator (Fig.~\ref{fig:AnMagMom}.c).
Owing to the non-perturbative character of the strong interaction at
low energies, the light-quark contribution cannot be reliably
calculated at present. This effect can be extracted from the
measurement of the cross-section $\sigma(e^+e^-\to \mbox{\rm
hadrons})$ and from the invariant-mass distribution of the final
hadrons in $\tau$ decays, which unfortunately provide slightly
different results \cite{DEHZ:03,DA:06,PI:07}:
\bel{eq:QED_pred}
a_\mu^{\mathrm{th}} \, = \,\left\{ \bat
(11\, 659\, 180.2\pm 5.6)\cdot 10^{-10}
&\qquad (e^+e^-\quad\mathrm{data})\, , \\
(11\, 659\, 199.7\pm 6.3)\cdot 10^{-10}
&\qquad (\tau\quad\mathrm{data})\, .
\ea\right.
\ee
The quoted uncertainties include also the smaller {\it
light-by-light scattering} contributions (Fig.~\ref{fig:AnMagMom}.d)
\cite{BP:07}. The difference between the SM prediction and the
experimental value \eqn{eq:a_mu} corresponds to $3.3\,\sigma$
($e^+e^-$) \ or \ $0.9\,\sigma$ \ ($\tau$). New precise $e^+e^-$ and
$\tau$ data sets are needed to settle the true value of
$a_\mu^{\mathrm{th}}$.

\subsection{Quantum chromodynamics}
\label{sec:QCDlagrangian}

\subsubsection{Quarks and colour}

\begin{figure}[htb]
\begin{center}
\includegraphics[width=5cm]{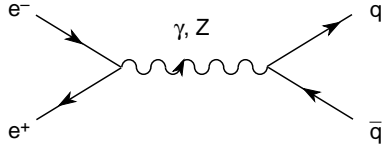}
\caption{Tree-level Feynman diagram for the \ $e^+e^-$ annihilation
into hadrons.} \label{fig:eeqqhad}
\end{center}
\end{figure}

The large number of known mesonic and baryonic states clearly
signals the existence of a deeper level of elementary constituents
of matter: {\it quarks}. Assuming that mesons are \ $M\equiv q\bar
q$ \ states, while baryons have three quark constituents, $B\equiv
qqq$, one can nicely classify the entire hadronic spectrum. However,
in order to satisfy the Fermi--Dirac statistics one needs to assume
the existence of a new quantum number, {\it colour}, such that each
species of quark may have $N_C=3$ different colours: $q^\alpha$,
$\alpha =1,2,3$ (red, green, blue). Baryons and mesons are then
described by the colour-singlet combinations
\be\label{eq:m_b_wf} B\, =\,
{1\over\sqrt{6}}\:\epsilon^{\alpha\beta\gamma}\, |q_\alpha q_\beta
q_\gamma\rangle \, , \qquad\qquad M\, =\,
{1\over\sqrt{3}}\:\delta^{\alpha\beta} \, |q_\alpha \bar q_\beta
\rangle \, . \ee
In order to avoid the existence of non-observed extra states with
non-zero colour, one needs to further postulate that all asymptotic
states are colourless, i.e., singlets under rotations in colour
space. This assumption is known as the {\it confinement hypothesis},
because it implies the non-observability of free quarks: since
quarks carry colour they are confined within colour-singlet bound
states.

A direct test of the colour quantum number can be obtained from the
ratio
\be\label{eq:R_ee}
R_{e^+e^-} \;\equiv\;
{\sigma(e^+e^-\to \mbox{\rm hadrons})\over\sigma(e^+e^-\to\mu^+\mu^-)} \, .
\ee
The hadronic production occurs through $e^+e^-\to\gamma^*, Z^*\to
q\bar q\to \mbox{\rm hadrons}$ (Fig.~2). Since quarks are assumed to
be confined, the probability to hadronize is just one; therefore,
summing over all possible quarks in the final state, we can estimate
the inclusive cross-section into hadrons. The electroweak production
factors which are common with the $e^+e^-\to\gamma^*,
Z^*\to\mu^+\mu^-$ process cancel in the ratio \eqn{eq:R_ee}. At
energies well below the $Z$ peak, the cross-section is dominated by
the $\gamma$-exchange amplitude; the ratio $R_{e^+e^-}$ is then
given by the sum of the quark electric charges squared:
\be\label{eq:R_ee_res}
R_{e^+e^-} \,\approx N_C \; \sum_{f=1}^{N_f} Q_f^2 \; = \;
\left\{
\begin{array}{cc}
\frac{2}{3}\, N_C = 2\, , \qquad & (N_f=3 \; :\; u,d,s)  \\[5pt]
\frac{10}{9}\, N_C = \frac{10}{3}\, , \qquad & (N_f=4 \; :\; u,d,s,c)  \\[5pt]
\frac{11}{9}\, N_C = \frac{11}{3} \, ,\qquad & (N_f=5 \; :\;
u,d,s,c,b)
\ea\right. . \ee
%

\begin{figure}[bth]
\begin{center}
\includegraphics[width=12cm]{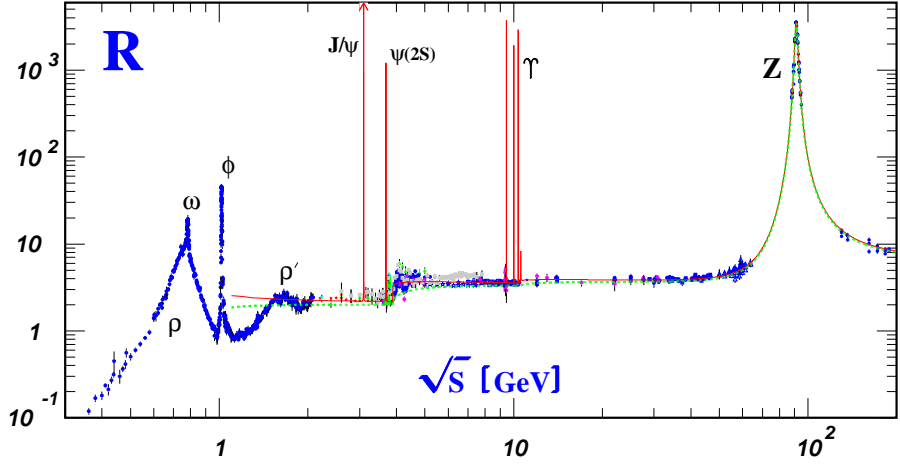} 
\caption{World data on the ratio $R_{e^+e^-}$ \protect\cite{PDG}.
The broken lines show the naive quark model approximation with
$N_C=3$. The solid curve is the 3-loop perturbative QCD prediction.}
\label{fig:Ree}
\end{center}
\end{figure}

The measured ratio is shown in Fig. \ref{fig:Ree}. Although
the simple
formula \eqn{eq:R_ee_res} cannot explain the complicated
structure around the different quark thresholds,
it gives the right average
value of the cross-section (away from thresholds),
provided that $N_C$ is taken to be three.
The agreement is better at larger energies.
Notice that strong
interactions have not been taken into account;
only the confinement hypothesis has been used.

Electromagnetic interactions are associated with the fermion
electric charges, while the quark flavours (up, down, strange,
charm, bottom, top) are related to electroweak phenomena. The strong
forces are flavour conserving and flavour independent. On the other
side, the carriers of the electroweak interaction ($\gamma$, $Z$,
$W^\pm$) do not couple to the quark colour. Thus it seems natural to
take colour as the charge associated with the strong forces and try
to build a quantum field theory based on it \cite{FG:72,FGL:73}.

\subsubsection{Non-Abelian gauge symmetry}

Let us denote $q^\alpha_f$ a quark field of colour $\alpha$ and flavour $f$.
To simplify the equations, let us adopt a vector notation
in colour space:
$q_f^T \,\equiv\, (q^1_f\, ,\, q^2_f\, ,\, q^3_f) $.
The free Lagrangian
\bel{eq:L_free}
\cL_0\, =\, \sum_f\;\bar q_f \, \left( i\gamma^\mu\partial_\mu - m_f\right) q_f
\ee
is invariant under arbitrary {\it global} $SU(3)_C$ transformations
in colour space,
\bel{eq:q_transf}
q^\alpha_f \;\longrightarrow\;
(q^\alpha_f)'\, =\, U^\alpha_{\phantom{\alpha}\beta}\; q^\beta_f \; ,
\qquad\qquad
U\, U^\dagger\, =\, U^\dagger U\, =\, 1 \; , \qquad\qquad
\det U\, = 1\, \; .
\ee
The $SU(3)_C$ matrices can be written in the form
\bel{eq:U_def}
U\, =\, \exp\left\{ i\, {\lambda^a\over 2}\,\theta_a\right\} \; ,
\ee
where $\frac{1}{2}\,\lambda^a$ ($a=1,2,\ldots,8$) denote the
generators of the fundamental representation of the
$SU(3)_C$ algebra,
and $\theta_a$ are arbitrary parameters. The matrices $\lambda^a$
are traceless and
satisfy the commutation relations
\bel{eq:commutation}
\left[\, {\lambda^a\over 2}\, ,\, {\lambda^b\over 2}\,\right]\, =\,
i\, f^{abc} \; {\lambda^c\over 2} \; ,
\ee
with $f^{abc}$ the $SU(3)_C$
structure constants, which are real and totally antisymmetric.
Some useful properties of $SU(3)$ matrices are collected in Appendix~B.

As in the QED case, we can now require the Lagrangian to be also invariant
under {\it local} $SU(3)_C$ transformations, $\theta_a = \theta_a(x)$. To
satisfy this requirement, we need to change the quark derivatives by covariant
objects. Since we have now eight independent gauge parameters, eight different
gauge bosons $G^\mu_a(x)$, the so-called {\it gluons}, are needed:
\bel{eq:D_cov}
D^\mu q_f \,\equiv\, \left[ \partial^\mu + i g_s\, {\lambda^a\over 2}\,
G^\mu_a(x)\right]
  \, q_f \, \equiv\, \left[ \partial^\mu + i g_s\, G^\mu(x)\right] \, q_f \, .
\ee
Notice that we have introduced the compact matrix notation
\bel{eq:G_matrix}
  [G^\mu(x)]_{\alpha\beta}\,\equiv\,
\left({\lambda^a\over 2}\right)_{\!\alpha\beta}\, G^\mu_a(x) \, .
\ee
%
\begin{figure}[tbh]
\begin{center}
\mbox{}\vskip .3cm
\includegraphics[width=11.5cm]{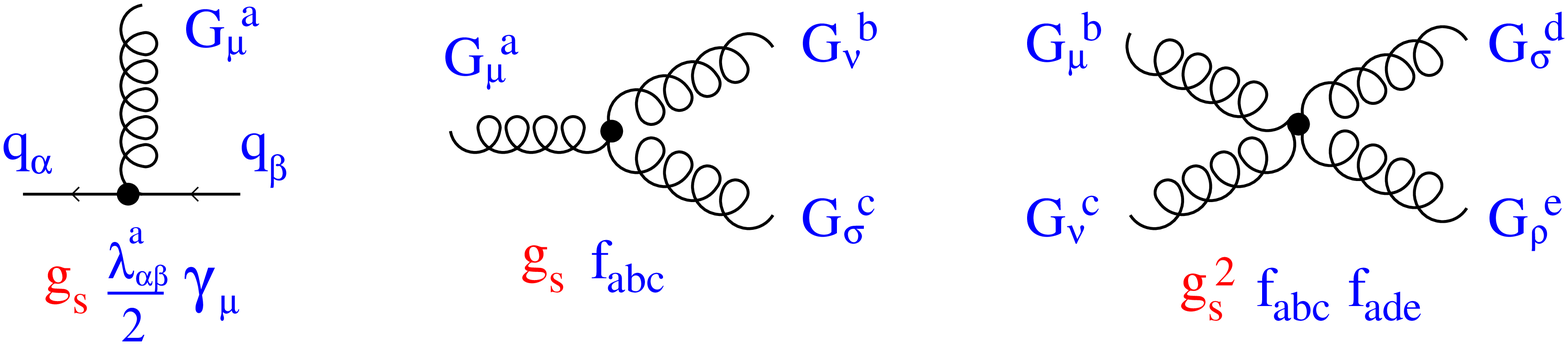}
\caption{Interaction vertices of the QCD Lagrangian.}
\label{fig:Lqcd}
\end{center}
\end{figure}
%
We want $D^\mu q_f$ to transform in exactly the same way as the
colour-vector $q_f$; this fixes the transformation properties of the
gauge fields:
\bel{eq:G_trans}
D^\mu \;\longrightarrow\; (D^\mu)'\, =\, U\, D^\mu\, U^\dagger \, ,
\qquad\qquad
G^\mu \;\longrightarrow\; (G^\mu)'\, =\, U\, G^\mu\, U^\dagger
  +{i\over g_s} \, (\partial^\mu U) \, U^\dagger \, .
\ee
Under an infinitesimal $SU(3)_C$ transformation,
\beqn\label{eq:inf_transf}
q^\alpha_f &\longrightarrow & (q^\alpha_f)'\, =\,
q^\alpha_f\, +\, i \,\left({\lambda^a\over 2}\right)_{\!\alpha\beta}
\,\delta\theta_a\; q^\beta_f \, ,
\no\\
G^\mu_a &\longrightarrow & (G^\mu_a)'\, =\,
G^\mu_a\, -\, {1\over g_s}\,\partial^\mu(\delta\theta_a)
\, \,- f^{abc} \,\delta\theta_b \, G^\mu_c \, .
\eeqn
The gauge transformation of the gluon fields is more complicated than the one
obtained in QED for the photon.
The non-commutativity of the $SU(3)_C$ matrices gives rise to an additional term
involving the gluon fields themselves.
For constant $\delta\theta_a$, the transformation rule for the gauge fields
is expressed in terms of the structure constants $f^{abc}$; thus,
the gluon fields belong to the adjoint representation of the colour group
(see Appendix~B).
Note also that there is a unique $SU(3)_C$
coupling $g_s$. In QED it was possible to assign arbitrary electromagnetic
charges to the
different fermions. Since the commutation relation \eqn{eq:commutation} is
non-linear, this freedom does not exist for $SU(3)_C$.

To build a gauge-invariant kinetic term for the gluon fields, we
introduce the corresponding field strengths:
\beqn\label{eq:G_tensor}
G^{\mu\nu}(x) & \equiv & -{i\over g_s}\, [D^\mu, D^\nu] \, = \,
  \partial^\mu G^\nu - \partial^\nu G^\mu + i g_s\, [G^\mu, G^\nu] \, \equiv \,
  {\lambda^a\over 2}\, G^{\mu\nu}_a(x) \, ,
\no\\
G^{\mu\nu}_a(x) & = & \partial^\mu G^\nu_a - \partial^\nu G^\mu_a
  - g_s\, f^{abc}\, G^\mu_b\, G^\nu_c \, .
\eeqn
Under a gauge transformation,
\bel{eq:G_tensor_transf}
G^{\mu\nu}\;\longrightarrow\; (G^{\mu\nu})'\, =\, U\, G^{\mu\nu}\, U^\dagger \, ,
\ee
and the colour trace \
Tr$(G^{\mu\nu}G_{\mu\nu}) = \frac{1}{2}\, G^{\mu\nu}_aG_{\mu\nu}^a$ \
remains invariant.

Taking the proper normalization for the gluon kinetic term, we finally have
the $SU(3)_C$ invariant Lagrangian of Quantum Chromodynamics (QCD):
\bel{eq:L_QCD}
\cL_{\rms QCD} \,\equiv\, -{1\over 4}\, G^{\mu\nu}_aG_{\mu\nu}^a
\, +\, \sum_f\;\bar q_f \, \left( i\gamma^\mu D_\mu - m_f\right)\, q_f \, .
\ee
It is worth while to decompose the Lagrangian into its different
pieces:
\beqn\label{eq:L_QCD_pieces}
\cL_{\rms QCD} & = &
 -\, {1\over 4}\, (\partial^\mu G^\nu_a - \partial^\nu G^\mu_a)\,
  (\partial_\mu G_\nu^a - \partial_\nu G_\mu^a)
\, +\, \sum_f\;\bar q^\alpha_f \, \left( i\gamma^\mu\partial_\mu - m_f\right)\,
q^\alpha_f
\qquad\no\\ && \mbox{}
 -\, g_s\, G^\mu_a\,\sum_f\;\bar q^\alpha_f\, \gamma_\mu\,
 \left({\lambda^a\over 2}\right)_{\!\alpha\beta}\, q^\beta_f
\\ && \mbox{}
 +\, {g_s\over 2}\, f^{abc}\, (\partial^\mu G^\nu_a - \partial^\nu G^\mu_a) \,
  G_\mu^b\, G_\nu^c \, - \,
{g_s^2\over 4} \, f^{abc} f_{ade} \, G^\mu_b\, G^\nu_c\, G_\mu^d\, G_\nu^e \, .
\no
\eeqn
The first line contains the correct kinetic terms for the different
fields, which give rise to the corresponding propagators. The colour
interaction between quarks and gluons is given by the second line;
it involves the $SU(3)_C$ matrices $\lambda^a$. Finally, owing to
the non-Abelian character of the colour group, the
$G^{\mu\nu}_aG_{\mu\nu}^a$ term generates the cubic and quartic
gluon self-interactions shown in the last line; the strength of
these interactions (Fig.~\ref{fig:Lqcd}) is given by the same
coupling $g_s$ which appears in the fermionic piece of the
Lagrangian.

\begin{figure}[t]\centering
\begin{minipage}[t]{.45\linewidth}\centering
\includegraphics[width=7cm,clip]{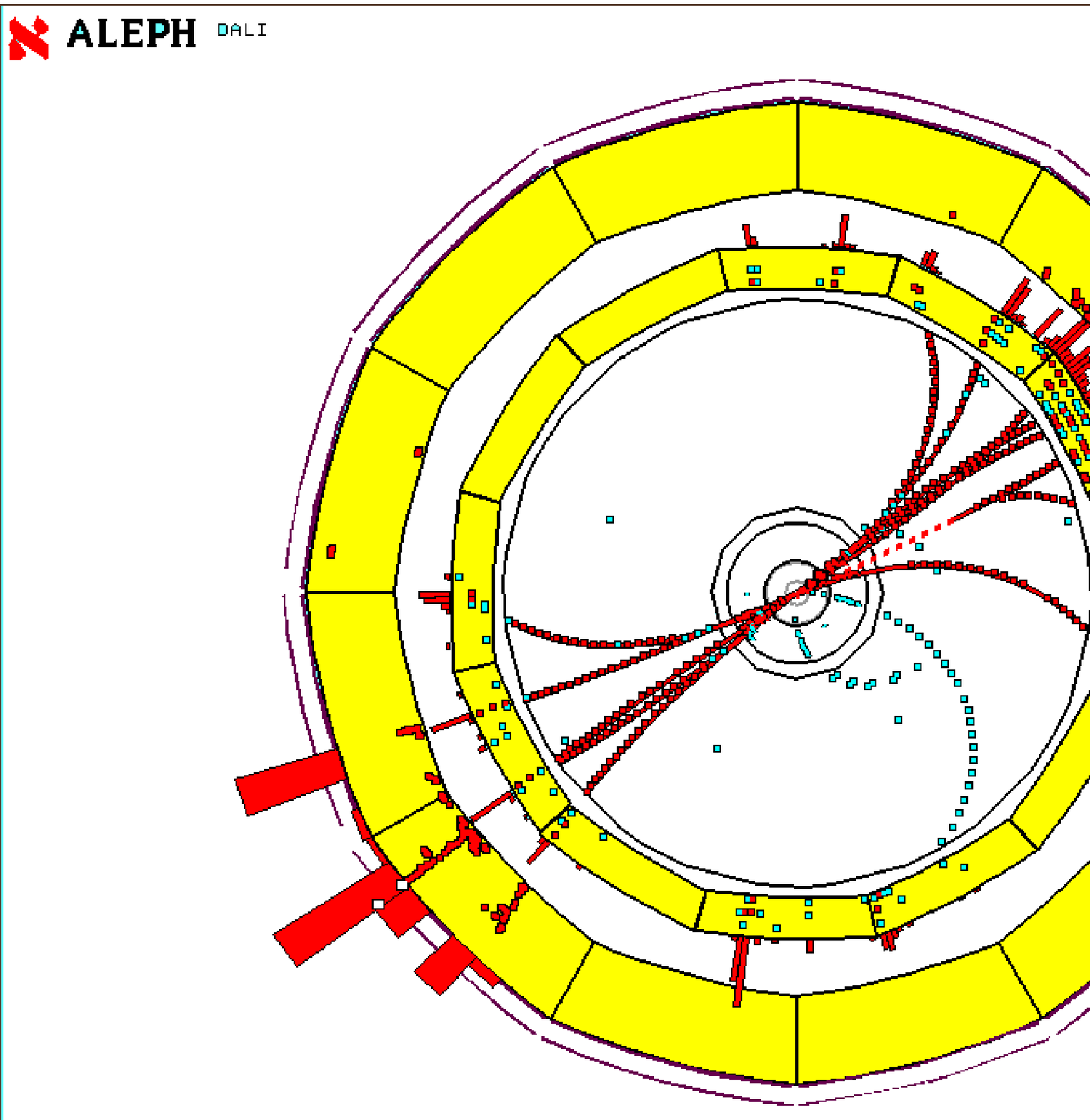} 
\end{minipage}
\hskip .75cm
\begin{minipage}[t]{.45\linewidth}\centering
\includegraphics[width=7cm,clip]{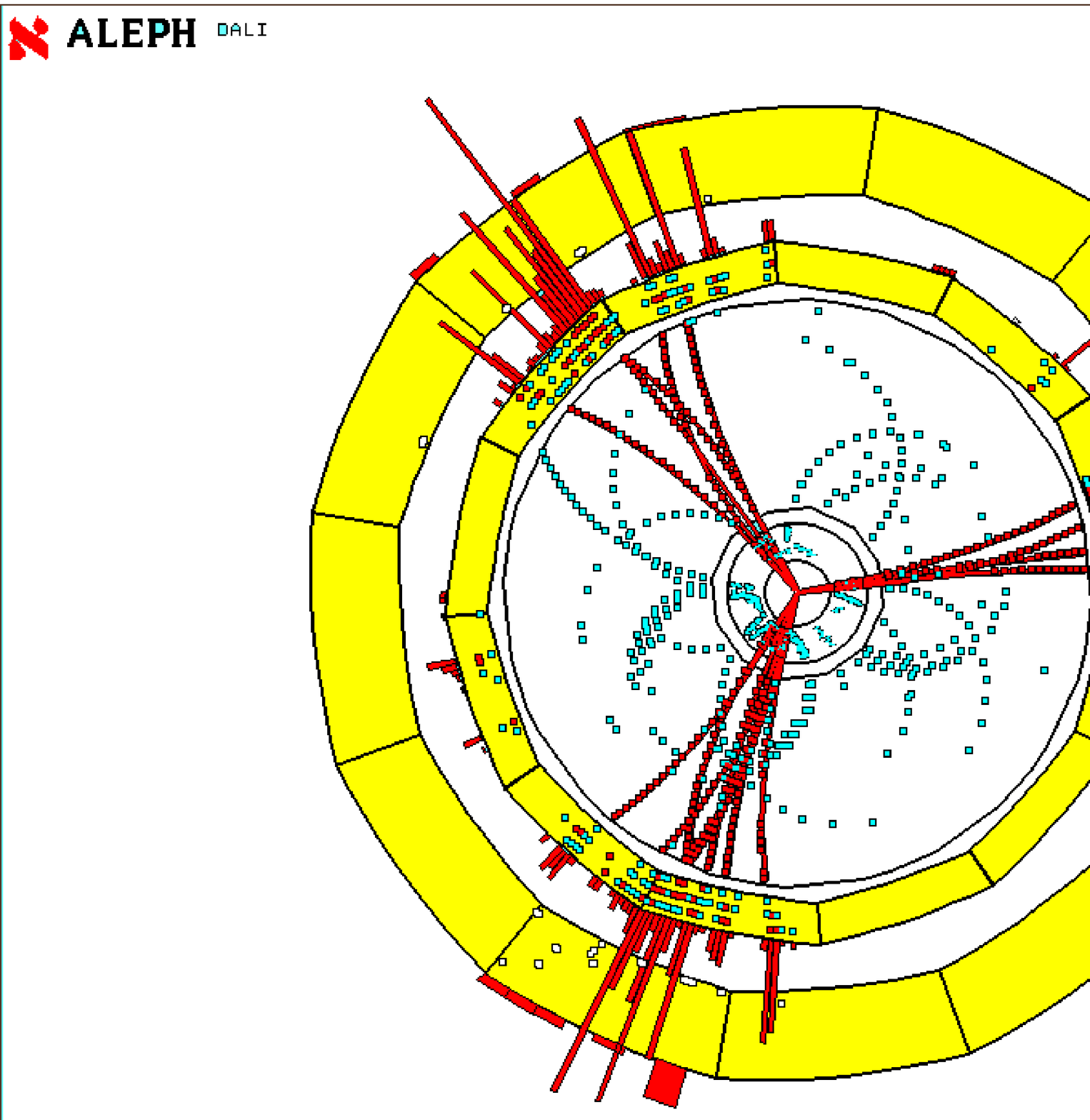} 
\end{minipage}
\caption{Two- and three-jet events from the hadronic $Z$ boson
decays \ $Z\to q\bar q$ \ and \ $Z\to  q\bar q G$ \ (ALEPH)
\protect\cite{ALEPHjets}.} \label{fig:ThreeJets}
\end{figure}

In spite of the rich physics contained in it, the Lagrangian
\eqn{eq:L_QCD} looks very simple because of its colour symmetry
properties. All interactions are given in terms of a single
universal coupling $g_s$, which is called the {\it strong coupling
constant}. The existence of self-interactions among the gauge fields
is a new feature that was not present in QED; it seems then
reasonable to expect that these gauge self-interactions could
explain properties like asymptotic freedom (strong interactions
become weaker at short distances) and confinement (the strong forces
increase at large distances), which do not appear in QED
\cite{PI:00}.

Without any detailed calculation, one can already extract
qualitative physical consequences from $\cL_{\rms QCD}$. Quarks can
emit gluons. At lowest order in $g_s$, the dominant process will be
the emission of a single gauge boson; thus, the hadronic decay of
the $Z$ should result in some $Z\to q\bar q G$ events, in addition
to the dominant $Z\to q\bar q$ decays.
Figure~\ref{fig:ThreeJets} clearly shows that 3-jet events, with the required
kinematics, indeed appear in the LEP data. Similar events show up in
$e^+e^-$ annihilation into hadrons, away from the $Z$ peak.
The ratio between 3-jet and 2-jet events provides a simple estimate of
the strength of the strong interaction at LEP energies ($s=M_Z^2$):
$\alpha_s\equiv g_s^2/(4\pi) \sim 0.12$.


\setcounter{equation}{0}
\section{Electroweak Unification}
\label{sec:EWmodel}

\subsection{Experimental facts}

Low-energy experiments have provided a large amount of information
about the dynamics underlying flavour-changing processes. The
detailed analysis of the energy and angular distributions in $\beta$
decays, such as \ $\mu^-\to e^-\bar\nu_e\,\nu_\mu$ \ or \ $n\to p\,
e^-\bar\nu_e\,$, made clear that only the left-handed (right-handed)
fermion (antifermion) chiralities participate in those weak
transitions; moreover, the strength of the interaction appears to be
universal. This is further corroborated through the study of other
processes like \ $\pi^-\to e^-\bar\nu_e$ \ or \ $\pi^-\to
\mu^-\bar\nu_\mu\,$, which show that neutrinos have left-handed
chiralities while anti-neutrinos are right-handed.

From neutrino scattering data, we learnt the existence of different
neutrino types ($\nu_e\not=\nu_\mu$) and that there are separately
conserved lepton quantum numbers which distinguish neutrinos from
antineutrinos; thus we observe the transitions \ $\bar\nu_e\, p\to
e^+ n\,$, \ $\nu_e\, n\to e^- p\,$, \ $\bar\nu_\mu\, p\to \mu^+ n$ \
or \ $\nu_\mu\, n\to \mu^- p\,$, but we do not see processes like \
$\nu_e\, p\not\to e^+ n\,$, \ $\bar\nu_e\, n\not\to e^- p\,$, \
$\bar\nu_\mu\, p\not\to e^+ n$ \ or \ $\nu_\mu\, n\not\to e^- p\,$.

Together with theoretical considerations related to unitarity (a
proper high-energy behaviour) and the absence of flavour-changing
neutral-current transitions ($\mu^-\not\to e^-e^-e^+$), the
low-energy information was good enough to determine the structure of
the modern electroweak theory \cite{PI:94}. The intermediate vector
bosons $W^\pm$ and $Z$ were theoretically introduced and their
masses correctly estimated, before their experimental discovery.
Nowadays, we have accumulated huge numbers of $W^\pm$ and $Z$ decay
events, which bring much direct experimental evidence of their
dynamical properties.

\subsubsection{Charged currents}

\begin{figure}[tbh]\centering
\begin{minipage}[t]{.4\linewidth}\centering
\includegraphics[width=6cm]{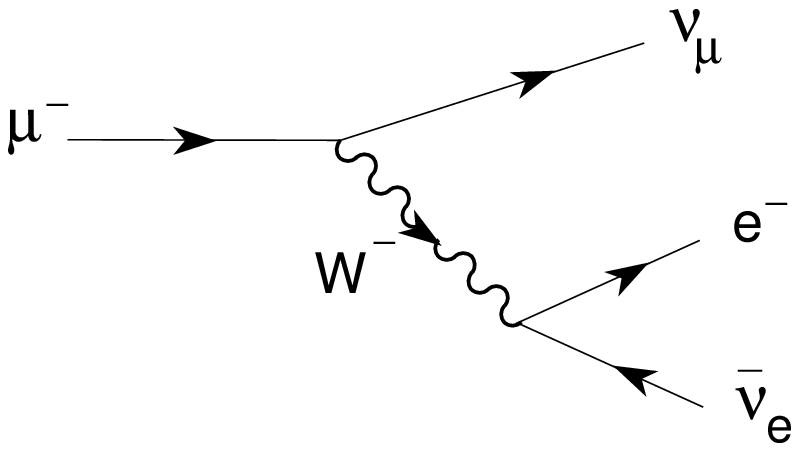}
\end{minipage}
\hskip .5cm
\begin{minipage}[t]{.4\linewidth}\centering
\includegraphics[width=5cm]{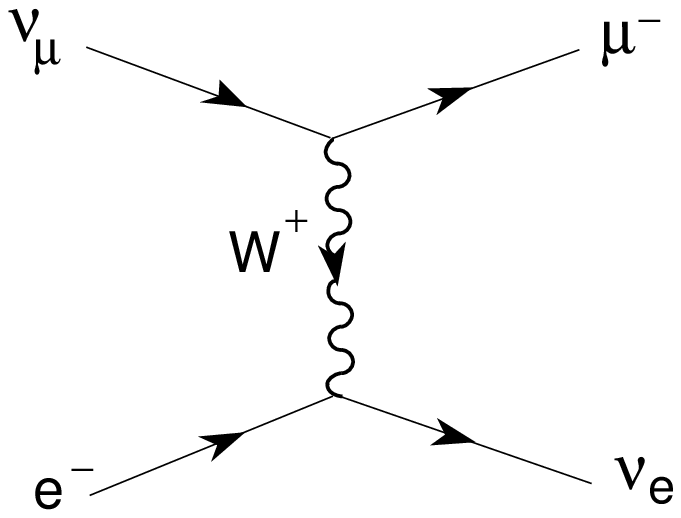}
\end{minipage}
\caption{Tree-level Feynman diagrams for \ $\mu^-\to e^-\bar\nu_e\,\nu_\mu$ \
and \ $\nu_\mu\, e^-\to\mu^-\nu_e$.}
\label{fig:ChargedCurrents}
\end{figure}

\noindent The interaction of quarks and leptons with the $W^\pm$
bosons (Fig.~\ref{fig:ChargedCurrents}) exhibits the following
features:

\bi

\item[--] Only left-handed fermions and right-handed antifermions couple to the
$W^\pm$. Therefore, there is a 100\% breaking of
parity $\cP$ (left $\leftrightarrow$ right) and charge conjugation $\cC$
(particle $\leftrightarrow$ antiparticle).
However, the combined transformation \ $\cC\cP$ \ is still a good
symmetry.

\item[--] The $W^\pm$ bosons couple to the fermionic doublets in Eq.~\eqn{eq:families},
where the electric charges of the two fermion partners differ in one
unit. The decay channels of the $W^-$ are then:
\be
W^-\,\to\, e^-\bar\nu_e\, ,\, \mu^-\bar\nu_\mu\, ,\,\tau^-\bar\nu_\tau\, ,\,
d\,'\,\bar u\, ,\, s\,'\,\bar c\, .
\ee
Owing to the very high mass of the top quark \cite{mt}, $m_t =
171~\mathrm{GeV}
> M_W = 80.4~\mathrm{GeV}$, its on-shell production through \
$W^-\to  b\,'\, \bar t$ \ is kinematically forbidden.

\item[--] All fermion doublets couple to the $W^\pm$ bosons with the same
universal strength.

\item[--] The doublet partners of the up, charm and top quarks appear to be
mixtures of the three quarks with charge $-\frac{1}{3}$:
\be
\left(\ba d\,'\\ s\,'\\ b\,'\ea\right) \, = \, \bV\;
\left(\ba d\\ s\\ b\ea\right)\; ,
\qquad\qquad
\bV\,\bV^\dagger\, =\,\bV^\dagger\, \bV\, =\, 1\, .
\ee
Thus, the weak eigenstates $d\,'\, ,\, s\,'\, ,\, b\,'\,$ are different than
the mass eigenstates $d\, ,\, s\, ,\, b\,$. They are related through the
$3\times 3$ unitary matrix $\bV$,
which characterizes flavour-mixing phenomena.

\item[--] The experimental evidence of neutrino oscillations shows that $\nu_e$,
$\nu_\mu$ and $\nu_\tau$ are also mixtures of mass eigenstates.
However, the neutrino masses are tiny: \ $\left|m^2_{\nu_3} -
m^2_{\nu_2}\right|\sim 2.5\cdot 10^{-3}\, \mathrm{eV}^2\,$, \
$m^2_{\nu_2} - m^2_{\nu_1}\sim 8\cdot 10^{-5}\, \mathrm{eV}^2\,$
\cite{PDG}.

\ei

\subsubsection{Neutral currents}

\begin{figure}[t]\centering
\begin{minipage}[t]{.4\linewidth}\centering
\includegraphics[width=5.5cm]{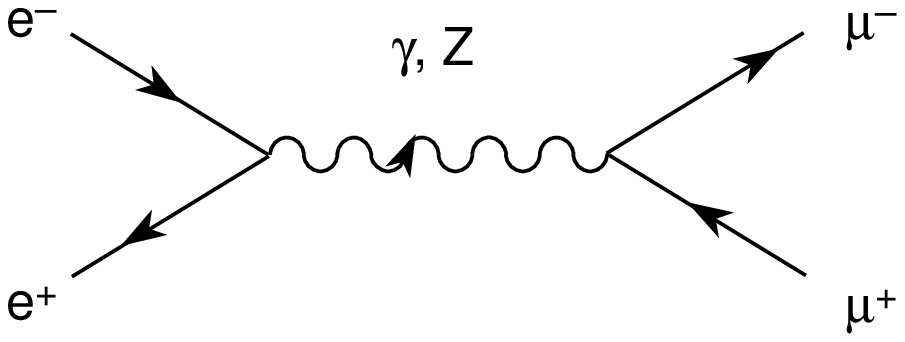}
\end{minipage}
\hskip 1cm
\begin{minipage}[t]{.4\linewidth}\centering
\includegraphics[width=5.5cm]{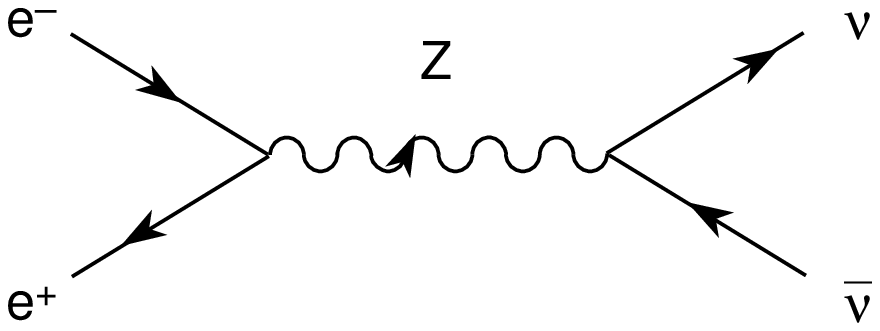}
\end{minipage}
\caption{Tree-level Feynman diagrams for \ $e^+ e^-\to\mu^+\mu^-$ \
and \ $e^+ e^-\to\nu\,\bar\nu$.}
\label{fig:NeutralCurrents}
\end{figure}

\noindent The neutral carriers of the electromagnetic and weak
interactions have fermionic couplings (Fig.~\ref{fig:NeutralCurrents})
with the following properties:

\bi

\item[--] All interacting vertices are flavour conserving. Both the $\gamma$
and the $Z$ couple to a fermion and its own antifermion, i.e.,
$\gamma\, f\,\bar f$ \ and \ $Z\, f\,\bar f$. Transitions of the
type \ $\mu\not\to e\gamma$ \ or \ $Z\not\to e^\pm\mu^\mp$ \ have
never been observed.

\item[--] The interactions depend on the fermion electric charge $Q_f$.
Fermions with the same $Q_f$ have exactly the same universal
couplings. Neutrinos do not have electromagnetic interactions
($Q_\nu = 0$), but they have a non-zero coupling to the $Z$ boson.

\item[--] Photons have the same interaction for both fermion chiralities, but
the $Z$ couplings are different for left-handed and right-handed
fermions. The neutrino coupling to the $Z$ involves only left-handed
chiralities.

\item[--] There are three different light neutrino species.

\ei

\subsection{The \ $\mathbf{SU(2)_L\otimes U(1)_Y}$ \ theory}
\label{sec:model}

Using gauge invariance, we have been able to determine the right QED
and QCD Lagrangians. To describe weak interactions, we need a more
elaborated structure, with several fermionic flavours and different
properties for left- and right-handed fields; moreover, the
left-handed fermions should appear in doublets, and we would like to
have massive gauge bosons $W^\pm$ and $Z$ in addition to the photon.
The simplest group with doublet representations is $SU(2)$. We want
to include also the electromagnetic interactions; thus we need an
additional $U(1)$ group. The obvious symmetry group to consider is
then
\bel{eq:group}
G\,\equiv\, SU(2)_L\otimes U(1)_Y \, ,
\ee
where $L$ refers to left-handed fields.
We do not specify, for the moment, the meaning of the
subindex $Y$ since, as we will see, the naive identification
with electromagnetism does not work.

For simplicity, let us consider a single family of quarks,
and introduce the notation
\bel{eq:psi_def}
\psi_1(x)\, =\,\left(\ba u \\ d \ea\right)_L \, , \qquad\quad
\psi_2(x) \, =\, u_R\, , \qquad\quad\psi_3(x) \, = \, d_R \, .
\ee
Our discussion will also be valid for the
lepton sector, with the identification
\bel{eq:psi_def_l}
\psi_1(x)\, =\,\left(\ba \nu_e \\ e^- \ea\right)_L\, , \qquad\quad
\psi_2(x) \, =\, \nu^{}_{eR} \, , \qquad\quad\psi_3(x) \, = \, e^-_R\, .
\ee

As in the QED and QCD cases, let us consider the free Lagrangian
\bel{eq:free_l}
\cL_0\; =\; i\,\bar u(x)\,\gamma^\mu\,\partial_\mu u(x)\, +\,
i\,\bar d(x)\,\gamma^\mu\,\partial_\mu d(x)
\; =\;\sum_{j=1}^3 \;i\,
\overline{\psi}_j(x)\,\gamma^\mu\,\partial_\mu\psi_j(x)\, .
\ee
$\cL_0$ is invariant under global $G$ transformations in flavour space:
\beqn\label{eq:G_transf}
\psi_1(x)&\toG & \psi'_1(x)\,\equiv
\,\exp{\left\{iy_1\beta\right\}}\; U_L\;\psi_1(x) \, ,
\no\\
\psi_2(x)& \toG & \psi'_2(x)\,\equiv\,
\exp{\left\{iy_2\beta\right\}}\;\psi_2(x) \, ,
\\
\psi_3(x)& \toG & \psi'_3(x)\,\equiv\,
\exp{\left\{iy_3\beta\right\}}\;\psi_3(x) \, ,
\no
\eeqn
where the $SU(2)_L$ transformation
\bel{eq:U_L}
U_L^{\phantom{\dagger}}
\,\equiv\,\exp{\left\{i\,\frac{\sigma_i}{2}\,\alpha^i\right\}}
\qquad\qquad\qquad (i=1,2,3)
\ee
only acts on the doublet field $\psi_1$. The parameters $y_i$ are
called hypercharges, since the $U(1)_Y$ phase transformation is
analogous to the QED one. The matrix transformation $U_L$ is
non-Abelian as in QCD. Notice that we have not included a mass term
in Eq.~\eqn{eq:free_l} because it would mix the left- and
right-handed fields [see Eq.~\eqn{eq:DiracL_LR}], therefore spoiling
our symmetry considerations.

We can now require the Lagrangian to be also invariant under local
$SU(2)_L\otimes U(1)_Y$ gauge transformations, i.e., with
$\alpha^i=\alpha^i(x)$ and $\beta=\beta(x)$. In order to satisfy
this symmetry requirement, we need to change the fermion derivatives
by covariant objects. Since we have now four gauge parameters,
$\alpha^i(x)$ and $\beta(x)$, four different gauge bosons are
needed:
\beqn\label{eq:cov_der}
D_\mu\psi_1(x)&\equiv &\left[\partial_\mu
+i\,g\, \widetilde W_\mu(x)+ i\,\gp\, y_1\, B_\mu(x)\right]
\,\psi_1(x)\, ,
\no\\
D_\mu\psi_2(x)&\equiv &\left[\partial_\mu
+ i\,\gp\, y_2\, B_\mu(x)\right]\,\psi_2(x)\, ,
\\[5pt]
D_\mu\psi_3(x)&\equiv &\left[\partial_\mu + i\,\gp\, y_3\,
B_\mu(x)\right]\,\psi_3(x)\, ,
\no \eeqn
where
\bel{eq:Wmatrix}
\widetilde W_\mu(x)\,\equiv\,{\sigma_i\over 2}\,W^i_\mu(x)
\ee
denotes a $SU(2)_L$ matrix field. Thus we have the correct number of
gauge fields to describe the $W^\pm$, $Z$ and $\gamma$.

We want $D_\mu\psi_j(x)$ to
transform in exactly the same way as the $\psi_j(x)$ fields;
this fixes the transformation properties of the gauge fields:
\beqn\label{eq:B_transf}
B_\mu(x)&\toG &B'_\mu(x)\,\equiv\, B_\mu(x) -{1\over \gp}\,
\partial_\mu\beta(x) ,
\\ \label{eq:W-transf}
\widetilde W_\mu&\toG &
\widetilde W'_\mu\,\equiv\,
U_L^{\phantom{\dagger}}(x)\, \widetilde W_\mu\, U_L^\dagger(x)
+ {i\over g}\, \partial_\mu U_L^{\phantom{\dagger}}(x)\, U_L^\dagger(x) ,
\eeqn
where \
$U_L(x)\equiv\exp{\left\{i\,\frac{\sigma_i}{2}\,\alpha^i(x)\right\}}$.
The transformation of $B_\mu$ is identical to the one obtained in
QED for the photon, while the $SU(2)_L$ $W^i_\mu$ fields transform
in a way analogous to the gluon fields of QCD. Note that the
$\psi_j$ couplings to $B_\mu$ are completely free as in QED, i.e.,
the hypercharges $y_j$ can be arbitrary parameters. Since the
$SU(2)_L$ commutation relation is non-linear, this freedom does not
exist for the $W^i_\mu$: there  is only a unique $SU(2)_L$ coupling
$g$.

The Lagrangian
\bel{eq:lagrangian} \cL\; =\; \sum_{j=1}^3 \;i\,
\overline{\psi}_j(x)\,\gamma^\mu\, D_\mu\psi_j(x)
\ee
is invariant under local $G$ transformations.
In order to build the gauge-invariant kinetic term for the
gauge fields, we introduce the corresponding field strengths:
\be\label{eq:b_mn}
B_{\mu\nu}  \;\equiv\;  \partial_\mu B_\nu - \partial_\nu B_\mu\, ,
\ee
\bel{eq:W_mn}
\widetilde W_{\mu\nu} \;\equiv\; -{i\over g}\,\left[
\left(\partial_\mu +i\,g\, \widetilde W_\mu\right)\, ,\,
\left(\partial_\nu +i\,g\, \widetilde W_\nu\right)\right]
\, =\,\partial_\mu \widetilde W_\nu -\partial_\nu \widetilde W_\mu
+i g\,\left[ W_\mu , W_\nu \right]\, ,
\ee
\bel{eq:Wi_mn}
\widetilde W_{\mu\nu} \;\equiv\; {\sigma_i\over 2}\, W^i_{\mu\nu}\, ,
\qquad\qquad
W^i_{\mu\nu}\, =\,\partial_\mu W^i_\nu - \partial_\nu W^i_\mu
- g\,\epsilon^{ijk}\, W^j_\mu\, W^k_\nu\, .
\ee
$B_{\mu\nu}$ remains invariant under $G$ transformations,
while $\widetilde W_{\mu\nu}$ transforms covariantly:
\bel{eq:W_mn_transf}
B_{\mu\nu}\,\toG\, B_{\mu\nu}\, ,
\qquad\qquad
\widetilde W_{\mu\nu}\,\toG\, U_L^{\phantom{\dagger}}\,
\widetilde W_{\mu\nu}\, U_L^\dagger
\, .
\ee
Therefore, the properly normalized kinetic Lagrangian is given by
\bel{eq:kinetic}
\cL_{\rms Kin} \, = \,
-{1\over 4}\, B_{\mu\nu}\, B^{\mu\nu} - {1\over 2}\,
\mbox{\rm Tr}\left[\widetilde W_{\mu\nu}\, \widetilde W^{\mu\nu}\right]
\, = \,
-{1\over 4}\, B_{\mu\nu}\, B^{\mu\nu} - {1\over 4}\,
W_{\mu\nu}^i\, W^{\mu\nu}_i\, .
\ee
Since the field strengths
$W^i_{\mu\nu}$ contain a quadratic piece, the Lagrangian
$\cL_{\rms Kin}$ gives rise to cubic and
quartic self-interactions among the gauge fields.
The strength of these interactions
is given by the same $SU(2)_L$ coupling $g$  which appears in the
fermionic piece of the Lagrangian.

The gauge symmetry forbids the writing of a mass term for the gauge
bosons. Fermionic masses are also not possible, because they would
communicate the left- and right-handed fields, which have different
transformation properties, and therefore would produce an explicit
breaking of the gauge symmetry. Thus, the $SU(2)_L\otimes U(1)_Y$
Lagrangian in Eqs.~\eqn{eq:lagrangian} and \eqn{eq:kinetic} only
contains massless fields.

\subsection{Charged-current interaction}
\label{subsec:cc}

\begin{figure}[tbh]\centering
\begin{minipage}[t]{.4\linewidth}\centering
\includegraphics[width=4cm]{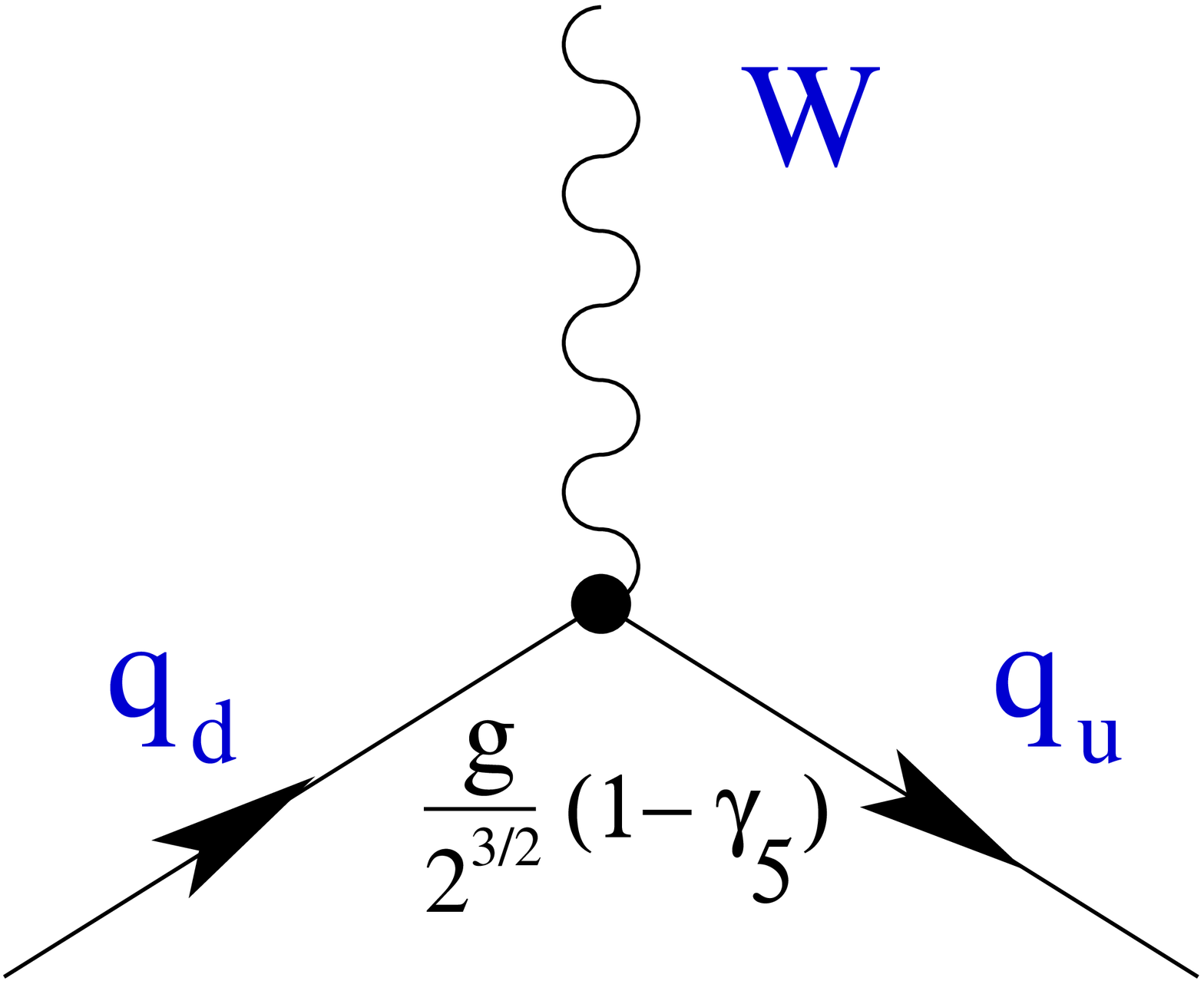}
\end{minipage}
\hskip 1cm
\begin{minipage}[t]{.4\linewidth}\centering
\includegraphics[width=4cm]{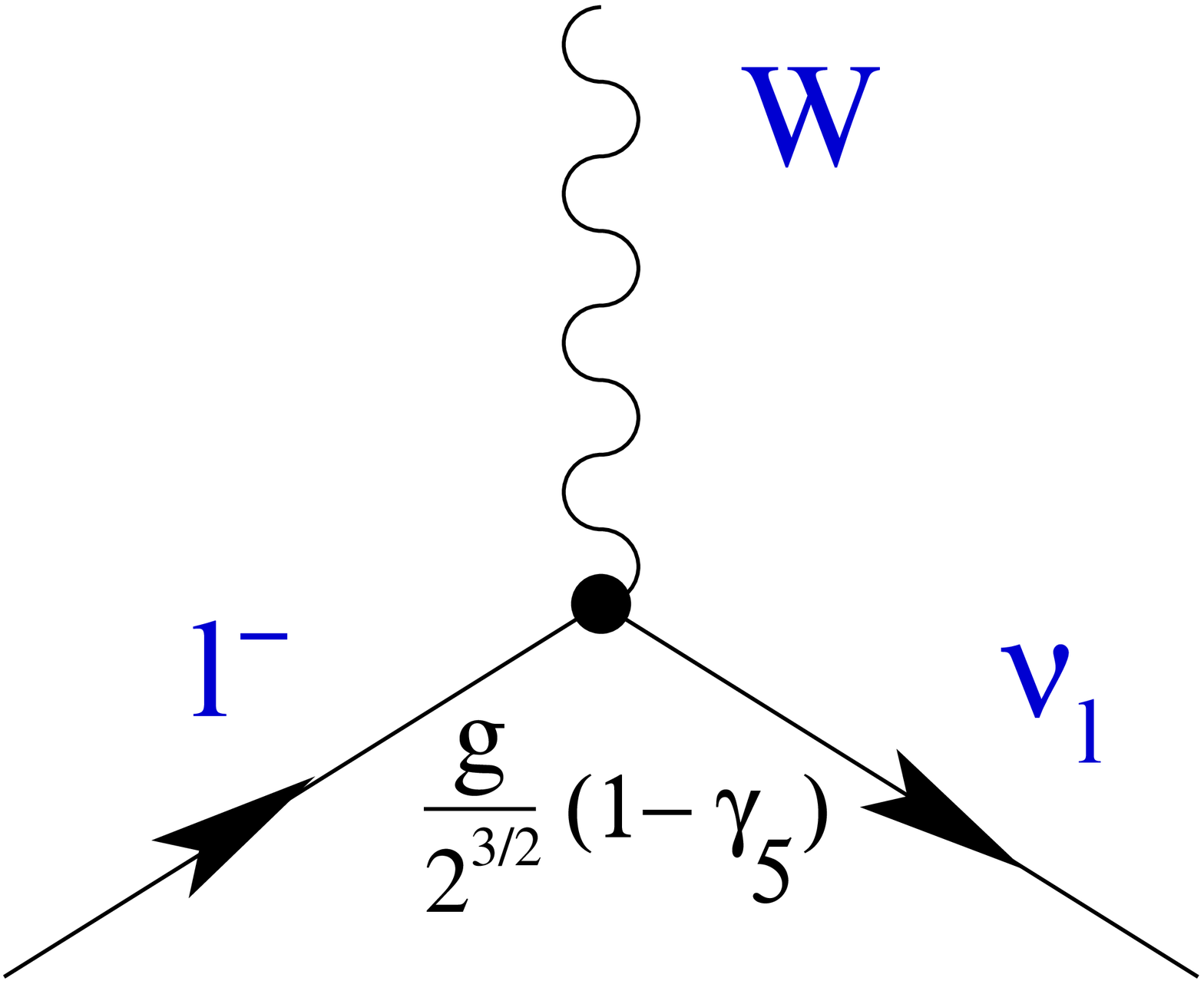}
\end{minipage}
\caption{Charged-current interaction vertices.}
\label{fig:L_CC}
\end{figure}

The Lagrangian \eqn{eq:lagrangian} contains interactions of the
fermion fields with the gauge bosons,
\bel{eq:int}
\cL\quad\longrightarrow\quad
-g\,\overline{\psi}_1\gamma^\mu\widetilde  W_\mu
\psi_1\, - \,\gp\,B_\mu\,
\sum_{j=1}^3\, y_j\,\overline{\psi}_j\gamma^\mu\psi_j \, .
\ee
The term containing the $SU(2)_L$ matrix
\bel{eq:W_matrix} \widetilde W_\mu\, =\, {\sigma^i\over 2}\,
W_\mu^i\, = \, {1\over 2} \,\left(\bat
W^3_\mu & \sqrt{2}\, W_\mu^\dagger \\[5pt] \sqrt{2}\, W_\mu & -W^3_\mu
\ea\right) \ee
gives rise to charged-current interactions with the boson field \
$W_\mu\equiv (W_\mu^1+i\, W_\mu^2)/\sqrt{2}$ \ and its
complex-conjugate \ $W^\dagger_\mu\equiv (W_\mu^1-i\,
W_\mu^2)/\sqrt{2}$ \ (Fig.~\ref{fig:L_CC}). For a single family of
quarks and leptons,
\bel{eq:W_interactions}
\cL_{\rms CC}\, = \, -{g\over 2\sqrt{2}}\,\left\{
W^\dagger_\mu\,\left[
\bar u\gamma^\mu(1-\gamma_5) d\, +\, \bar\nu_e\gamma^\mu(1-\gamma_5) e
\right]\, + \, \mbox{\rm h.c.}\right\}\, .
\ee
The universality of the quark and lepton interactions is now a
direct consequence of the assumed gauge symmetry. Note, however,
that Eq.~\eqn{eq:W_interactions} cannot describe the observed
dynamics, because the gauge bosons are massless and, therefore, give
rise to long-range forces.

\subsection{Neutral-current interaction}
\label{subsec:nc_int}

\begin{figure}[tbh]\centering
\begin{minipage}[t]{.4\linewidth}\centering
\includegraphics[width=4cm]{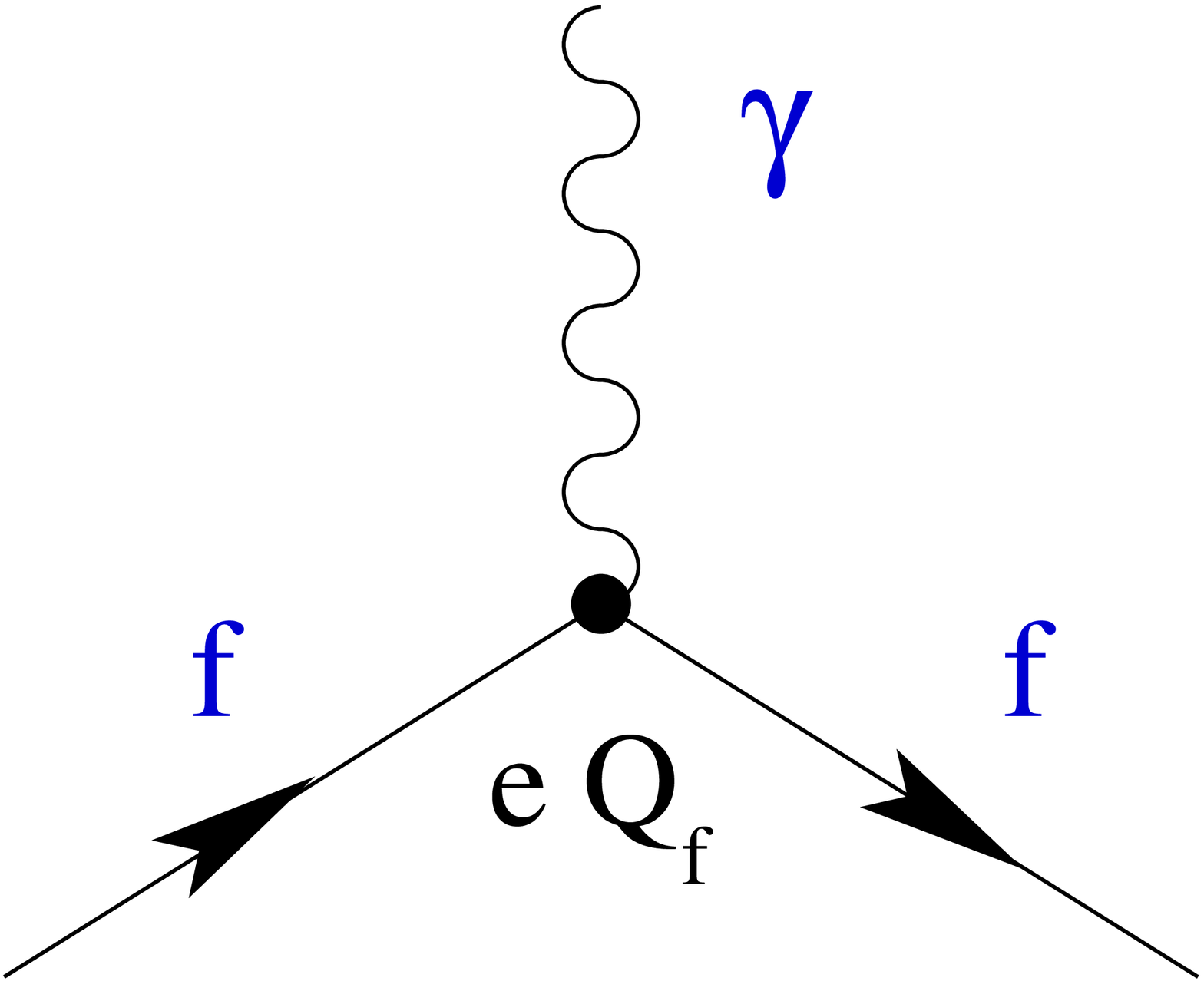}
\end{minipage}
\hskip 1cm
\begin{minipage}[t]{.4\linewidth}\centering
\includegraphics[width=4cm]{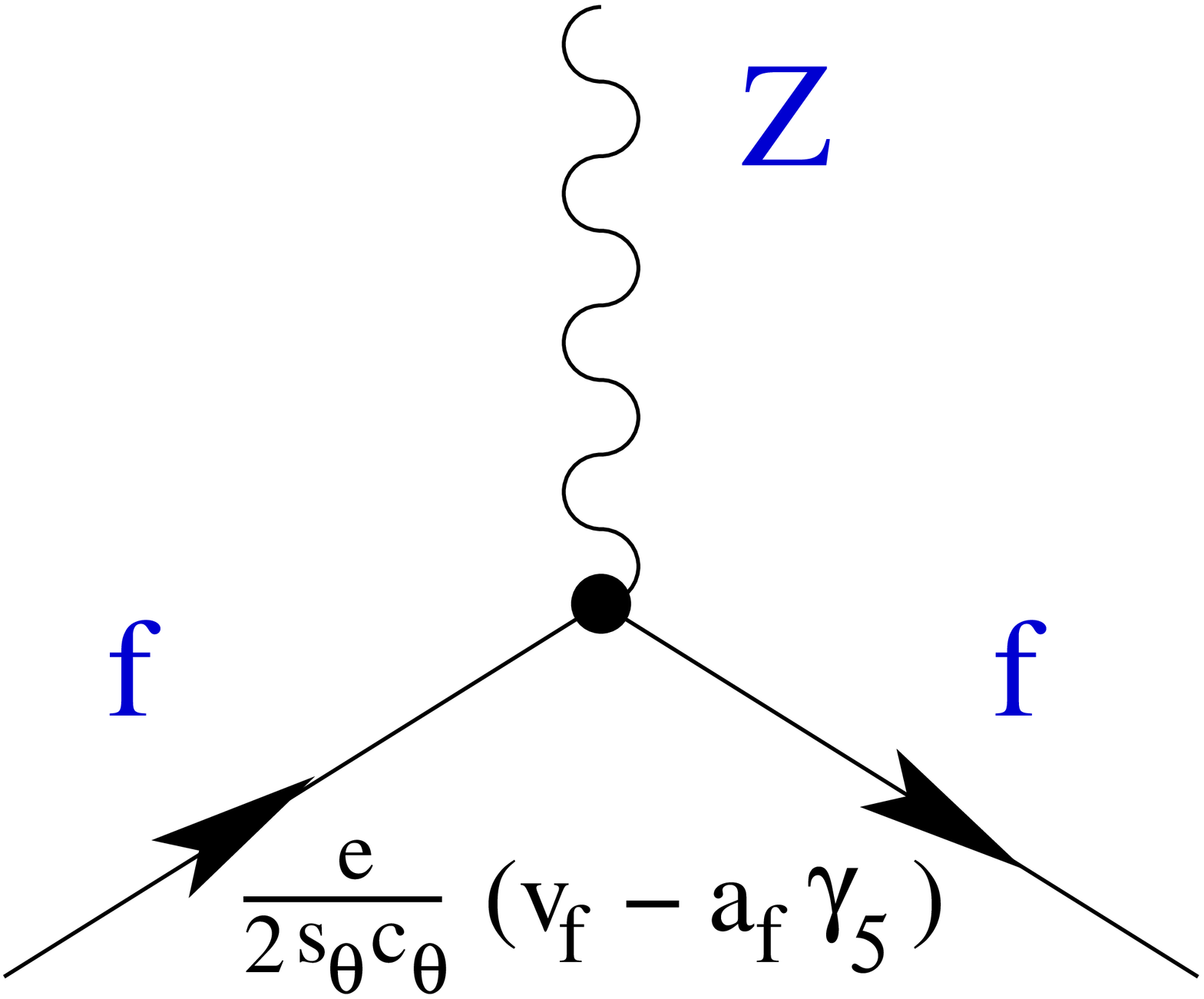}
\end{minipage}
\caption{Neutral-current interaction vertices.}
\label{fig:L_NC}
\end{figure}

Equation~\eqn{eq:int} contains also interactions with the neutral
gauge fields $W^3_\mu$ and $B_\mu$. We would like to identify these
bosons with the $Z$ and the $\gamma$. However, since the photon has
the same interaction with both fermion chiralities, the singlet
gauge boson $B_\mu$ cannot be equal to the electromagnetic field.
That would require \ $y_1 = y_2 = y_3$ \ and \ $\gp y_j = e\, Q_j$,
which cannot be simultaneously true.

Since both fields are neutral, we can try with
an arbitrary combination of them:
\bel{eq:Z_g_mixing}
\left(\ba W_\mu^3 \\ B_\mu\ea\right) \,\equiv\,
\left(\bat
\cos{\theta_W} & \sin{\theta_W} \\ -\sin{\theta_W} & \cos{\theta_W}
\ea\right) \, \left(\ba Z_\mu \\ A_\mu\ea\right)\, .
\ee
The physical $Z$ boson has a mass different from zero, which is
forbidden by the local gauge symmetry. We will see in the next
section how it is possible to generate non-zero boson masses,
through the SSB mechanism. For the moment, we just assume that
something breaks the symmetry, generating the $Z$ mass, and that the
neutral mass eigenstates are a mixture of the triplet and singlet
$SU(2)_L$ fields. In terms of the fields $Z$ and $\gamma$, the
neutral-current Lagrangian is given by
\be\label{eq:L_NC}
\cL_{\rms NC}\, =\, -
\sum_j\,\overline{\psi}_j\,\gamma^\mu\left\{ A_\mu\,
\left[ g\, {\sigma_3\over 2}\,\sin{\theta_W} + \gp\,y_j\,\cos{\theta_W}\right]
\, +\,
Z_\mu\,
\left[ g\, {\sigma_3\over 2}\,\cos{\theta_W} - \gp\,y_j\,\sin{\theta_W}\right]
\right\}\,\psi_j\, .
\ee
In order to get QED from the $A_\mu$ piece, one needs to impose the
conditions:
\bel{eq:unification}
g\,\sin{\theta_W} \, = \, \gp\,\cos{\theta_W}\, = \, e\, ,
\qquad\qquad\qquad
 Y \, = \, Q-T_3\, ,
\ee
where \ $T_3\equiv\sigma_3/2$ \ and \
$Q$ \ denotes the electromagnetic charge operator
\bel{eq:Q}
Q_1\,\equiv\,\left(\bat Q_{u/\nu}& 0 \\ 0 & Q_{d/e}\ea\right)\, ,
\qquad\qquad Q_2\, =\,Q_{u/\nu}\, ,\qquad\qquad Q_3\, =\,Q_{d/e}
\, .
\ee
The first equality relates the $SU(2)_L$ and $U(1)_Y$ couplings
to the electromagnetic coupling, providing the wanted unification
of the electroweak interactions. The second identity fixes
the fermion hypercharges in terms of their electric charge and
weak isospin quantum numbers:
$$  
\begin{array}{lccccc}
\mbox{Quarks:}\quad &
y_1\, =\, Q_u-{1\over 2}\, =\, Q_d+{1\over 2}\, =\, {1\over 6}\, ,
&\quad & y_2\, =\, Q_u\, =\, {2\over 3}\, ,
&\quad & y_3\, =\, Q_d\, =\, -{1\over 3}\, ,
\\[10pt]
\mbox{Leptons:}\quad &
y_1\, =\, Q_\nu-{1\over 2}\, =\, Q_e+{1\over 2}\, =\, -{1\over 2}\, ,
&\quad & y_2\, =\, Q_\nu\, =\, 0\, ,
&\quad & y_3\, =\, Q_e\, =\, -1\, .
\end{array}
$$  
A hypothetical right-handed neutrino would have both electric charge
and weak hypercharge equal to zero. Since it would not couple either
to the $W^\pm$ bosons, such a particle would not have any kind of
interaction (sterile neutrino). For aesthetic reasons, we shall then
not consider right-handed neutrinos any longer.

Using the relations \eqn{eq:unification},
the neutral-current Lagrangian can be written as
\bel{eq:L_NCb}
\cL_{\rms NC}\, = \,
\cL_{\rms QED}\, + \,
\cL_{\rms NC}^Z\, ,
\ee
where
\bel{eq:L_QED}
\cL_{\rms QED}\, = \,- e\,A_\mu\,\sum_j\,
\overline{\psi}_j\gamma^\mu Q_j\psi_j
\,\equiv\, - e\, A_\mu\, J^\mu_{\rms em}
\ee
is the usual QED Lagrangian and
\bel{eq:Z_NC}
\cL_{\rms NC}^Z \, =\,
- {e\over 2 \sin{\theta_W}\cos{\theta_W}} \,
J^\mu_Z\, Z_\mu
\ee
contains the interaction of the $Z$ boson with the neutral fermionic
current
\bel{eq:J_NC}
J^\mu_Z \,\equiv\, \sum_j\,\overline{\psi}_j\gamma^\mu
\left(\sigma_3-2\sin^2{\theta_W}Q_j\right)\psi_j
\, = \, J^\mu_3 - 2 \sin^2{\theta_W}\, J^\mu_{\rms em}\, .
\ee
In terms of the more usual fermion fields, $\cL_{\rms NC}^Z$ has the
form (Fig.~\ref{fig:L_NC})
\bel{eq:Z_Lagrangian}
\cL_{\rms NC}^Z\, = \,
- {e\over 2\sin{\theta_W}\cos{\theta_W}} \,Z_\mu\,\sum_f\,
\bar f \gamma^\mu (v_f-a_f\gamma_5) \, f\, ,
\ee
where \
$a_f =  T_3^f$ \ and \
$v_f = T_3^f \left( 1 - 4 |Q_f| \sin^2{\theta_W}\right)$.
Table~\ref{tab:nc_couplings} shows the neutral-current
couplings of the different fermions.

\begin{table}[t]
\begin{center}
{\renewcommand{\arraystretch}{1.5}
\caption{Neutral-current couplings.\label{tab:nc_couplings}}
\vspace{0.2cm}
\begin{tabular}{c@{\hspace{.75cm}}cccc}  
\hline\hline
 & $u$ & $d$ & $\;\nu_e\;$ & $e$
 \\ \hline
 $\, 2\, v_f\, $ & $1 -{8\over 3} \sin^2{\theta_W}$ &
 $-1 +{4\over 3} \sin^2{\theta_W}$ & $\,\, 1\,\,$ &
 $-1 +4 \sin^2{\theta_W}$
 \\
 $2\, a_f$ & $1$ & $-1$ & $1$ & $-1$
\\ \hline\hline
\end{tabular}}
\end{center}
\end{table}

\subsection{Gauge self-interactions}
\label{subsec:self-interactions}

\begin{figure}[tbh]\centering
\includegraphics[width=15.cm]{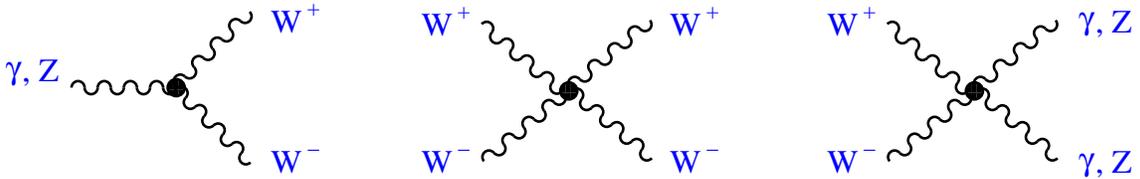}
\caption{Gauge boson self-interaction vertices.}
\label{fig:L_GSI}
\end{figure}

In addition to the usual kinetic terms, the Lagrangian
\eqn{eq:kinetic} generates cubic and quartic self-interactions among
the gauge bosons (Fig.~\ref{fig:L_GSI}):
\beqn\label{eq:cubic}
\cL_3 &\!\!\!\! = &\!\!\!\!
i e \cot{\theta_W}\left\{
\left(\partial^\mu W^\nu -\partial^\nu W^\mu\right)
 W^\dagger_\mu Z_\nu -
\left(\partial^\mu W^{\nu\dagger} -\partial^\nu W^{\mu\dagger}\right)
 W_\mu Z_\nu +
W_\mu W^\dagger_\nu\left(\partial^\mu Z^\nu -\partial^\nu Z^\mu\right)
\right\}
\no\\[10pt] &&\!\!\!\!\mbox{}
+i e \left\{
\left(\partial^\mu W^\nu -\partial^\nu W^\mu\right)
 W^\dagger_\mu A_\nu -
\left(\partial^\mu W^{\nu\dagger} -\partial^\nu W^{\mu\dagger}\right)
 W_\mu A_\nu +
W_\mu W^\dagger_\nu\left(\partial^\mu A^\nu -\partial^\nu A^\mu\right)
\right\}  ;
\no\\[8pt] &&\\[8pt] 
\cL_4 &\!\!\!\! = &\!\!\!\!\mbox{}
-{e^2\over 2\sin^2{\theta_W}}\left\{
\left(W^\dagger_\mu W^\mu\right)^2 - W^\dagger_\mu W^{\mu\dagger}
W_\nu W^\nu \right\}
- e^2 \cot^2{\theta_W}\,\left\{
W_\mu^\dagger W^\mu Z_\nu Z^\nu - W^\dagger_\mu Z^\mu W_\nu Z^\nu
\right\}
\no\\[10pt] &\!\!\!\! &\!\!\!\!\mbox{}
- e^2 \cot{\theta_W}\left\{
2 W_\mu^\dagger W^\mu Z_\nu A^\nu - W^\dagger_\mu Z^\mu W_\nu A^\nu
- W^\dagger_\mu A^\mu W_\nu Z^\nu
\right\}
\no\\[10pt] &\!\!\!\! &\!\!\!\!\mbox{}
- e^2\,\left\{
W_\mu^\dagger W^\mu A_\nu A^\nu - W^\dagger_\mu A^\mu W_\nu A^\nu
\right\} .
\no\eeqn
Notice that at least a pair of charged $W$ bosons are always present.
The $SU(2)_L$ algebra does not generate any
neutral vertex with only photons and $Z$ bosons.


\setcounter{equation}{0}
\section{Spontaneous Symmetry Breaking}
\label{sec:ssb}

\begin{figure}[tbh]\centering
\begin{minipage}[t]{.47\linewidth}
\includegraphics[width=7.25cm,clip]{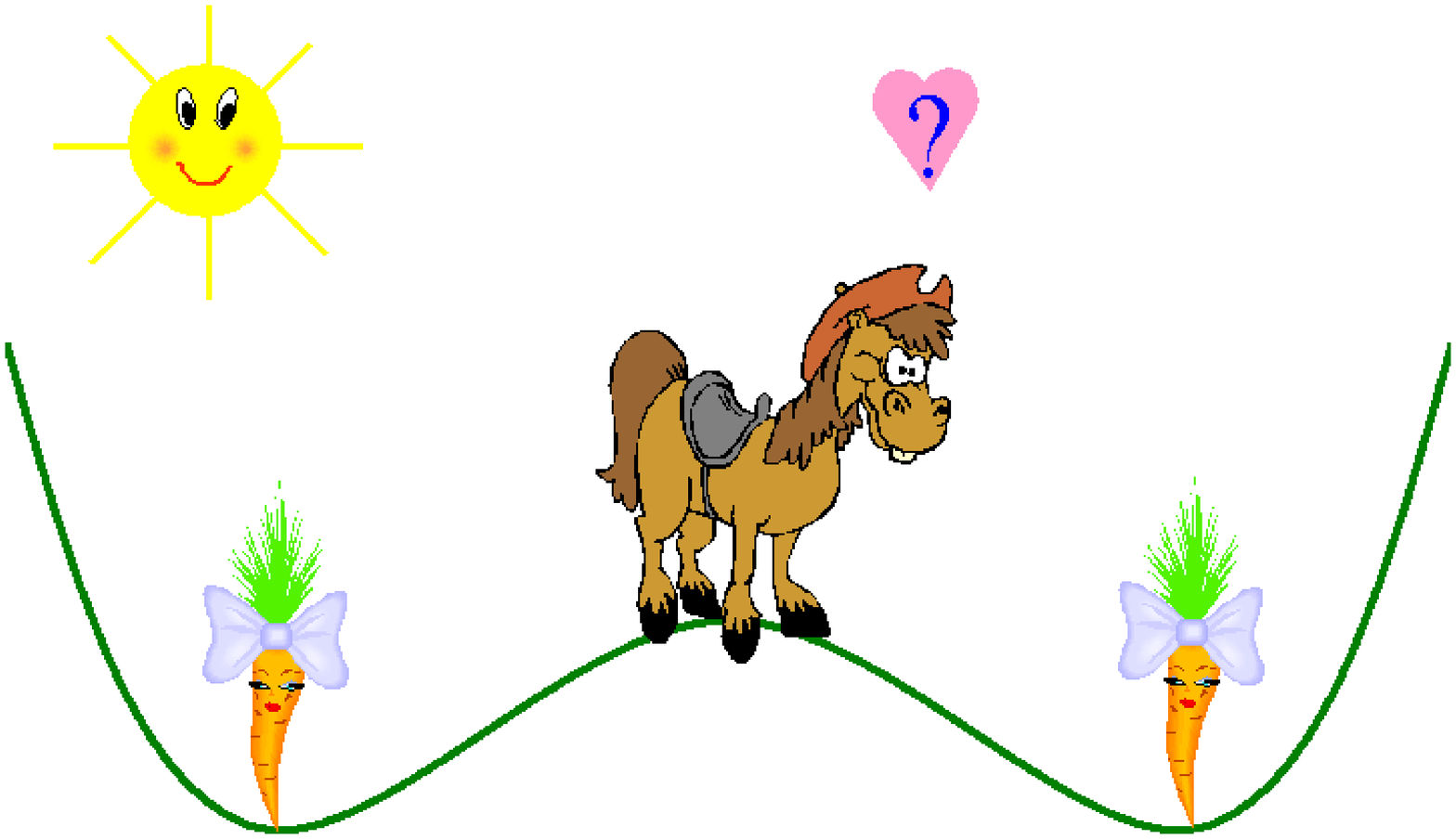}
\end{minipage}
\hfill
\begin{minipage}[t]{.47\linewidth}
\includegraphics[width=7.25cm,clip]{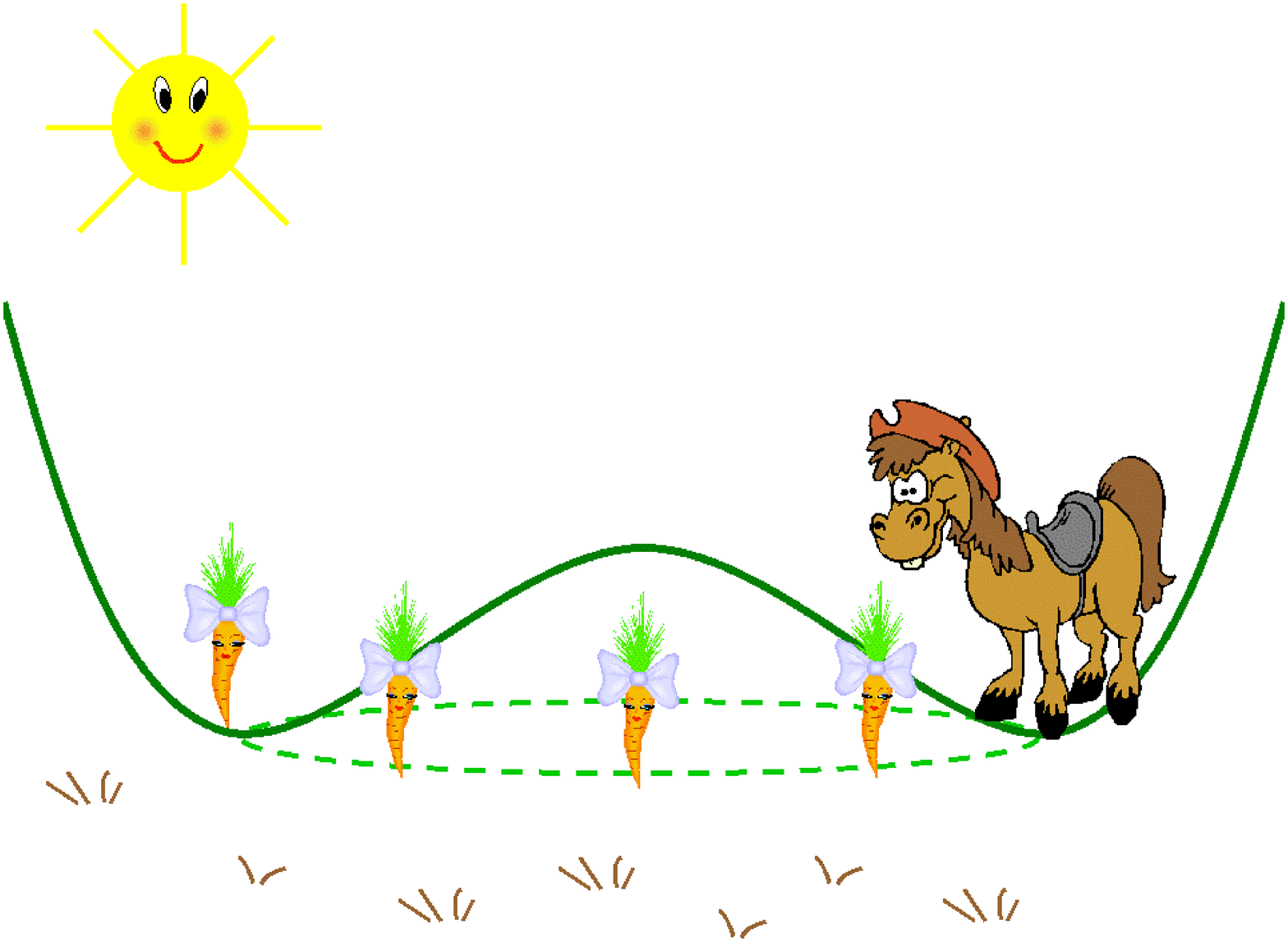}
\end{minipage}
\caption{Although Nicol\'as likes the symmetric food configuration,
he must break the symmetry deciding which carrot is more appealing.
In three dimensions, there is a continuous valley where Nicol\'as
can move from one carrot to the next without effort.}
\label{fig:Nicolas}
\end{figure}

So far, we have been able to derive charged- and neutral-current
interactions of the type needed to describe weak decays; we have
nicely incorporated QED into the same theoretical framework and,
moreover, we have got additional self-interactions of the gauge
bosons, which are generated by the non-Abelian structure of the
$SU(2)_L$ group. Gauge symmetry also guarantees that we have a
well-defined renormalizable Lagrangian. However, this Lagrangian has
very little to do with reality. Our gauge bosons are massless
particles; while this is fine for the photon field, the physical
$W^\pm$ and $Z$ bosons should be quite heavy objects.

In order to generate masses, we need to break the gauge symmetry in
some way; however, we also need a fully symmetric Lagrangian to
preserve renormalizability. This dilemma may be solved by the
possibility of getting non-symmetric results from an invariant
Lagrangian.

Let us consider a Lagrangian, which

\begin{enumerate}
\item Is invariant under a group $G$ of transformations.

\item Has a degenerate set of states with minimal energy,
which transform under $G$ as the members of a given multiplet.
\end{enumerate}

\noindent If one of those states is arbitrarily selected as the
ground state of the system, the symmetry is said to be spontaneously
broken.

A well-known physical example is provided by a ferromagnet: although
the Hamiltonian is invariant under rotations, the ground state has
the spins aligned into some arbitrary direction; moreover, any
higher-energy state, built from the ground state by a finite number
of excitations, would share this anisotropy. In a Quantum Field
Theory, the ground state is the vacuum; thus the SSB mechanism will
appear when there is a symmetric Lagrangian, but a non-symmetric
vacuum.

The horse in Fig.~\ref{fig:Nicolas} illustrates in a very simple way
the phenomenon of SSB. Although the left and right carrots are
identical, Nicol\'as must take a decision if he wants to get food.
What is important is not whether he goes left or right, which are
equivalent options, but that the symmetry gets broken. In two
dimensions (discrete left-right symmetry), after eating the first
carrot Nicol\'as would need to make an effort to climb the hill in
order to reach the carrot on the other side; however, in three
dimensions (continuous rotation symmetry) there is a marvelous flat
circular valley along which Nicol\'as can move from one carrot to
the next without any effort.

The existence of flat directions connecting the degenerate states of minimal
energy is a general property of the SSB of continuous symmetries.
In a Quantum Field Theory it implies the existence of massless degrees of freedom.

\subsection{Goldstone theorem}
\label{subsec:goldstone}

\begin{figure}[tbh]\centering
\begin{minipage}[c]{.4\linewidth}\centering
\includegraphics[width=5.3cm]{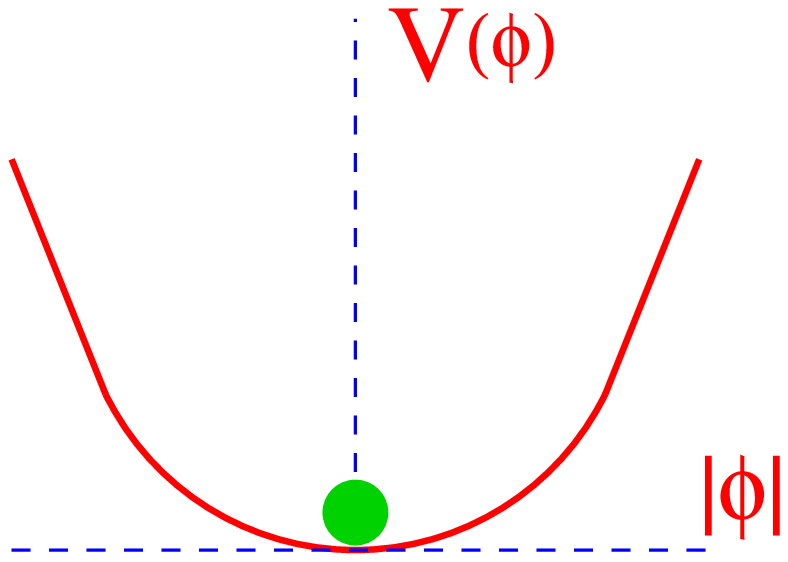}
\vskip .5cm\mbox{}
\end{minipage}
\hskip 1cm
\begin{minipage}[c]{.4\linewidth}\centering
\includegraphics[width=6cm]{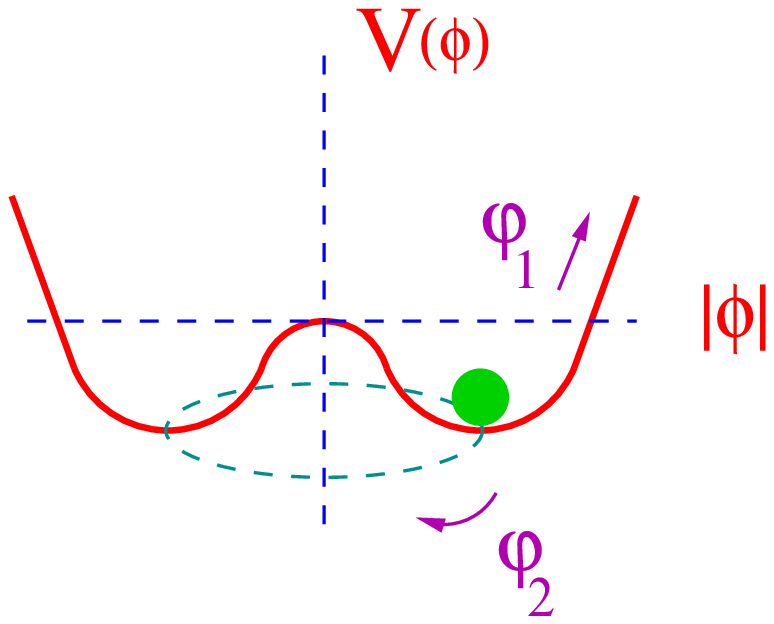}
\end{minipage}
\vskip -.3cm
\caption{Shape of the scalar potential for \ $\mu^2>0$ \ (left) \
and \ $\mu^2<0$ \ (right). In the second case there is a continuous
set of degenerate vacua, corresponding to different phases \ $\theta$,
connected through a massless field excitation \ $\varphi_2$.}
\label{fig:HiggsPot}
\end{figure}

Let us consider a complex scalar field $\phi(x)$, with
Lagrangian
\bel{eq:L_phi}
\cL\, = \, \partial_\mu\phi^\dagger \partial^\mu\phi - V(\phi)\, ,
\qquad\qquad
V(\phi)\, = \, \mu^2 \phi^\dagger\phi + h
\left(\phi^\dagger\phi\right)^2\, .
\ee
$\cL$ is invariant under global phase transformations
of the scalar field
\bel{eq:phi_transf}
\phi(x)\,\longrightarrow\,\phi'(x)\,\equiv\,
\exp{\left\{i\theta\right\}}\,\phi(x) \, .
\ee

In order to have a ground state the potential should be bounded from
below, i.e., $h>0$. For the quadratic piece there are two
possibilities, shown in Fig.~\ref{fig:HiggsPot}:
\vskip .1cm
\begin{enumerate}

\item \mbox{\boldmath $\mu^2>0$:} \
The potential has only the trivial minimum $\phi=0$.
It describes a massive scalar particle with mass $\mu$
and quartic coupling $h$.
\vskip .25cm

\item \mbox{\boldmath $\mu^2<0$:} \
The minimum is obtained for those field configurations
satisfying
\bel{eq:minimum}
|\phi_0|\, = \, \sqrt{{-\mu^2\over 2 h}} \,\equiv\, {v\over\sqrt{2}}
\, > \, 0 \, ,
\qquad\qquad\qquad V(\phi_0)\, =\, -{h\over 4} v^4\,  .
\ee
Owing to the $U(1)$ phase-invariance of the Lagrangian,
there is an infinite number of degenerate states of
minimum energy,
$\phi_0(x) = {v\over\sqrt{2}}\, \exp{\left\{i\theta\right\}}$.
By choosing a particular solution, $\theta=0$ for example, as the
ground state, the symmetry gets spontaneously broken.
If we parametrize the excitations over the ground state as
\bel{eq:perturbations}
\phi(x)\, \equiv \, {1\over\sqrt{2}}\,\left[
v + \varphi_1(x) + i\, \varphi_2(x)\right]\, ,
\ee
where $\varphi_1$ and $\varphi_2$ are real fields,
the potential takes the form
\bel{eq:pot}
V(\phi)\, = \, V(\phi_0) -\mu^2\varphi_1^2 +
h\, v \,\varphi_1 \left(\varphi_1^2+\varphi_2^2\right) +
{h\over 4} \left(\varphi_1^2+\varphi_2^2\right)^2\, .
\ee
Thus, $\varphi_1$ describes a massive state of mass
$m_{\varphi_1}^2 = -2\mu^2$, while $\varphi_2$ is massless.

\end{enumerate}
\vskip .1cm

The first possibility ($\mu^2>0$) is just the usual situation with a
single ground state. The other case, with SSB, is more interesting.
The appearance of a massless particle when $\mu^2<0$ is easy to
understand: the field $\varphi_2$ describes excitations around a
flat direction in the potential, i.e., into states with the same
energy as the chosen ground state. Since those excitations do not
cost any energy, they obviously correspond to a massless state.

The fact that there are massless excitations associated with the SSB
mechanism is a completely general result, known as the Goldstone
theorem \cite{goldstone}: if a Lagrangian is invariant under a
continuous symmetry group $G$, but the vacuum is only invariant
under a subgroup $H\subset G$, then there must exist as many
massless spin-0 particles (Goldstone bosons) as broken generators
(i.e., generators of $G$ which do not belong to $H$).

\subsection{The Higgs--Kibble mechanism}
\label{subsec:Higgs-Kibble}

At first sight, the Goldstone theorem has very little to do with our
mass problem; in fact, it makes it worse since we want massive
states and not massless ones. However, something very interesting
happens when there is a local gauge symmetry \cite{HI:66,KI:67}.

Let us consider \cite{WE:67}
an $SU(2)_L$ doublet of complex scalar fields
\bel{eq:scalar_multiplet}
\phi(x)\,\equiv\,\left(\ba \phi^{(+)}(x)\\ \phi^{(0)}(x)\ea\right)\, .
\ee
The gauged scalar Lagrangian of the Goldstone model in
Eq.~(\ref{eq:L_phi}),
\bel{eq:LS}
\cL_S \, = \, \left(D_\mu\phi\right)^\dagger D^\mu\phi
-\mu^2\phi^\dagger\phi - h \left( \phi^\dagger\phi\right)^2
\qquad\qquad\qquad (h>0\, ,\, \mu^2<0)\, ,
\ee
\bel{eq:DS}
D^\mu\phi \, = \,
\left[\partial^\mu
+i\, g\,\widetilde W^\mu
+ i\,\gp\, y_\phi\, B^\mu\right]\,\phi\; ,
\qquad\qquad\qquad
y_\phi\, =\, Q_\phi - T_3 \, =\, \frac{1}{2}\, ,
\ee
is invariant under local $SU(2)_L\otimes U(1)_Y$ transformations.
The value of the scalar hypercharge is fixed by the requirement of
having the correct couplings between $\phi(x)$ and $A^\mu(x)$; i.e.,
the photon does not couple to $\phi^{(0)}$, and $\phi^{(+)}$ has the
right electric charge.

The potential is very similar to the one considered before.
There is a infinite set of degenerate states with minimum
energy, satisfying
\bel{eq:vev}
\big|\langle 0|\phi^{(0)}|0\rangle\big|
\, = \, \sqrt{{-\mu^2\over 2 h}} \,\equiv\, {v\over\sqrt{2}} \, .
\ee
Note that we have made explicit the association of the classical
ground state with the quantum vacuum. Since the electric charge is a
conserved quantity, only the neutral scalar field can acquire a
vacuum expectation value. Once we choose a particular ground state,
the $SU(2)_L\otimes U(1)_Y$ symmetry gets spontaneously broken to
the electromagnetic subgroup $U(1)_{\rms QED}$, which by
construction still remains a true symmetry of the vacuum. According
to the Goldstone theorem three massless states should then appear.

Now, let us parametrize the scalar doublet in the general form
\bel{eq:parametrization}
\phi(x)\, = \, \exp{\left\{
i\, {\sigma_i\over 2}\,\theta^i(x)\right\}}
\; {1\over\sqrt{2}}\,
\left(\ba 0 \\ v + H(x) \ea\right)\, ,
\ee
with four real fields $\theta^i(x)$ and $H(x)$. The crucial point is
that the local $SU(2)_L$ invariance of the Lagrangian allows us to
rotate away any dependence on $\theta^i(x)$. These three fields are
precisely the would-be massless Goldstone bosons associated with the
SSB  mechanism.

The covariant derivative \eqn{eq:DS} couples the scalar
multiplet to the $SU(2)_L\otimes U(1)_Y$ gauge bosons.
If one takes the physical (unitary) gauge \ $\theta^i(x)=0\,$,
the kinetic piece of the scalar Lagrangian \eqn{eq:LS}
takes the form:
\bel{eq:unitary_gauge}
\left(D_\mu\phi\right)^\dagger D^\mu\phi\quad\toTheta\quad
{1\over 2}\, \partial_\mu H \partial^\mu H
+ (v+H)^2 \, \left\{ {g^2\over 4}\, W_\mu^\dagger W^\mu
+ {g^2\over 8\cos^2{\theta_W}}\, Z_\mu Z^\mu \right\}\, .
\ee
The vacuum expectation value of the neutral scalar has generated a
quadratic term for the $W^\pm$ and the $Z$, i.e., those gauge bosons
have acquired masses:
\bel{eq:boson_masses}
M_Z\,\cos{\theta_W}\, = \, M_W \, = \,  \frac{1}{2}\, v\, g \, .
\ee

Therefore, we have found a clever way of giving masses to the
intermediate carriers of the weak force. We just add $\cL_S$ to our
$SU(2)_L\otimes U(1)_Y$ model. The total Lagrangian is invariant
under gauge transformations, which guarantees the renormalizability
of the associated Quantum Field Theory \cite{TH:71}. However, SSB
occurs. The three broken generators give rise to three massless
Goldstone bosons which, owing to the underlying local gauge
symmetry, can be eliminated from the Lagrangian. Going to the
unitary gauge, we discover that the $W^\pm$ and the $Z$ (but not the
$\gamma$, because $U(1)_{\rms QED}$ is an unbroken symmetry) have
acquired masses, which are moreover related as indicated in
Eq.~\eqn{eq:boson_masses}. Notice that Eq.~(\ref{eq:Z_g_mixing}) has
now the meaning of writing the gauge fields in terms of the physical
boson fields with definite mass.

It is instructive to count the number of degrees of freedom
(d.o.f.). Before the SSB mechanism, the Lagrangian contains massless
$W^\pm$ and $Z$ bosons, i.e., $3\times 2 = 6$ d.o.f., due to the two
possible polarizations of a massless spin-1 field, and four real
scalar fields. After SSB, the three Goldstone modes are `eaten' by
the weak gauge bosons, which become massive and, therefore, acquire
one additional longitudinal polarization. We have then $3\times 3=9$
d.o.f. in the gauge sector, plus the remaining scalar particle $H$,
which is called the Higgs boson. The total number of d.o.f. remains
of course the same.

\subsection{Predictions}
\label{subsec:predictions}

We have now all the needed ingredients to describe the electroweak
interaction within a well-defined Quantum Field Theory.
Our theoretical framework implies the existence of massive
intermediate gauge bosons, $W^\pm$ and $Z$. Moreover,
the Higgs-Kibble mechanism has produced a precise
prediction\footnote{
Note, however, that the relation
$M_Z\cos{\theta_W} =  M_W$ has a more general validity.
It is a direct consequence of the symmetry properties of
$\cL_S$ and does not depend on its detailed dynamics.}
for the $W^\pm$ and $Z$ masses, relating them to the vacuum
expectation value of the scalar field through
Eq.~\eqn{eq:boson_masses}. Thus, $M_Z$ is predicted to be bigger
than $M_W$ in agreement with the measured masses
\cite{LEPEWWG,LEPEWWG_SLD:06}:
\bel{eq:WZmass} M_Z = 91.1875\pm 0.0021\:\mathrm{GeV}\, ,
\qquad\qquad M_W = 80.398\pm 0.025\:\mathrm{GeV}\, . \ee
From these experimental numbers, one obtains the electroweak
mixing angle
\bel{eq:thetaW} \sin^2{\theta_W}\, =\, 1 - {M_W^2\over M_Z^2} \, =\,
0.223\, . \ee

We can easily get and independent estimate of $\sin^2{\theta_W}$
from the decay $\mu^-\to e^-\bar\nu_e\,\nu_\mu$. The momentum
transfer \ $q^2 = (p_\mu - p_{\nu_\mu})^2 = (p_e + p_{\nu_e})^2
\lesssim m_\mu^2$ \ is much smaller than $M_W^2$. Therefore, the $W$
propagator in Fig.~\ref{fig:ChargedCurrents} shrinks to a point and
can be well approximated through a local four-fermion interaction,
i.e.,
\bel{eq:4fermion}
{g^2\over M_W^2 - q^2}\,\approx\, {g^2\over M_W^2}
\, =\, {4\pi\alpha\over\sin^2{\theta_W} M_W^2}
\,\equiv\, 4\sqrt{2}\, G_F\, .
\ee
The measured muon lifetime, $\tau_\mu = (2.197019\pm 0.000021)\cdot
10^{-6}$~s \cite{MuLan:07}, provides a very precise determination of the
Fermi coupling constant $G_F$:
\bel{eq:muonlifetime}
{1\over \tau_\mu}\, =\, \Gamma_\mu\, =\,
{G_F^2 m_\mu^5\over 192\,\pi^3}\, f(m_e^2/m_\mu^2)
\,\left( 1 +\delta_{\mathrm{RC}}\right)\, ,
\qquad
f(x)\,\equiv\, 1-8x+8x^3-x^4-12x^2\log{x}\, .
\ee
Taking into account the radiative corrections
$\delta_{\mathrm{RC}}$, which are known to $O(\alpha^2)$
\cite{MS:88,vRS:99}, one gets \cite{MuLan:07}:
\bel{eq:G_F}
G_F\, =\, (1.166371\pm 0.000006)\cdot 10^{-5}~\mathrm{GeV}^{-2}\, .
\ee
The measured values of $\alpha^{-1} = 137.035999710\, (96)$, $M_W$
and $G_F$ imply
\bel{eq:thetaW2}
\sin^2{\theta_W}\, =\, 0.215\, ,
\ee
in very good agreement with Eq.~\eqn{eq:thetaW}. We shall see later
that the small difference between these two numbers can be
understood in terms of higher-order quantum corrections. The Fermi
coupling gives also a direct determination of the electroweak scale,
i.e., the scalar vacuum expectation value:
\bel{eq:v}
v \, =\, \left(\sqrt{2}\, G_F\right)^{-1/2}\, =\, 246\:\mbox{\rm GeV}\, .
\ee

\subsection{The Higgs boson}
\label{subsec:Higgs}

\begin{figure}[htb]
\begin{center}
\includegraphics[width=9.25cm]{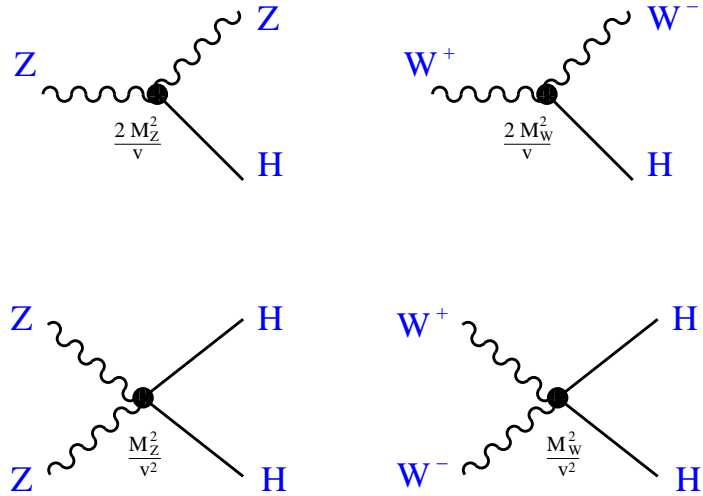}
\caption{Higgs couplings to the gauge bosons.}
\label{fig:HiggsWZcoup}
\end{center}
\end{figure}

The scalar Lagrangian in Eq.~\eqn{eq:LS} has introduced a new scalar
particle into the model: the Higgs $H$. In terms of the physical
fields (unitary gauge), $\cL_S$ takes the form
\bel{eq:H_lag}
\cL_S\, = \, {1\over 4}\, h\, v^4\, + \,\cL_H \, + \, \cL_{HG^2} \, ,
\ee
where
\beqn\label{eq:H_int}
&&\cL_H \, = \, {1\over 2}\, \partial_\mu H \partial^\mu H -
{1\over 2}\, M_H^2\, H^2
- {M_H^2\over 2 v}\, H^3 - {M_H^2\over 8 v^2}\, H^4\, ,
\\[5pt] \label{eq:HGG}
\cL_{HG^2} &\! = &\! M_W^2\, W_\mu^\dagger W^\mu\,
\left\{ 1+ {2\over v}\, H
+ {H^2\over v^2}\right\}\, + \,
{1\over 2}\, M_Z^2\, Z_\mu Z^\mu \,\left\{ 1+{2\over v}\, H
+ {H^2\over v^2}\right\}
\eeqn
and the Higgs mass is given by
\bel{H_mass}
M_H\, = \, \sqrt{-2\mu^2}\, =\, \sqrt{2 h}\, v \, .
\ee
The Higgs interactions (Fig.~\ref{fig:HiggsWZcoup}) have a very
characteristic form: they are always proportional to the mass
(squared) of the coupled boson. All Higgs couplings are determined
by $M_H$, $M_W$, $M_Z$ and the vacuum expectation value $v$.

So far the experimental searches for the Higgs have only provided
a lower bound on its mass, corresponding to the exclusion
of the kinematical range accessible at LEP and the Tevatron
\cite{PDG}:   
\bel{eq:MH_low}
M_H\, >\, 114.4\:\mathrm{GeV}\qquad (95\%\;\mathrm{C.L.})\, .
\ee

\subsection{Fermion masses}
\label{subsec:f_masses}

\begin{figure}[htb]
\begin{center}
\includegraphics[width=4cm]{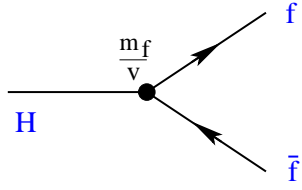}
\caption{Fermionic coupling of the Higgs boson.}
\label{fig:Hffcoup}
\end{center}
\end{figure}

A fermionic mass term \
$\cL_m = -m \,\overline{\psi}\psi = - m \left(\overline{\psi}_L\psi_R
+  \overline{\psi}_R\psi_L\right)$ \
is not allowed, because it breaks the gauge symmetry.
However, since we have introduced an additional scalar doublet into
the model, we can write the following gauge-invariant fermion-scalar coupling:
\bel{eq:yukawa}
\cL_Y\, =\, -c_1\, \left(\bar u , \bar d\right)_L
\left(\ba \phi^{(+)}\\ \phi^{(0)}\ea\right)\, d_R \, - \,
c_2\,\left(\bar u , \bar d\right)_L
\left(\ba \phi^{(0)*}\\ -\phi^{(-)}\ea\right)\, u_R \, - \,
c_3\,\left(\bar \nu_e , \bar e\right)_L
\left(\ba \phi^{(+)}\\ \phi^{(0)}\ea\right)\, e_R
\, +\, \mbox{\rm h.c.}\, ,
\ee
where the second term involves the $\cC$-conjugate scalar field
$\phi^c\equiv i\,\sigma_2\,\phi^*$. In the unitary gauge (after
SSB), this Yukawa-type Lagrangian takes the simpler form
\bel{eq:y_m}
\cL_Y\, =\, -{1\over\sqrt{2}} \,(v+H)\,\left\{ c_1 \,\bar d d + c_2
\,\bar u u + c_3 \,\bar e e\right\}\, .
\ee
Therefore, the SSB mechanism generates also fermion masses:
\bel{eq:f_masses}
 m_d\, = \, c_1\, {v\over\sqrt{2}} \; , \qquad
 m_u\, = \, c_2 \, {v\over\sqrt{2}} \; , \qquad
 m_e\, = \, c_3\, {v\over\sqrt{2}} \; .
\ee

Since we do not know the parameters $c_i$, the values of the fermion
masses are arbitrary. Note, however, that all Yukawa couplings are
fixed in terms of the masses (Fig.~\ref{fig:Hffcoup}):
\bel{eq:y_mass}
\cL_Y\, =\, -\left( 1 + {H\over v}\right)
 \,\left\{ m_d \,\bar d d + m_u \,\bar u u
+ m_e \,\bar e e\right\}\, .
\ee
%


\setcounter{equation}{0}
\section{Electroweak Phenomenology}
\label{sec:phenom}

In the gauge and scalar sectors, the SM Lagrangian contains only
four parameters: $g$, $\gp$, $\mu^2$ and $h$. One could trade them
by $\alpha$, $\theta_W$, $M_W$ and $M_H$. Alternatively, we can
choose as free parameters:
\beqn\label{eq:SM_inputs}
G_F & = & (1.166\, 371 \pm 0.000\, 006) \cdot 10^{-5}\:\mathrm{GeV}^{-2}
\quad\mbox{\protect\cite{MuLan:07}}\, ,
\no\\
\alpha^{-1} & = & 137.035\, 999\, 710\pm 0.000\, 000\, 096
\quad\mbox{\protect\cite{GHKNO:06}}\, ,
\\
M_Z & = & (91.1875\pm 0.0021)\,\mathrm{GeV}
\quad\mbox{\protect\cite{LEPEWWG,LEPEWWG_SLD:06}} \no\eeqn
and the Higgs mass $M_H$. This has the advantage of using the three
most precise experimental determinations to fix the interaction.
The relations
\bel{eq:A_def}
\sin^2{\theta_W}\,  =\,  1 - {M_W^2\over M_Z^2}\, ,
\qquad\qquad\qquad
M_W^2 \sin^2{\theta_W}\, =\, {\pi\alpha\over\sqrt{2}\, G_F}\,
\ee
determine then \ $\sin^2{\theta_W} = 0.212$ \ and \ $M_W= 80.94\;\mathrm{GeV}$.
The predicted $M_W$ is in
good agreement with the measured value in \eqn{eq:WZmass}.

\begin{figure}[tbh]\centering
\begin{minipage}[c]{.4\linewidth}\centering
\includegraphics[width=4cm]{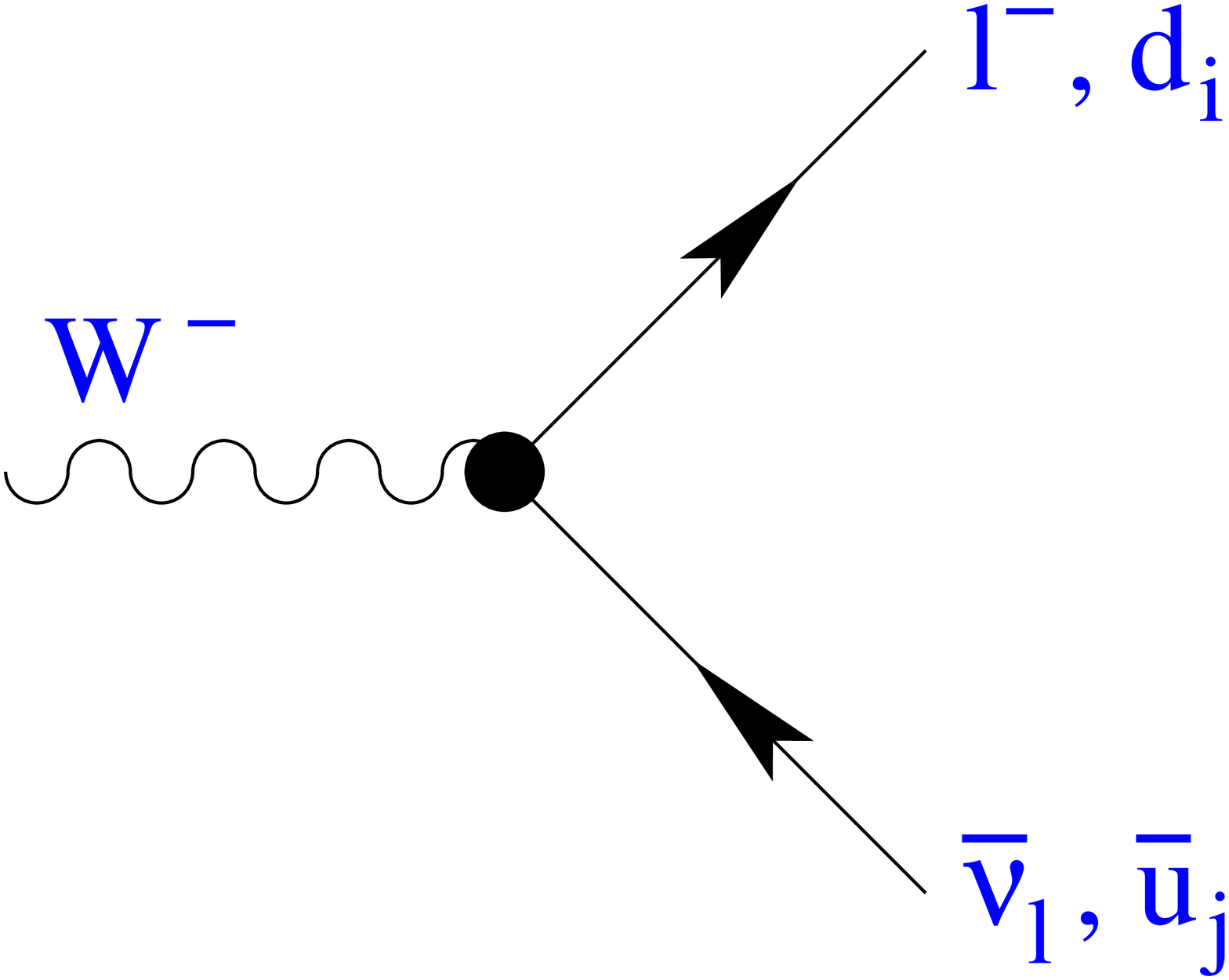}
\end{minipage}
\hskip 1cm
\begin{minipage}[c]{.4\linewidth}\centering
\includegraphics[width=3.4cm]{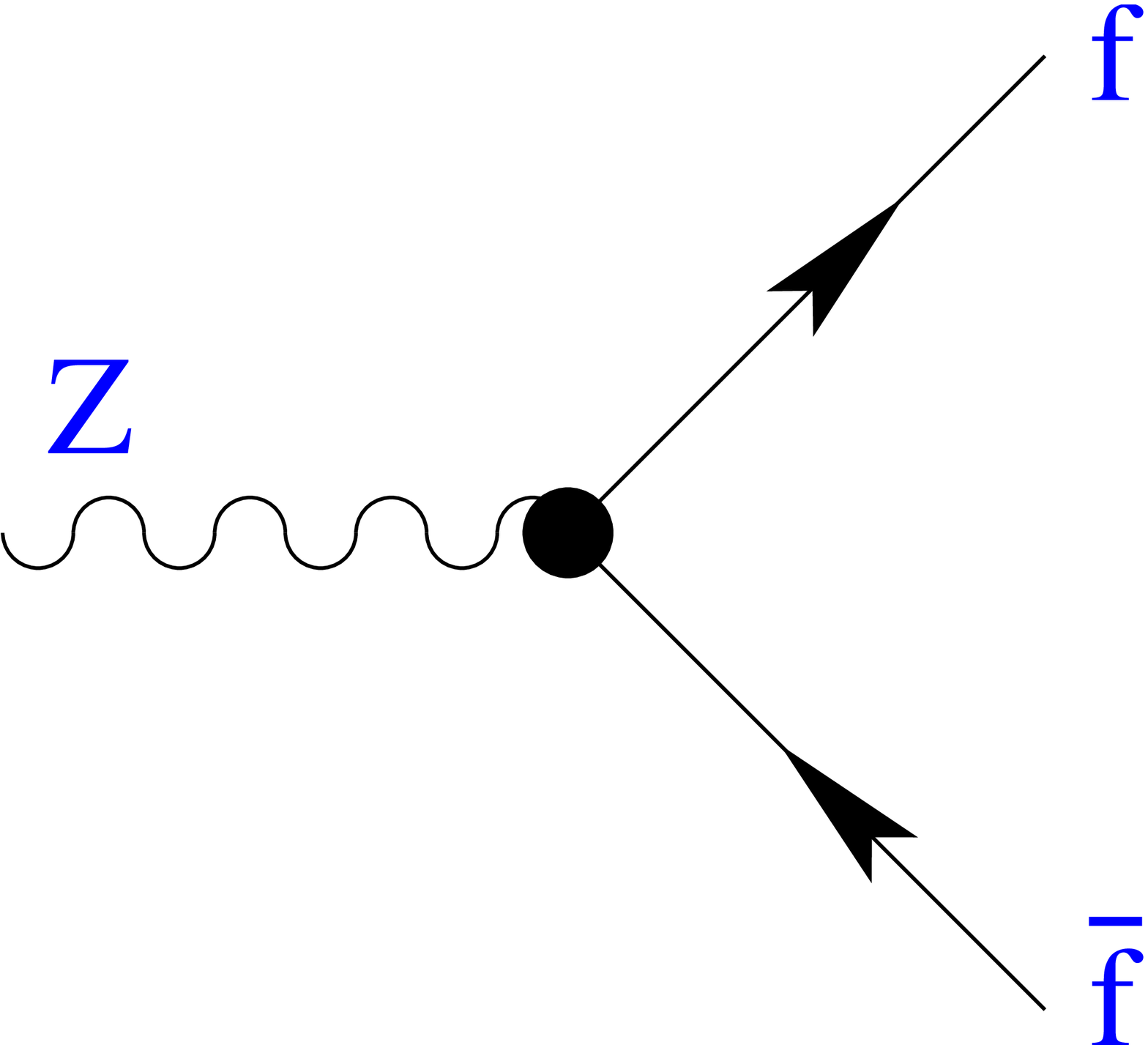}
\end{minipage}
\caption{Tree-level Feynman diagrams contributing to the $W^\pm$ and $Z$ decays.}
\label{fig:WZdecay}
\end{figure}

At tree level (Fig.~\ref{fig:WZdecay}), the decay widths of the weak
gauge bosons can be easily computed. The $W$ partial widths,
\bel{eq:W_width}
\Gamma\left(W^-\to \bar\nu_l l^-\right)\,
=\, {G_F M_W^3\over 6\pi\sqrt{2}} \, ,
\qquad\qquad\qquad
\Gamma\left(W^-\to \bar u_i d_j\right)\,
=\, N_C\; |\mathbf{V}_{\! ij}|^2\; {G_F M_W^3\over 6\pi\sqrt{2}}
\, ,
\ee
are equal for all leptonic decay modes (up to small kinematical mass
corrections). The quark modes involve also the
colour quantum number $N_C=3$ and the mixing factor $\mathbf{V}_{\! ij}$
relating weak and mass eigenstates, $d\,'_i = \mathbf{V}_{\! ij}\, d_j$.
The $Z$ partial widths are different for each decay mode, since
its couplings depend on the fermion charge:
\bel{eq:Z_width}
\Gamma\left( Z\to \bar f f\right)\, =\, N_f\,
{G_F M_Z^3\over 6\pi\sqrt{2}} \, \left(|v_f|^2 + |a_f|^2\right)
\, ,
\ee
where $N_l=1$ and $N_q=N_C$. Summing over all possible final fermion
pairs, one predicts the total widths \ $\Gamma_W=2.09$ GeV \ and \
$\Gamma_Z=2.48$ GeV, in excellent agreement with the experimental
values $\Gamma_W=(2.147\pm 0.060)$ GeV \ and \ $\Gamma_Z=(2.4952\pm
0.0023)$ GeV \cite{LEPEWWG,LEPEWWG_SLD:06}.

The universality of the $W$ couplings implies
\bel{eq:W_lep}
\mathrm{Br}(W^-\to\bar\nu_l\, l^-)\, = \,
{1\over 3 + 2 N_C}\, = \, 11.1 \% \, ,
\ee
where we have taken into account that the decay into the top quark
is kinematically forbidden. Similarly, the leptonic decay widths of
the $Z$ are predicted to be \ $\Gamma_l\equiv\,\Gamma(Z\to l^+l^-) =
84.85\;\mathrm{MeV}$. As shown in Table~\ref{tab:leptonic_modes},
these predictions are in good agreement with the measured leptonic
widths, confirming the universality of the $W$ and $Z$ leptonic
couplings. There is, however, an excess of the branching ratio
$W\to\tau\,\bar\nu_\tau$ \ with respect to \ $W\to e\,\bar\nu_e$ \
and \ $W\to\mu\,\bar\nu_\mu\,$, which represents a $2.8\,\sigma$
effect \cite{LEPEWWG,LEPEWWG_SLD:06}.

The universality of the leptonic $W$ couplings can also be tested
indirectly, through weak decays mediated by charged-current
interactions. Comparing the measured decay widths of leptonic or
semileptonic decays which only differ by the lepton flavour, one can
test experimentally that the $W$ interaction is indeed the same,
i.e., that \ $g_e = g_\mu = g_\tau \equiv g\, $. As shown in
Table~\ref{tab:univtm}, the present data verify
the universality of the leptonic charged-current couplings to the
0.2\% level.

\begin{table}[htb]
\begin{center}
{\renewcommand{\arraystretch}{1.3}
\caption{Measured values of \
$\mbox{\rm Br} (W^-\to\bar\nu_l\; l^-)$ \ and \ $\Gamma(Z\to l^+l^-)$
\protect\cite{LEPEWWG,LEPEWWG_SLD:06}. The average of the three
leptonic modes is shown in the last column (for a massless charged
lepton $l$). \label{tab:leptonic_modes}} \vspace{0.2cm}
\begin{tabular}{c@{\hspace{.75cm}}cccc}    
 \hline\hline & $e$ & $\mu$ & $\tau$ & $l$
 \\ \hline
Br($W^-\to\bar\nu_l l^-$) \, (\%) & $10.65\pm 0.17$ & $10.59\pm
0.15$ & $11.44\pm 0.22$ & $10.84\pm 0.09$
\\
$\Gamma(Z\to l^+l^-)$ \, (MeV) & $83.92\pm 0.12$ & $83.99\pm 0.18$ &
$84.08\pm 0.22$ & $83.985\pm 0.086$
\\ \hline\hline
\end{tabular}}
\end{center}
\end{table}

\begin{table}[htb]
\caption{Experimental determinations of the ratios \ $g_l/g_{l'}$
\cite{PI:07,Kl3}}
\begin{center}
\renewcommand{\arraystretch}{1.3}
\begin{tabular}{c@{\hspace{.5cm}}cccc}
 \hline\hline &
 $\Gamma_{\tau\to\nu_\tau e\,\bar\nu_e}/\Gamma_{\mu\to\nu_\mu e\,\bar\nu_e}$ &
 $\Gamma_{\tau\to\nu_\tau\pi}/\Gamma_{\pi\to\mu\,\bar\nu_\mu}$ &
 $\Gamma_{\tau\to\nu_\tau K}/\Gamma_{K\to\mu\,\bar\nu_\mu}$ &
 $\Gamma_{W\to\tau\,\bar\nu_\tau}/\Gamma_{W\to\mu\,\bar\nu_\mu}$
 \\ \hline
 $|g_\tau/g_\mu|$ & $1.0004\pm 0.0022$ & $0.996\pm 0.005$ &
 $0.979\pm 0.017$ & $1.039\pm 0.013$
 \\ \hline\hline &
 $\Gamma_{\tau\to\nu_\tau\mu\,\bar\nu_\mu}/
 \Gamma_{\tau\to\nu_\tau e\,\bar\nu_e}$ &
 $\Gamma_{\pi\to\mu\,\bar\nu_\mu} /\Gamma_{\pi\to e\,\bar\nu_e}$ &
 $\Gamma_{K\to\mu\,\bar\nu_\mu} /\Gamma_{K\to e\,\bar\nu_e}$ &
 $\Gamma_{K\to\pi\mu\,\bar\nu_\mu} /\Gamma_{K\to\pi e\,\bar\nu_e}$
 \\ \hline
 $|g_\mu/g_e|$ & $1.0000\pm 0.0020$ & $1.0017\pm 0.0015$ &
 $1.012\pm 0.009$ & $1.0002\pm 0.0026$
 \\ \hline\hline &
 $\Gamma_{W\to\mu\,\bar\nu_\mu} /\Gamma_{W\to e\,\bar\nu_e}$ &
 \multicolumn{1}{||c}{} &
 $\Gamma_{\tau\to\nu_\tau \mu\,\bar\nu_\mu}/\Gamma_{\mu\to\nu_\mu e\,\bar\nu_e}$
 & $\Gamma_{W\to\tau\,\bar\nu_\tau}/\Gamma_{W\to e\,\bar\nu_e}$
 \\ \hline
 $|g_\mu/g_e|$ & $0.997\pm 0.010$ &
 \multicolumn{1}{||c}{$|g_\tau/g_e|$} &
 $1.0004\pm 0.0023$ & $1.036\pm 0.014$\hfill
 \\ \hline\hline
\end{tabular}
\end{center}
\label{tab:univtm}
\end{table}

Another interesting quantity is the $Z$ decay width into invisible modes,
\bel{eq:Z_inv}
 {\Gamma_{\rms inv}\over\Gamma_l}\,\equiv\,
{N_\nu\;\Gamma(Z\to\bar\nu\,\nu)\over\Gamma_l}\, = \,
{2\, N_\nu\over (1 - 4\, \sin^2{\theta_W})^2 + 1}\; ,
\ee
which is usually normalized to the charged leptonic width. The
comparison with the measured value, $\Gamma_{\rms inv}/\Gamma_l =
5.942 \pm 0.016$  \cite{LEPEWWG,LEPEWWG_SLD:06}, provides very
strong experimental evidence for the existence of three different
light neutrinos.

\subsection{Fermion-pair production at the $Z$ peak}
\label{subsec:Zpeak}

\begin{figure}[tbh]\centering
\includegraphics[width=10.5cm]{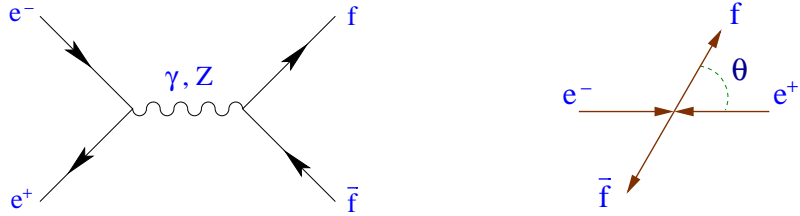}
\caption{Tree-level contributions to \ $e^+e^-\to\bar f f \,$ \
and kinematical configuration in the centre-of-mass system.}
\label{fig:eeZff}
\end{figure}

Additional information can be obtained from the study of the process
\ $e^+e^-\to\gamma,Z\to\bar f f \,$ (Fig.~\ref{fig:eeZff}).
For unpolarized $e^+$ and $e^-$ beams, the differential 
cross-section can be written, at lowest order, as
\bel{eq:dif_cross}
{d\sigma\over d\Omega}\, = \, {\alpha^2\over 8 s} \, N_f \,
         \left\{ A \, (1 + \cos^2{\theta}) \, + B\,  \cos{\theta}\,
     - \, h_f \left[ C \, (1 + \cos^2{\theta}) \, +\, D \cos{\theta}
         \right] \right\} ,
\ee
where \ $h_f=\pm 1$ \ denotes the sign of the helicity of the
produced fermion $f$, and $\theta$ is the scattering angle between
$e^-$ and $f$ in the centre-of-mass system. Here,
\beqn\label{eq:A}
A & = & 1 + 2\, v_e v_f \,\mathrm{Re}(\chi)
 + \left(v_e^2 + a_e^2\right) \left(v_f^2 + a_f^2\right) |\chi|^2\,,
\no\\ \label{eq:B}
B & = & 4\, a_e a_f \,\mathrm{Re}(\chi) + 8\, v_e a_e v_f a_f\,  |\chi|^2\, ,
\no\\ \label{eq:C}
C & = & 2\, v_e a_f \,\mathrm{Re}(\chi) + 2\, \left(v_e^2 + a_e^2\right) v_f
      a_f\, |\chi|^2\, ,
\no\\ \label{eq:D}
D & = & 4\, a_e v_f \,\mathrm{Re}(\chi) + 4\, v_e a_e \left(v_f^2 +
      a_f^2\right) |\chi|^2\,  ,
\eeqn
and  $\chi$  contains the $Z$  propagator
\bel{eq:Z_propagator}
\chi \, = \, {G_F M_Z^2 \over 2 \sqrt{2} \pi \alpha }
     \; {s \over s - M_Z^2 + i s \Gamma_Z  / M_Z } \, .
\ee

The coefficients $A$, $B$, $C$ and $D$ can be experimentally
determined by measuring the total cross-section, the
forward--backward asymmetry, the polarization asymmetry, and the
forward--backward polarization asymmetry, respectively:
$$
\sigma(s)\, =\,
{4 \pi \alpha^2 \over 3 s } \, N_f \, A \, ,
\qquad\qquad\qquad\qquad
\cA_{\rms FB}(s)\,\equiv\, {N_F - N_B \over N_F + N_B}
   \, =\,  {3 \over 8} {B \over A}\, ,
$$

\bel{eq:A_pol}
\cA_{\rms Pol}(s) \,\equiv\,
{\sigma^{(h_f =+1)}
- \sigma^{(h_f =-1)} \over \sigma^{(h_f =+1)} + \sigma^{(h_f = -1)}}
\, =\,  - {C \over A} \, ,
\ee

$$
\cA_{\rms FB,Pol}(s)\, \equiv\,
{N_F^{(h_f =+1)} -
N_F^{(h_f = -1)} - N_B^{(h_f =+1)} + N_B^{(h_f = -1)} \over
 N_F^{(h_f =+1)} + N_F^{(h_f = -1)} + N_B^{(h_f =+1)} + N_B^{(h_f = -1)}}
\, =\,  -{3 \over 8} {D \over A}\, .
$$

\noindent Here, $N_F$ and $N_B$ denote the number of $f$'s emerging
in the forward and backward hemispheres, respectively, with respect
to the electron direction. The measurement of the final fermion
polarization can be done for $f=\tau$ by measuring the distribution
of the final $\tau$ decay products.

For $s = M_Z^2$, the real part of the $Z$ propagator vanishes and
the photon-exchange terms can be neglected in comparison with the
$Z$-exchange contributions ($\Gamma_Z^2 / M_Z^2 << 1$). Equations
\eqn{eq:A_pol} become then,
$$
\sigma^{0,f} \,\equiv\,\sigma(M_Z^2)\, =\,
 {12 \pi  \over M_Z^2 } \; {\Gamma_e \Gamma_f\over\Gamma_Z^2}\, ,
\qquad\qquad\qquad
\cA_{\rms FB}^{0,f}\,\equiv\,\cA_{FB}(M_Z^2)\, =\, {3 \over 4}\,
\cP_e \cP_f \, ,
$$
\bel{eq:A_pol_Z}
\cA_{\rms Pol}^{0,f} \,\equiv\,
  \cA_{\rms Pol}(M_Z^2)\, =\, \cP_f \, ,
\qquad\qquad\qquad
\cA_{\rms FB,Pol}^{0,f} \,\equiv\,
\cA_{\rms FB,Pol}(M_Z^2)\, =\,  {3 \over 4}\, \cP_e  \, ,
\ee
where $\Gamma_f$ is the $Z$ partial decay width into the $\bar f f$ final state,
and
\bel{eq:P_f}
\cP_f \,\equiv \, - A_f \,\equiv \,
{ - 2\, v_f a_f \over v_f^2 + a_f^2}
\ee
is the average longitudinal polarization of the fermion $f$,
which only depends on the ratio of the vector and axial-vector couplings.

With polarized $e^+e^-$ beams, which have been available at SLC, one
can also study the left--right asymmetry between the cross-sections
for initial left- and right-handed electrons, and the corresponding
forward--backward left--right asymmetry:
\bel{eq:A_LR}
\cA_{\rms LR}^0\,\equiv\,
\cA_{\rms LR}(M_Z^2)
  \, = \, {\sigma_L(M_Z^2)
- \sigma_R(M_Z^2) \over \sigma_L(M_Z^2) + \sigma_R(M_Z^2)}
\, = \, - \cP_e \,  ,
\qquad\quad
\cA_{\rms FB,LR}^{0,f} \,\equiv\,
\cA_{\rms FB,LR}(M_Z^2)\, =\, - {3 \over 4}\, \cP_f \, .
\ee
At the $Z$ peak, $\cA_{\rms LR}^0$ measures
the average initial lepton polarization, $\cP_e$,
without any need for final particle identification,
while $\cA_{\rms FB,LR}^{0,f}$ provides a direct determination
of the final fermion polarization.

$\cP_f$ is a very sensitive function of $\sin^2{\theta_W}$. Small
higher-order corrections can produce large variations on the
predicted lepton polarization because $|v_l| =\frac{1}{2}\,
|1-4\,\sin^2{\theta_W}|\ll 1$. Therefore, $\cP_l$ provides an
interesting window to search for electroweak quantum effects.

\subsection{QED and QCD corrections}
\label{subsec:QED_QCD_loops}

\begin{figure}[tbh]\centering
\begin{minipage}[c]{5.25cm}\centering
\includegraphics[width=5.cm]{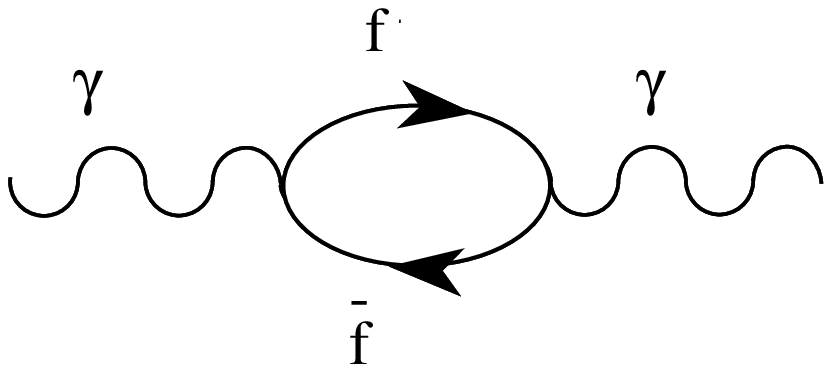}
\end{minipage}
\hskip 1.5cm
\begin{minipage}[c]{8.25cm}\centering
\includegraphics[width=8cm]{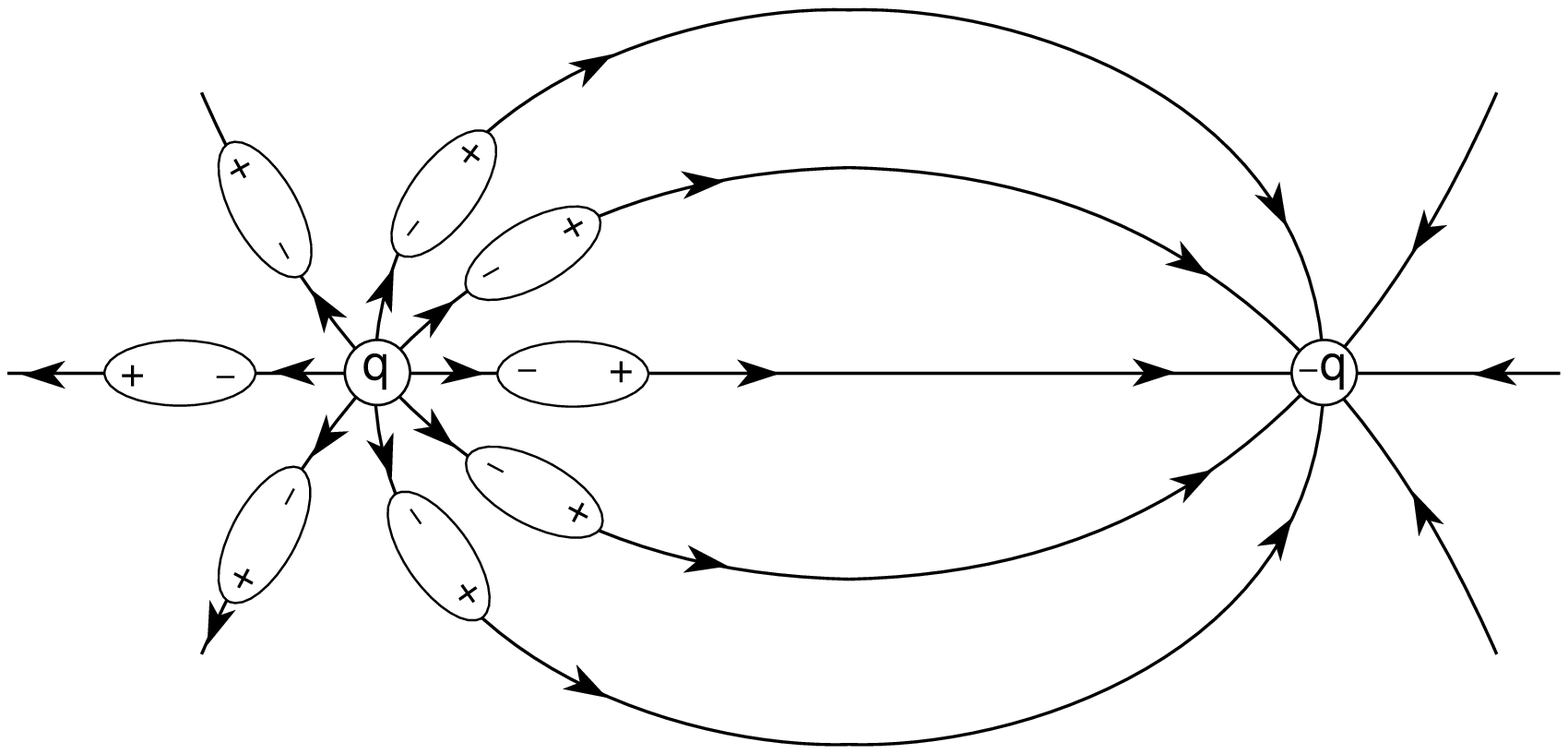}
\end{minipage}
\caption{The photon vacuum polarization (left)
generates a charge screening effect, making $\alpha(s)$ smaller
at larger distances.}
\label{fig:PhotonPolarization}
\end{figure}

Before trying to analyse the relevance of higher-order electroweak
contributions, it is instructive to consider the numerical impact of
the well-known QED and QCD corrections.
The photon propagator gets vacuum polarization corrections, induced
by virtual fermion--antifermion pairs. This kind of QED loop
corrections can be taken into account through a redefinition of the
QED coupling, which depends on the energy scale. The resulting QED
running coupling $\alpha(s)$ decreases at large distances. This can
be intuitively understood as the charge screening generated by the
virtual fermion pairs (Fig.~\ref{fig:PhotonPolarization}). The
physical QED vacuum behaves as a polarized dielectric medium.
The huge difference between the electron and $Z$ mass scales makes
this quantum correction relevant at LEP energies
\cite{GHKNO:06,LEPEWWG,LEPEWWG_SLD:06}:
\begin{equation}\label{eq:alpha}
\alpha(m_e^2)^{-1}\; =\; 137.035\, 999\, 710\, (96) \;\; > \;\;
\alpha(M_Z^2)^{-1}\; =\; 128.93\pm 0.05\; .
\end{equation}

The running effect generates an important change in Eq.~\eqn{eq:A_def}.
Since $G_F$ is measured at low energies, while $M_W$ is a
high-energy parameter, the relation between both quantities is
modified by vacuum-polarization contributions.
Changing $\alpha$ by $\alpha(M_Z^2)$,
one gets the corrected predictions:
\bel{eq:new_pred}
\sin^2{\theta_W}\, = \, 0.231 \, ,
\qquad\qquad\qquad
M_W\, = \, 79.96\,\mathrm{GeV} \, .
\ee
The experimental value of $M_W$ is in the range between the two
results obtained with either $\alpha$ or $\alpha(M_Z^2)$, showing
its sensitivity to quantum corrections. The effect is more
spectacular in the leptonic asymmetries at the $Z$ peak. The small
variation of $\sin^2{\theta_W}$ from 0.212 to 0.231 induces a large
shift on the vector $Z$ coupling to charged leptons from \ $v_l=
-0.076$ \ to \ $-0.038\,$, changing the predicted average lepton
polarization $\cP_l$ by a factor of two.

So far, we have treated quarks and leptons on an equal footing.
However, quarks are strong-interacting particles. The gluonic
corrections to the decays $Z\to\bar q q$ and $W^-\to\bar u_i d_j$
can be directly incorporated into the formulae given before by
taking an `effective' number of colours:
\bel{N_C_eff}
N_C \quad \Longrightarrow \quad
N_C\,\left\{ 1 + {\alpha_s\over\pi} + \ldots\right\}\,
\approx\, 3.115 \, ,
\ee
where we have used the value of $\alpha_s$ at $s=M_Z^2$, \
$\alpha_s(M_Z^2)= 0.119\pm 0.002\, $ \cite{PDG,BE:07}.

Note that the strong coupling also `runs'. However, the gluon
self-interactions generate an anti-screening effect, through
gluon-loop corrections to the gluon propagator, which spread out the
QCD charge \cite{PI:00}. Since this correction is larger than the screening of
the colour charge induced by virtual quark--antiquark pairs, the net
result is that the strong coupling decreases at short distances.
Thus, QCD has the required property of asymptotic freedom: quarks
behave as free particles when $Q^2\to\infty$ \cite{GW:73,PO:73}.

QCD corrections increase the probabilities of the $Z$ and the $W^\pm$
to decay into hadronic modes. Therefore, their leptonic branching fractions
become smaller. The effect can be easily estimated from Eq.~\eqn{eq:W_lep}.
The probability of the decay $W^-\to \bar\nu_e\, e^-$ gets reduced from
11.1\% to 10.8\%, improving the agreement with the measured value in
Table~\ref{tab:leptonic_modes}.

\subsection{Higher-order electroweak corrections}
\label{subsec:loops}

\begin{figure}[tbh]\centering
\includegraphics[width=10cm]{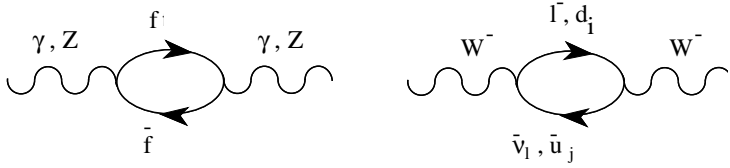}
\caption{Self-energy corrections to the gauge boson propagators.}
\label{fig:Oblique}
\end{figure}

Quantum corrections offer the possibility to be sensitive to heavy
particles, which cannot be kinematically accessed, through their
virtual loop effects. In QED and QCD the vacuum polarization
contribution of a heavy fermion pair is suppressed by inverse powers
of the fermion mass. At low energies, the information on the heavy
fermions is then lost. This `decoupling' of the heavy fields happens
in theories with only vector couplings and an exact gauge symmetry
\cite{AC:75}, where the effects generated by the heavy particles can
always be reabsorbed into a redefinition of the low-energy
parameters.

The SM involves, however, a broken chiral gauge symmetry. This has
the very interesting implication of avoiding the decoupling theorem
\cite{AC:75}. The vacuum polarization contributions induced by a
heavy top generate corrections to the $W^\pm$ and $Z$ propagators
(Fig.~\ref{fig:Oblique}), which increase quadratically with the top
mass \cite{VE:77}. Therefore, a heavy top does not decouple. For
instance, with $m_t = 171$ GeV, the leading quadratic correction to
the second relation in Eq.~\eqn{eq:A_def} amounts to a sizeable $3\%
$ effect. The quadratic mass contribution originates in the strong
breaking of weak isospin generated by the top and bottom quark
masses, i.e., the effect is actually proportional to $m_t^2-m_b^2$.

Owing to an accidental $SU(2)_C$ symmetry of the scalar sector
(the so-called custodial symmetry), the
virtual production of Higgs particles does not generate any
quadratic dependence on the Higgs mass at one loop \cite{VE:77}.
The dependence on $M_H$ is only logarithmic.
The numerical size of the corresponding correction in
Eq.~\eqn{eq:A_def} varies from a 0.1\% to a 1\% effect
for $M_H$ in the range from 100 to 1000 GeV.

\begin{figure}[htb]\centering
\includegraphics[width=9cm]{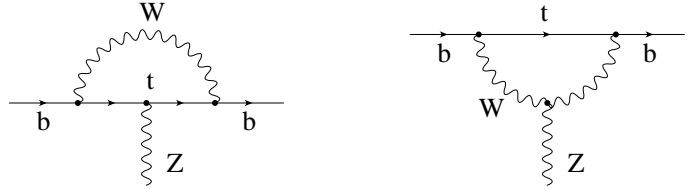}
\caption{One-loop corrections to the $Z\bar b b$ vertex,
involving a virtual top.}
\label{fig:Zbb}
\end{figure}

Higher-order corrections to the different electroweak couplings are
non-universal and usually smaller than the self-energy
contributions. There is one interesting exception, the $Z \bar bb$
vertex (Fig.~\ref{fig:Zbb}), which is sensitive to the top quark
mass \cite{BPS:88}. The $Z\bar f f$ vertex gets one-loop corrections
where a virtual $W^\pm$ is exchanged between the two fermionic legs.
Since the $W^\pm$ coupling changes the fermion flavour, the decays
$Z\to \bar d d, \bar s s, \bar b b$ \ get contributions with a top
quark in the internal fermionic lines, i.e., $Z\to\bar t t\to \bar
d_i d_i$. Notice that this mechanism can also induce the
flavour-changing neutral-current decays $Z\to \bar d_i d_j$ with
$i\not=j$. These amplitudes are suppressed by the small CKM mixing
factors $|\bV^{\phantom{*}}_{\! tj}\bV^*_{\! ti}|^2$. However, for
the $Z\to\bar b b$ vertex, there is no suppression because $|\bV_{\!
tb}|\approx 1$.

The explicit calculation \cite{BPS:88,ABR:86,BH:88,LS:90} shows the
presence of hard $m_t^2$ corrections to the $Z\to\bar b b$ vertex.
This effect can be easily understood \cite{BPS:88} in non-unitary
gauges where the unphysical charged scalar $\phi^{(\pm)}$ is
present. The fermionic couplings of the charged scalar are
proportional to the fermion masses; therefore the exchange of a
virtual $\phi^{(\pm)}$ gives rise to a $m_t^2$ factor. In the
unitary gauge, the charged scalar has been `eaten' by the $W^\pm$
field; thus the effect comes now from the exchange of a longitudinal
$W^\pm$, with terms proportional to $q^\mu q^\nu$ in the propagator
that generate fermion masses.
Since the $W^\pm$ couples only to left-handed fermions, the induced
correction is the same for the vector and axial-vector $Z\bar b b$
couplings and, for $m_t = 171$~GeV, amounts to a 1.6\% reduction of
the $Z\to \bar b b$ decay width \cite{BPS:88}.

The `non-decoupling' present in the $Z\bar b b$ vertex is quite
different from the one happening in the boson self-energies. The
vertex correction is not dependent on the Higgs mass. Moreover,
while any kind of new heavy particle coupling to the gauge bosons
would contribute to the $W$ and $Z$ self-energies, the possible new
physics contributions to the $Z\bar b b$ vertex are much more
restricted and, in any case, different. Therefore, the independent
experimental measurement of the two effects is very valuable in
order to disentangle possible new physics contributions from the SM
corrections. In addition, since the `non-decoupling' vertex effect
is related to $W_L$-exchange, it is sensitive to the SSB mechanism.

\subsection{SM electroweak fit}
\label{subsec:EWfit}

\begin{figure}[tbh]\centering
\begin{minipage}[c]{.45\linewidth}\centering
\includegraphics[width=7.cm]{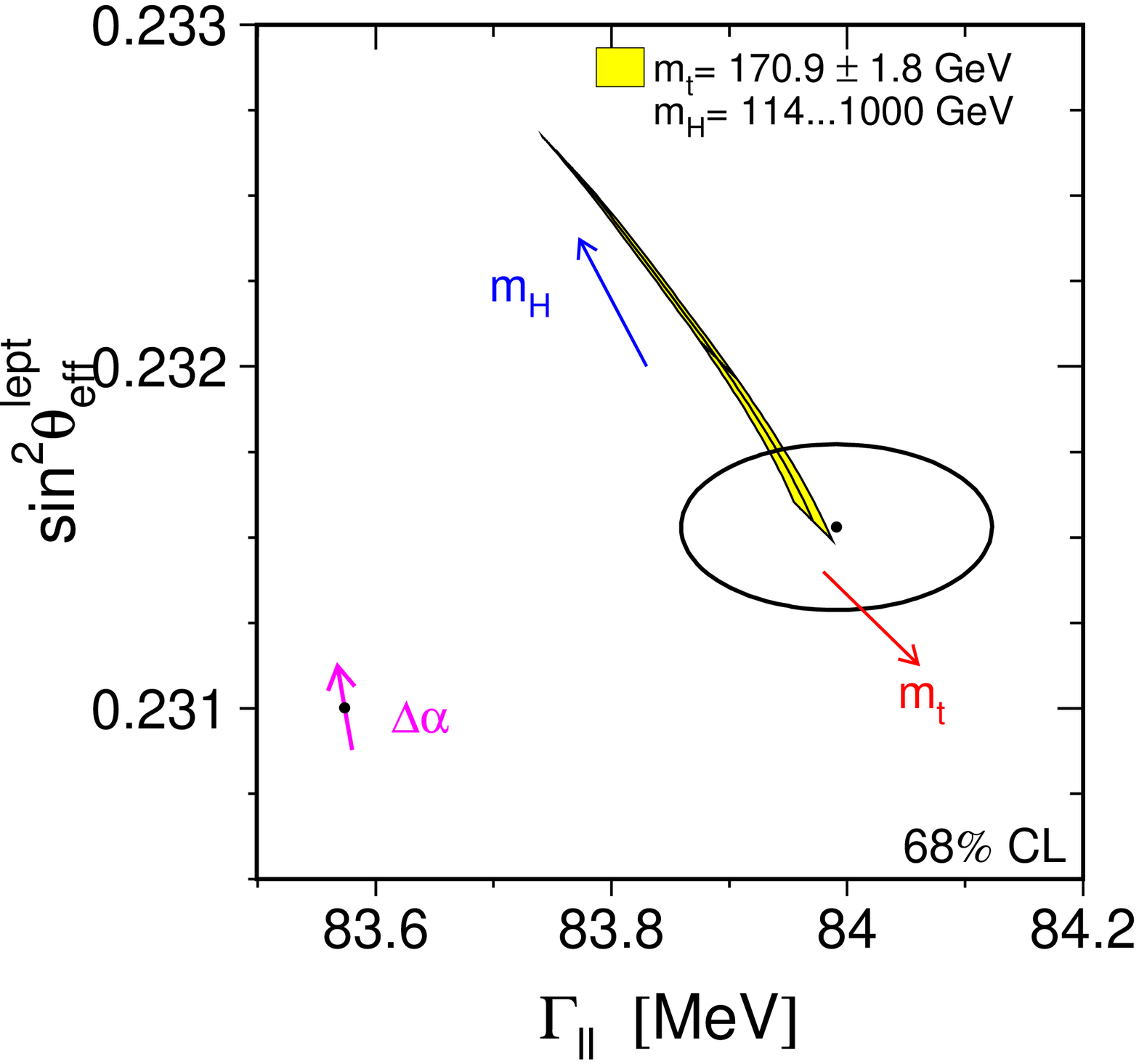} 
\end{minipage}
\hskip 1cm
\begin{minipage}[c]{.45\linewidth}\centering
\includegraphics[width=6.5cm]{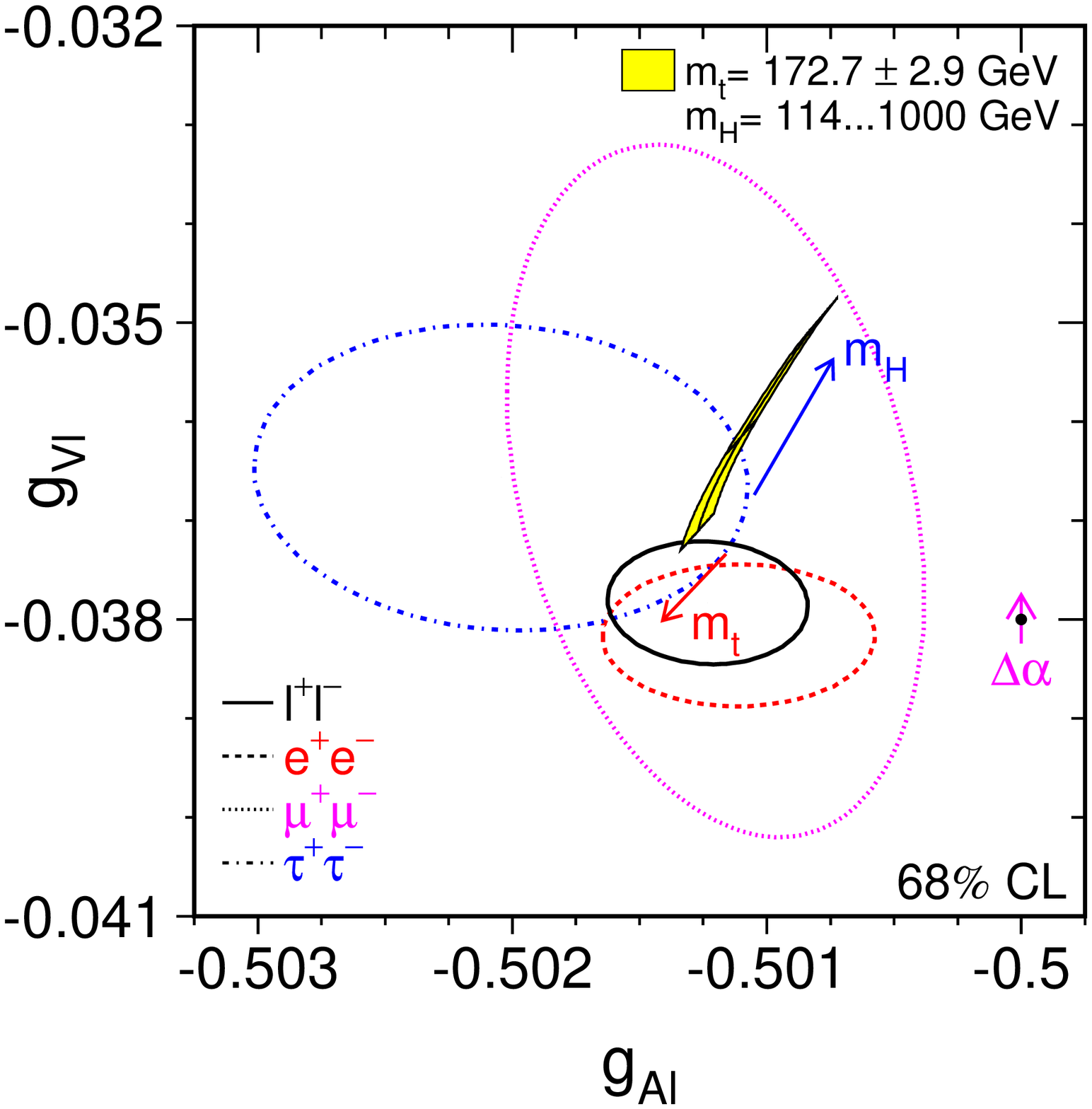}  
\end{minipage}
\caption{Combined LEP and SLD measurements of $\sweff$ and
$\Gamma_l$ (left) and the corresponding effective vector and
axial-vector couplings $v_l$ and $a_l$ (right). The shaded region
shows the SM prediction. The arrows point in the direction of
increasing values of $m_t$ and $M_H$. The point shows the predicted
values if, among the electroweak radiative corrections, only the
photon vacuum polarization is included. Its arrow indicates the
variation induced by the uncertainty in $\alpha(M_Z^2)$
\cite{LEPEWWG,LEPEWWG_SLD:06}.} \label{fig:Zcouplings}
\end{figure}

The leptonic asymmetry measurements from LEP and SLD
can all be combined to determine the ratios $v_l/a_l$
of the vector and axial-vector couplings of the
three charged leptons,
or equivalently the effective electroweak mixing angle
\bel{eq:sW-eff}
\sweff\,\equiv\,
\frac{1}{4}\,\left(1 -\frac{v_l}{a_l}\right)\, .
\ee
The sum $(v_l^2 + a_l^2)$ is derived from the leptonic decay widths
of the $Z$, i.e., from Eq.~\eqn{eq:Z_width} corrected with a
multiplicative factor \ $\left(1 + {3\over4}\,
{\alpha\over\pi}\right)$ to account for final-state QED corrections.
The signs of $v_l$ and $a_l$ are fixed by requiring $a_e<0$.

\begin{figure}[tbh]\centering
\begin{minipage}[t]{.46\linewidth}\centering
\includegraphics[width=6.25cm]{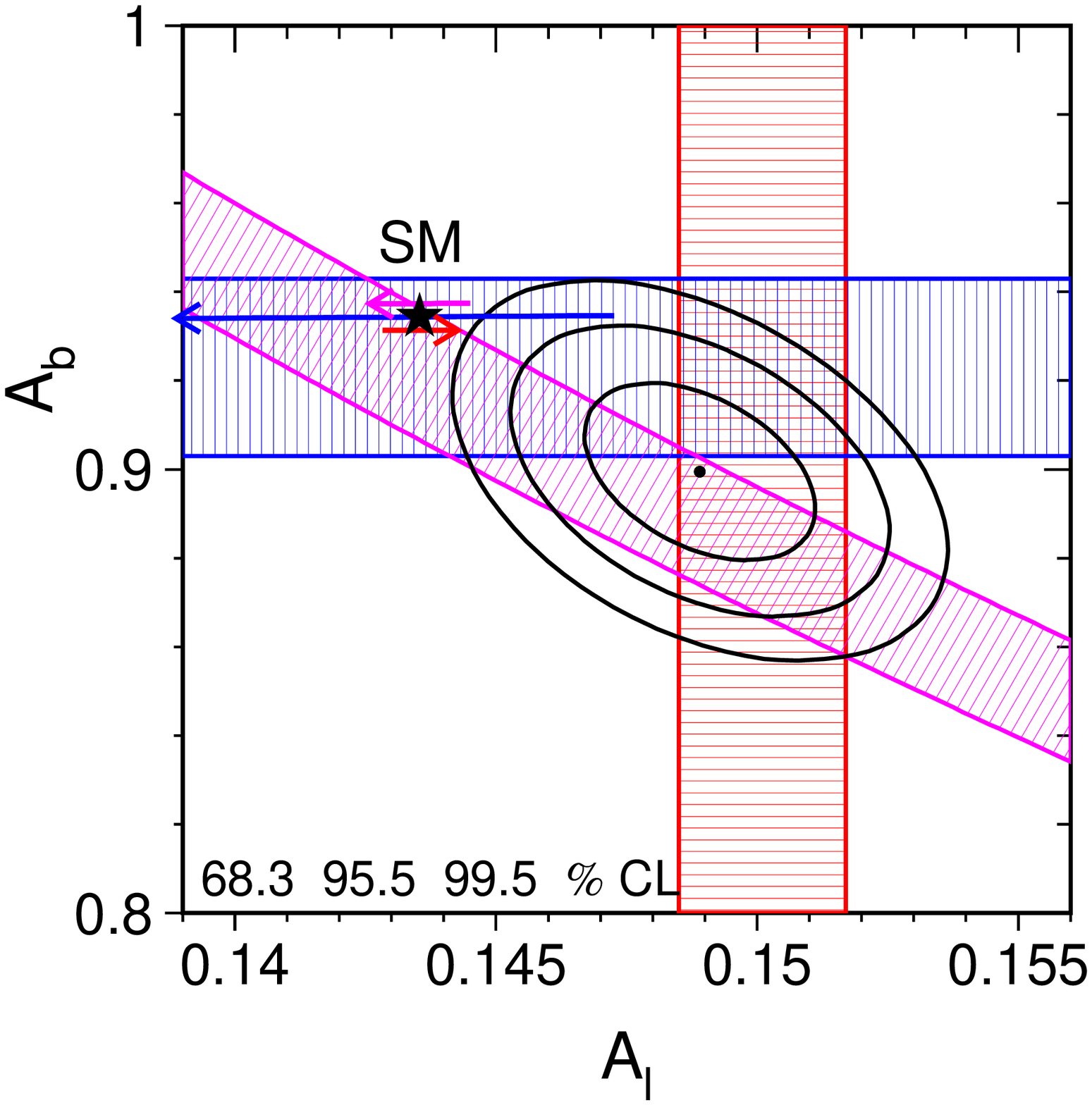}
\caption{Measurements of $A_l$, $A_b$ (SLD) and $\cA_{\rms
FB}^{0,b}$. 
The arrows pointing to the left (right) show the variations of the
SM prediction with $M_H = 300\,{}^{+700}_{-186}\:\mathrm{GeV}$ ($m_t
= 172.7\pm 2.9\:\mathrm{GeV}$). The small arrow oriented to the left
shows the additional uncertainty from $\alpha(M_Z^2)$
\cite{LEPEWWG,LEPEWWG_SLD:06}.} \label{fig:Al_Ab}
\end{minipage}
\hfill
\begin{minipage}[t]{.46\linewidth}\centering
\includegraphics[width=6.5cm,clip]{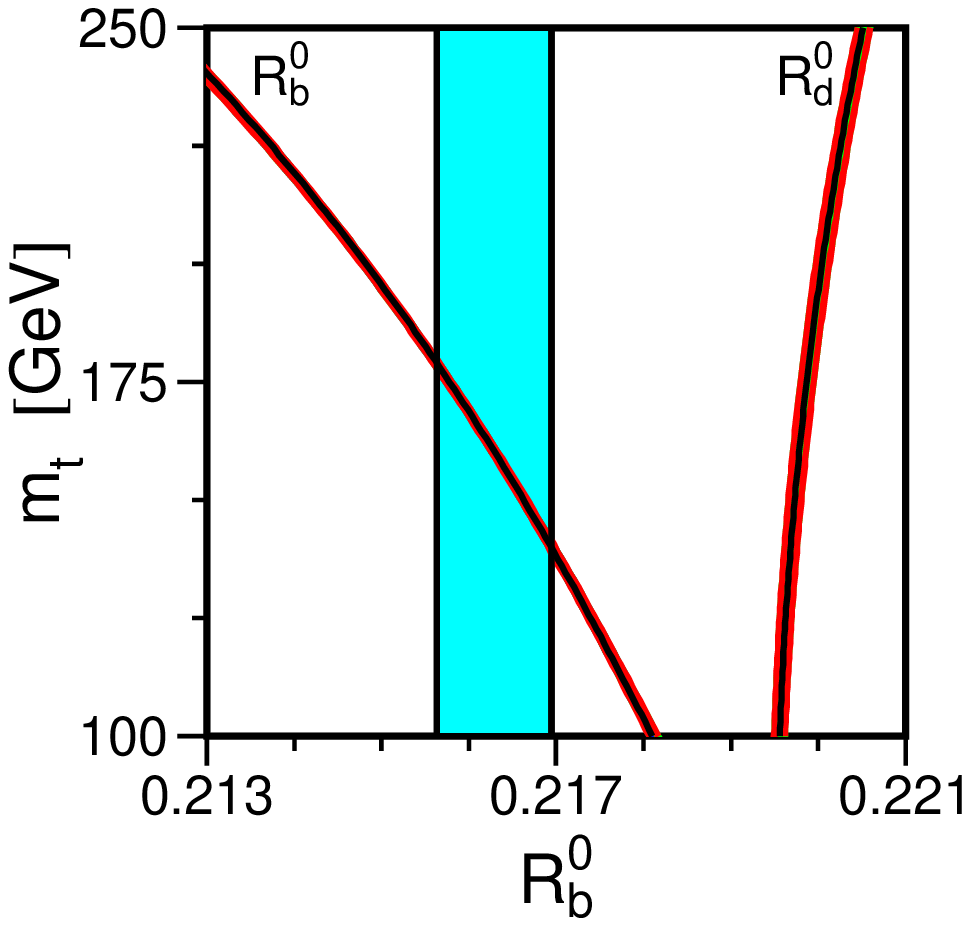}
\caption{The SM prediction of the ratios \ $R_b$ \ and \ $R_d$ \
[$R_q\equiv\Gamma(Z\to\bar q q)/\Gamma(Z\to\mathrm{hadrons})$], as a
function of the top mass. The measured value of $R_b$ (vertical
band) provides a determination of $m_t$
\cite{LEPEWWG,LEPEWWG_SLD:06}.} \label{fig:Rb}
\end{minipage}
\end{figure}

The resulting 68\% probability contours are shown in
Fig.~\ref{fig:Zcouplings}, which provides strong evidence
of the electroweak radiative corrections. The good agreement
with the SM predictions, obtained for low values of the
Higgs mass, is lost if only the QED vacuum
polarization contribution is taken into account, as indicated
by the point with an arrow.
Notice that the uncertainty induced by the input value of
$\alpha(M_Z^2)^{-1}=128.93\pm 0.05$ is sizeable. The measured
couplings of the three charged leptons confirm lepton universality
in the neutral-current sector. The solid contour combines the three
measurements assuming universality.

The neutrino couplings can also be determined from the invisible $Z$
decay width, by assuming three identical neutrino generations with
left-handed couplings, and fixing the sign from neutrino scattering
data. Alternatively, one can use the SM prediction for $\Gamma_{\rms
inv}$ to get a determination of the number of light neutrino
flavours \cite{LEPEWWG,LEPEWWG_SLD:06}:
\bel{eq:Nnu} N_\nu = 2.9840\pm 0.0082\, . \ee

Figure~\ref{fig:Al_Ab} shows the measured values of $A_l$ and $A_b$,
together with the joint constraint obtained from $\cA_{\rms
FB}^{0,b}$ (diagonal band). The direct measurement of $A_b$ at SLD
agrees well with the SM prediction; however, a much lower value is
obtained from the ratio $\frac{4}{3}\, \cA_{\rms FB}^{0,b}/A_l$.
This is the most significant discrepancy observed in the $Z$-pole
data.
Heavy quarks ($\frac{4}{3}\, \cA_{\rms FB}^{0,b}/A_b$) seem to
prefer a high value of the Higgs mass, while leptons ($A_l$) favour
a light Higgs. The combined analysis prefers low values of $M_H$,
because of the influence of $A_l$.

The strong sensitivity of the ratio $R_b\equiv\Gamma(Z\to\bar b
b)/\Gamma(Z\to\mathrm{hadrons})$ to the top quark mass is shown in
Fig.~\ref{fig:Rb}. Owing to the $|V_{td}|^2$ suppression, such a
dependence is not present in the analogous ratio $R_d$. Combined
with all other electroweak precision measurements at the $Z$ peak,
$R_b$ provides a determination of $m_t$ in good agreement with the
direct and most precise measurement at the Tevatron. This is shown
in Fig.~\ref{fig:MW_Mt_MH}, which compares the information on $M_W$
and $m_t$ obtained at LEP1 and SLD, with the direct measurements
performed at LEP2 and the Tevatron. A similar comparison for $m_t$
and $M_H$ is also shown. The lower bound on $M_H$ obtained from
direct searches excludes a large portion of the 68\% C.L. allowed
domain from precision measurements.
%
\begin{figure}[hbt]\centering
\begin{minipage}[c]{.45\linewidth}\centering
\includegraphics[width=6.25cm]{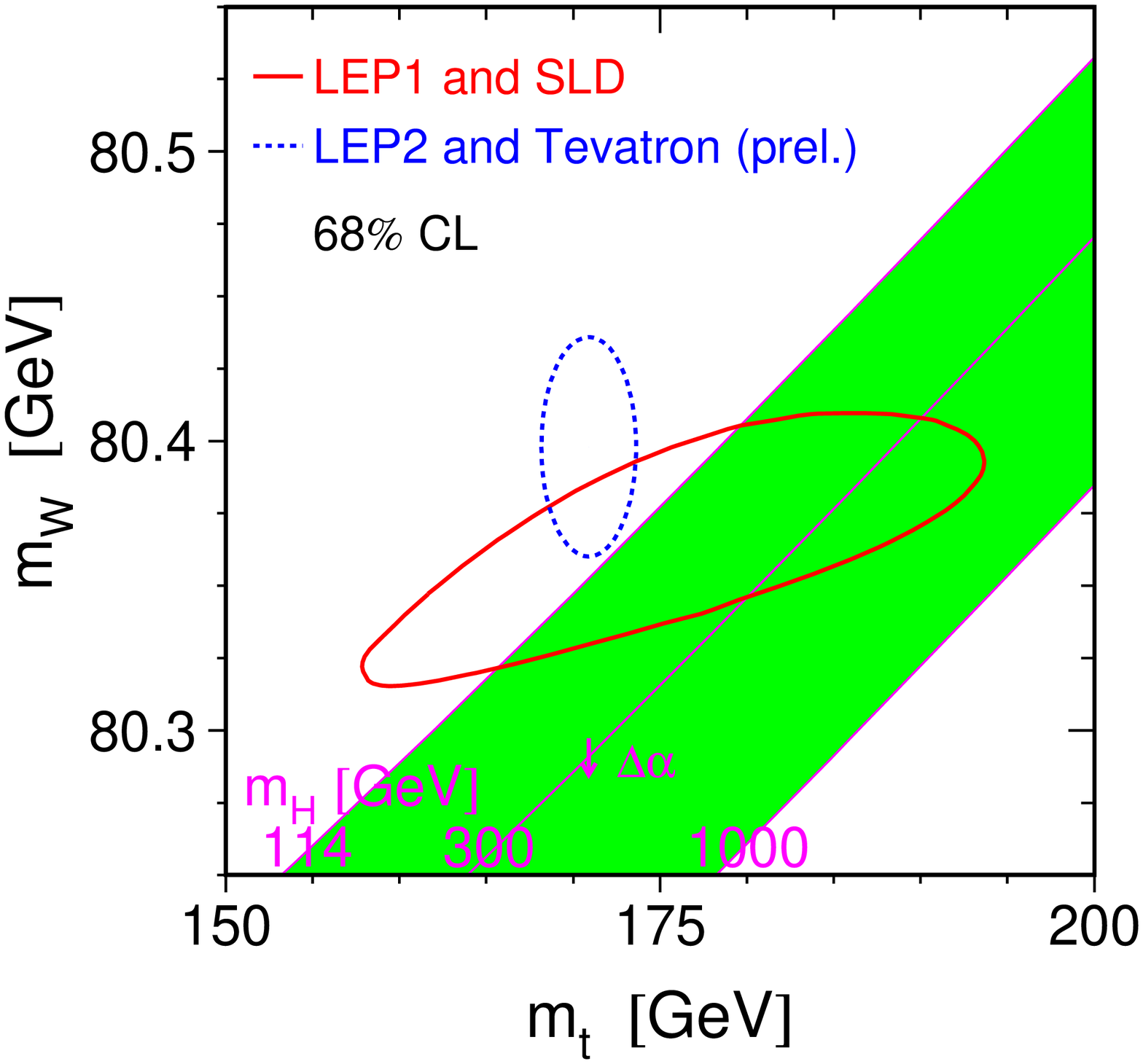}
\end{minipage}
\hskip 1cm
\begin{minipage}[c]{.45\linewidth}\centering
\includegraphics[width=6.25cm]{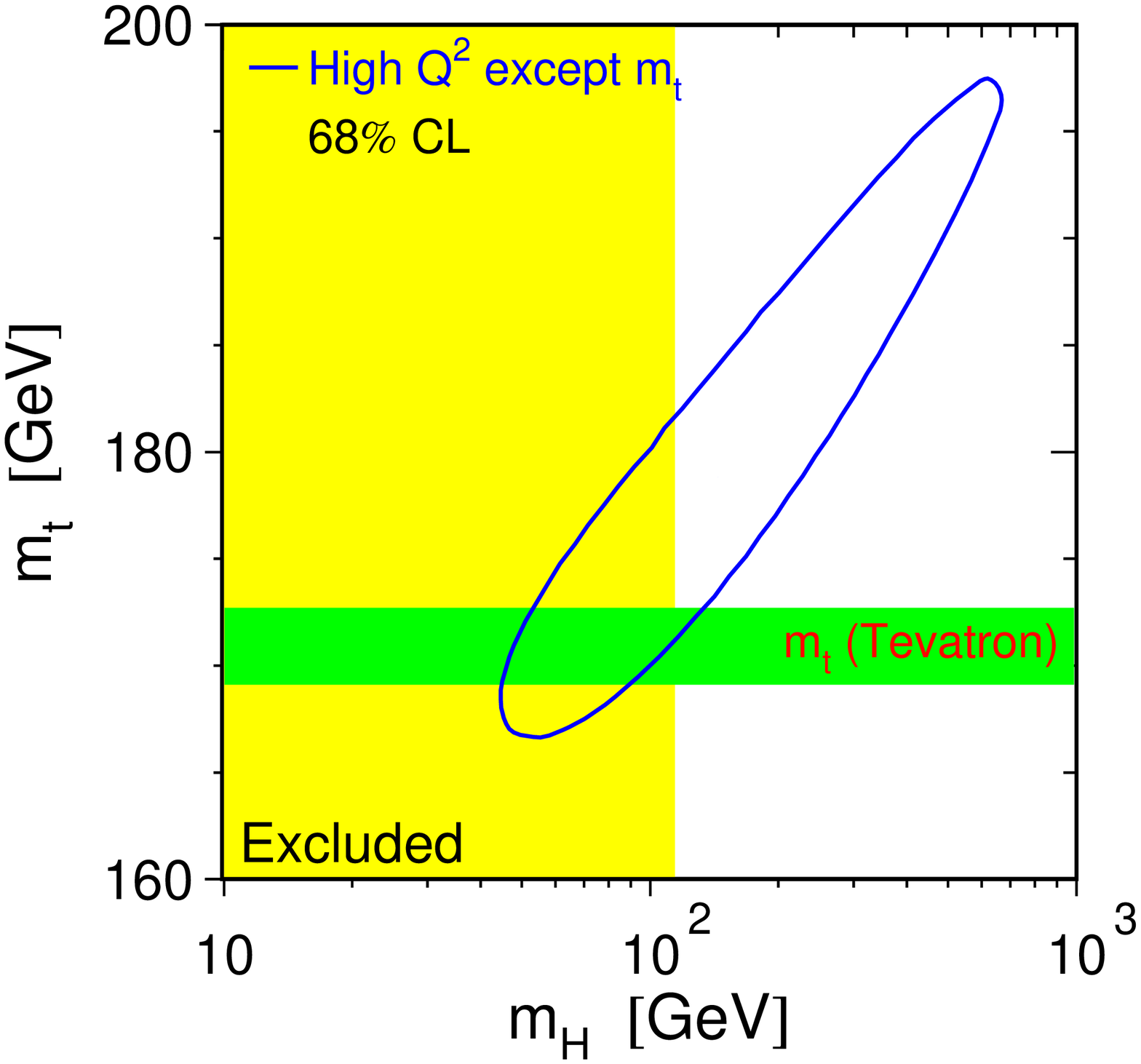}
\end{minipage}
\caption{Comparison (left) of the direct measurements of $M_W$ and
$m_t$ (LEP2 and Tevatron data) with the indirect determination
through electroweak radiative corrections (LEP1 and SLD). Also shown
in the SM relationship for the masses as function of $M_H$. The
figure on the right makes the analogous comparison for $m_t$ and
$M_H$ \cite{LEPEWWG,LEPEWWG_SLD:06}.} \label{fig:MW_Mt_MH}
\end{figure}

\begin{figure}[tbh]\centering
\begin{minipage}[t]{.45\linewidth}\centering
\includegraphics[width=7.25cm]{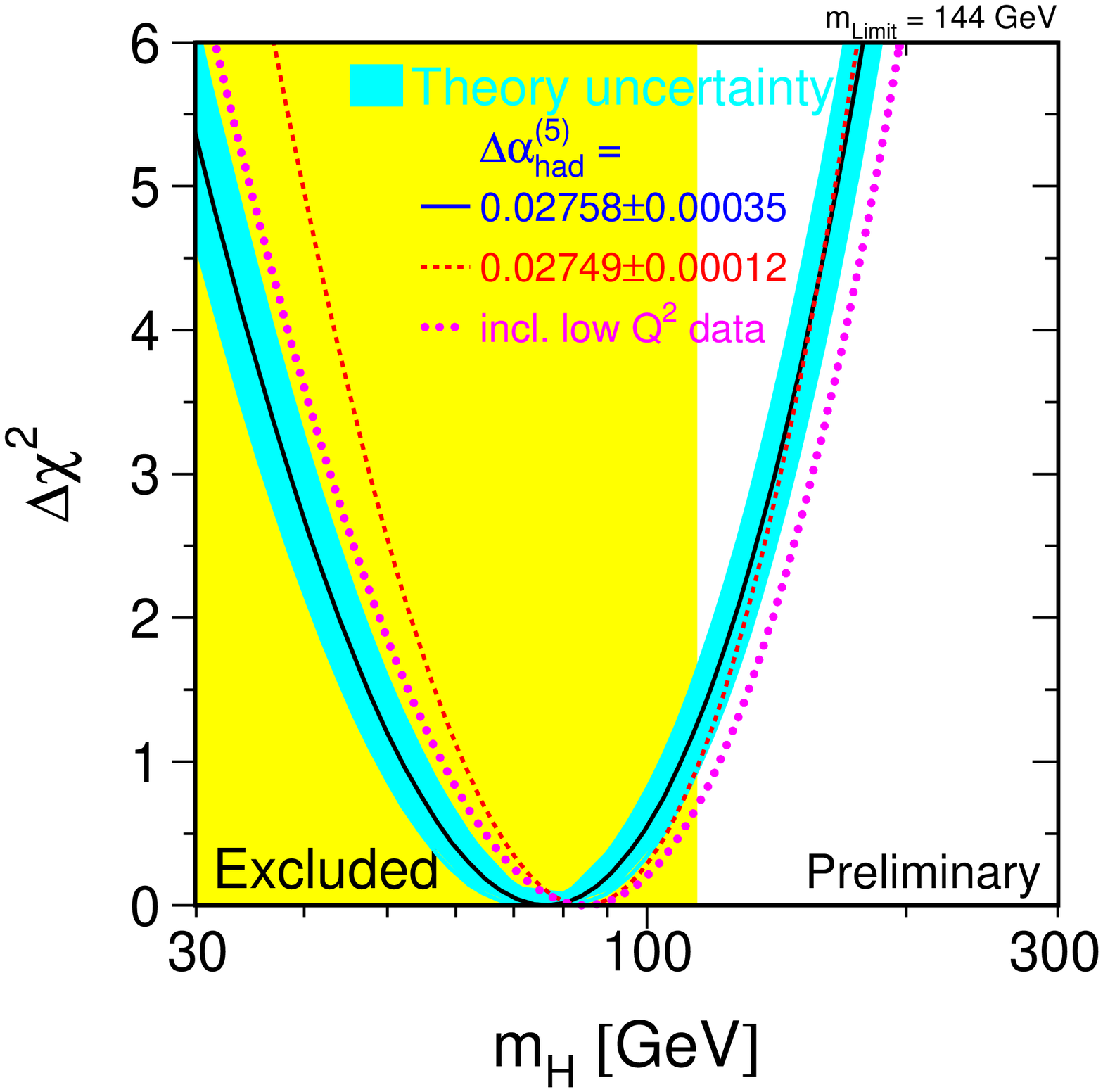}
\caption{$\Delta\chi^2 = \chi^2-\chi^2_{\rms min}$ versus $M_H$,
from the global fit to the electroweak data. The vertical band
indicates the 95\% exclusion limit from direct searches
\cite{LEPEWWG,LEPEWWG_SLD:06}.} \label{fig:MH_chi2}
\end{minipage}
\hfill
\begin{minipage}[t]{.45\linewidth}\centering
\includegraphics[width=6.5cm]{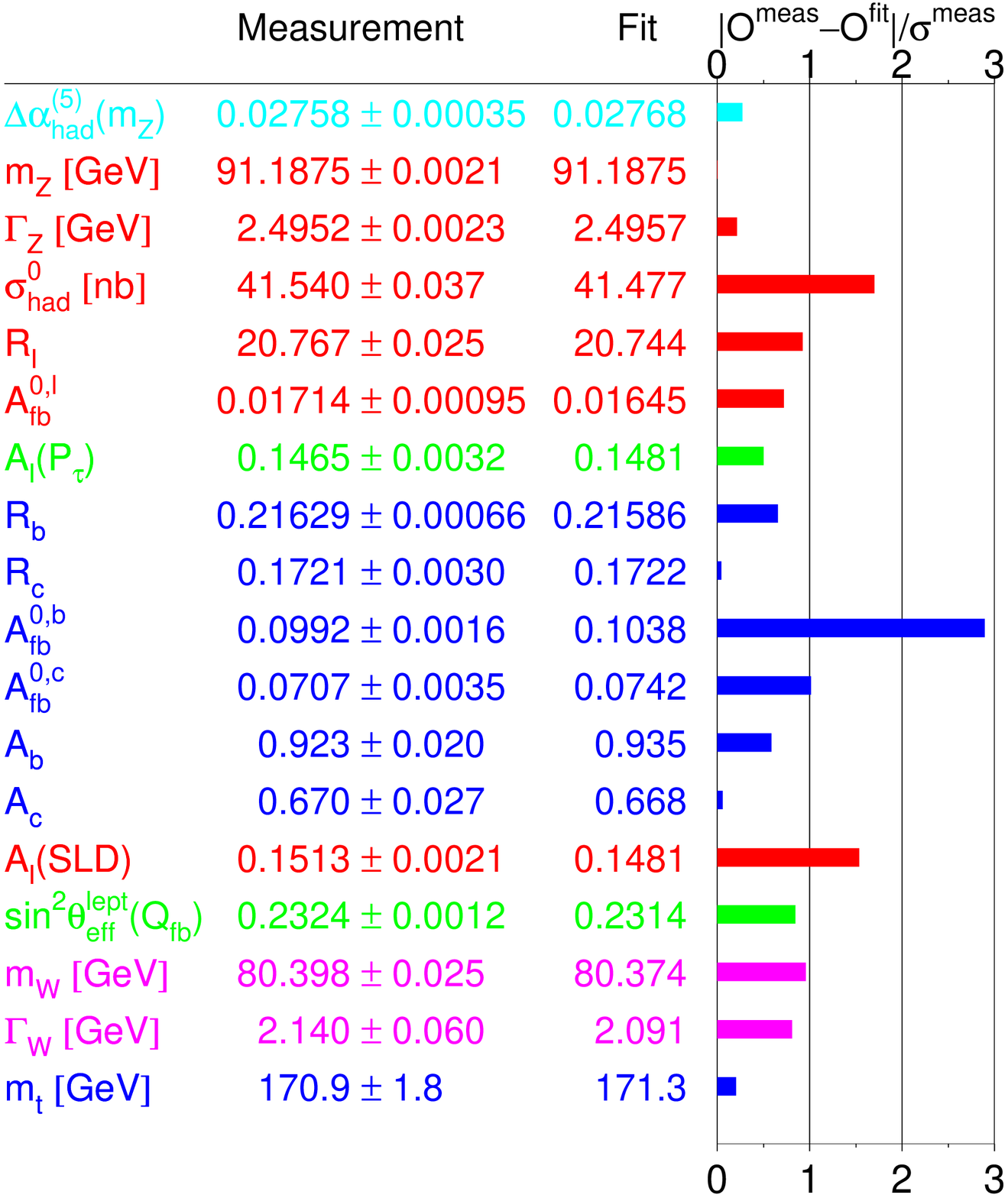}
\caption{Comparison between the measurements included in the
combined analysis of the SM and the results from the global
electroweak fit \cite{LEPEWWG,LEPEWWG_SLD:06}.} \label{fig:pulls}
\end{minipage}
\end{figure}

Taking all direct and indirect data into account, one obtains the
best constraints on $M_H$. The global electroweak fit results in the
$\Delta\chi^2 = \chi^2-\chi^2_{\rms min}$ curve shown in
Fig.~\ref{fig:MH_chi2}. The lower limit on $M_H$ obtained from
direct searches is close to the point of minimum $\chi^2$. At 95\%
C.L., one gets \cite{LEPEWWG,LEPEWWG_SLD:06}
\bel{eq:MH_limits} 114.4\;\mathrm{GeV}\; <\; M_H\; <\;
144\;\mathrm{GeV}. \ee
The fit provides also a very accurate value of the strong coupling
constant, $\alpha_s(M_Z^2) = 0.1186\pm 0.0027$, in very good
agreement with the world average value $\alpha_s(M_Z^2) = 0.119\pm
0.002$ \cite{PDG,BE:07}. The largest discrepancy between theory and
experiment occurs for $\cA_{\rms FB}^{0,b}$, with the fitted value
being nearly $3\,\sigma$ larger than the measurement. As shown in
Fig.~\ref{fig:pulls}, a good agreement is obtained for all other
observables.

\subsection{Gauge self-interactions}

\begin{figure}[tbh]\centering
\includegraphics[width=15cm]{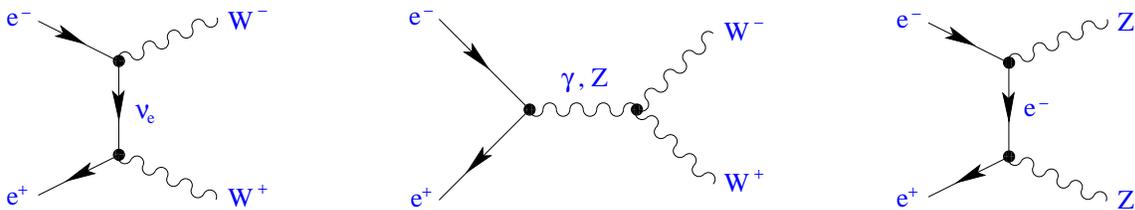}
\caption{Feynman diagrams contributing to \ $e^+e^-\!\to W^+W^-$
\ and \ $e^+e^-\!\to ZZ$.}
\label{fig:eeWW}
\end{figure}

At tree level, the $W$-pair production process \ $e^+e^-\to W^+W^-$
\ involves three different contributions (Fig.~\ref{fig:eeWW}),
corresponding to the exchange of $\nu_e$, $\gamma$ and $Z$. The
cross-section measured at LEP2 agrees very well with the SM
predictions. As shown in Fig.~\ref{fig:sigma_eeWW_ZZ}, the
$\nu_e$-exchange contribution alone would lead to an unphysical
growing of the cross-section at large energies and, therefore, would
imply a violation of unitarity. Adding the $\gamma$-exchange
contribution softens this behaviour, but a clear disagreement with
the data persists. The $Z$-exchange mechanism, which involves the
$ZWW$ vertex, appears to be crucial in order to explain the data.

\begin{figure}[tbh]\centering
\begin{minipage}[c]{.45\linewidth}\centering
\includegraphics[width=7.25cm]{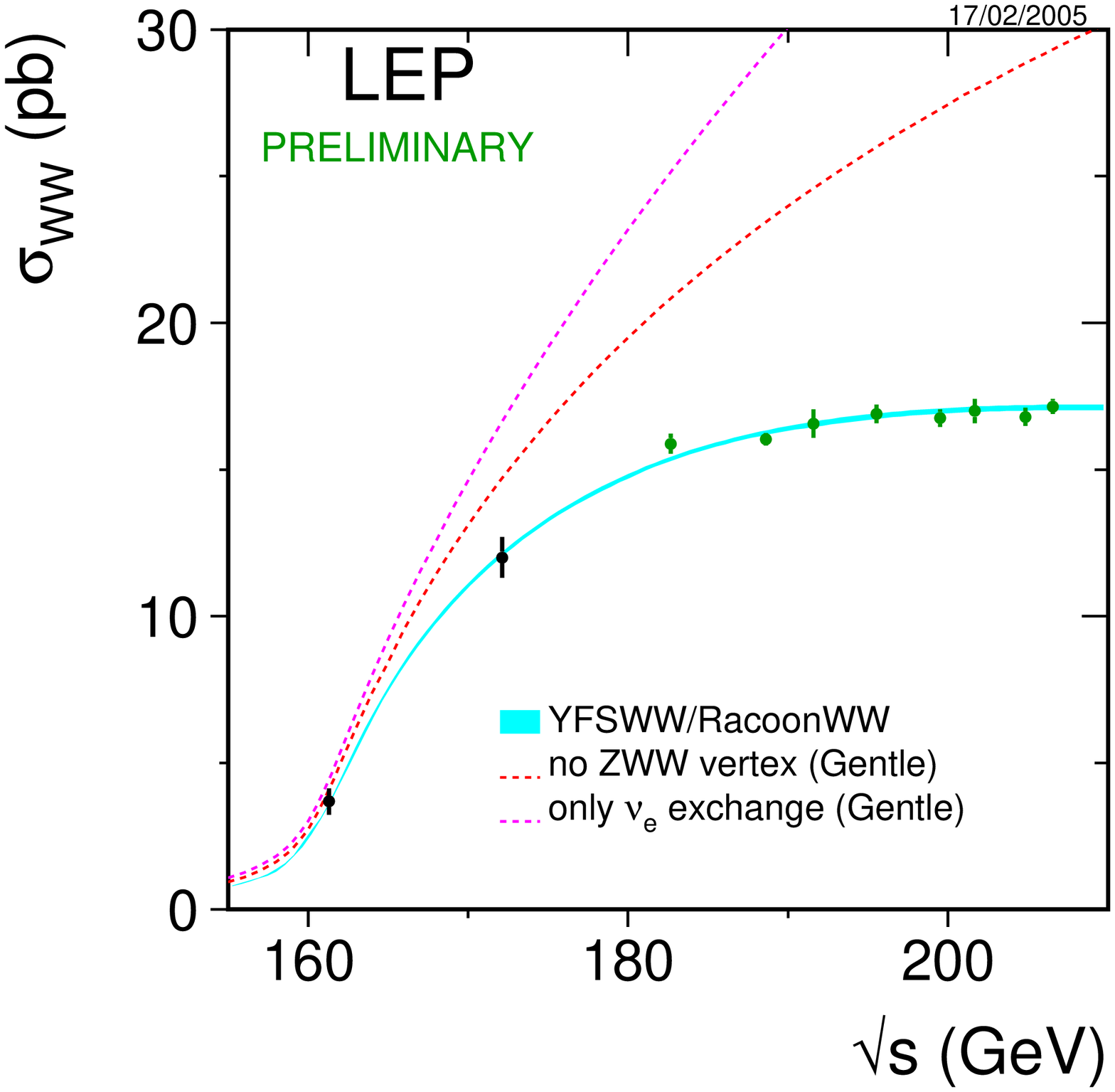}
\end{minipage}
\hfill
\begin{minipage}[c]{.45\linewidth}\centering
\includegraphics[width=7.25cm]{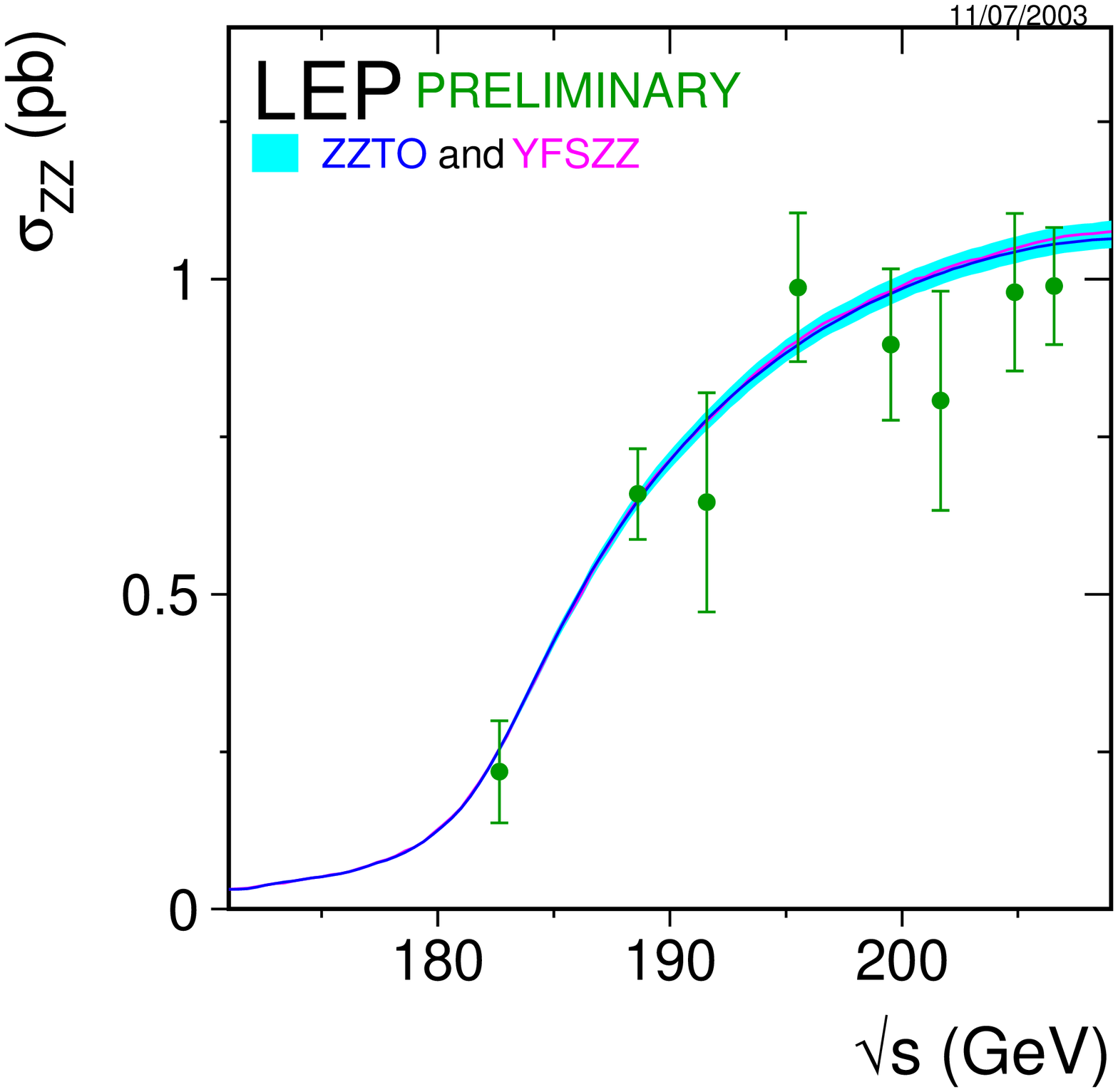}
\end{minipage}
\caption{Measured energy dependence of \ $\sigma(e^+e^-\to W^+W^-)$
\ (left) and \ $\sigma(e^+e^-\to ZZ)$ \ (right). The three curves
shown for the $W$-pair production cross-section correspond to only
the $\nu_e$-exchange contribution (upper curve), $\nu_e$ exchange
plus photon exchange (middle curve) and all contributions including
also the $ZWW$ vertex (lower curve). Only the $e$-exchange mechanism
contributes to $Z$--pair production \cite{LEPEWWG,LEPEWWG_SLD:06}.}
\label{fig:sigma_eeWW_ZZ}
\end{figure}

Since the $Z$ is electrically neutral, it does not interact
with the photon. Moreover, the SM does not include any
local $ZZZ$ vertex. Therefore, the \ $e^+e^-\to ZZ$ \
cross-section only involves the contribution from $e$ exchange.
The agreement of the SM predictions with the experimental
measurements in both production channels, $W^+W^-$ and $ZZ$,
provides a test of the gauge self-interactions.
There is a clear signal of the presence of a $ZWW$ vertex, with the
predicted strength, and no evidence for any $\gamma ZZ$ or
$ZZZ$ interactions. The gauge structure of the
$SU(2)_L\otimes U(1)_Y$ theory is nicely confirmed
by the data.

\subsection{Higgs decays}

\begin{figure}[tbh]\centering
\begin{minipage}[c]{.48\linewidth}\centering
\includegraphics[width=8cm]{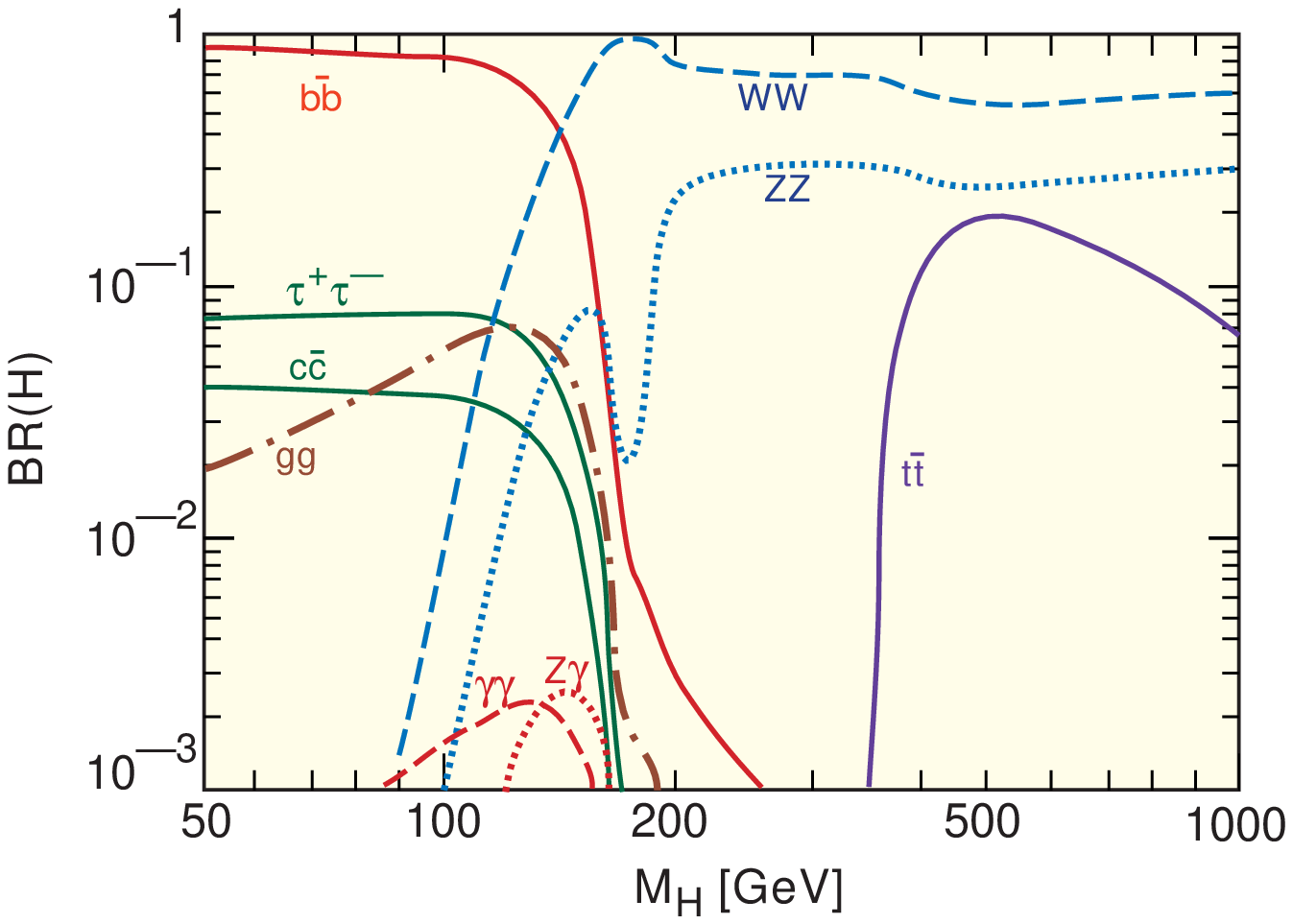}
\end{minipage}
\hfill
\begin{minipage}[c]{.48\linewidth}\centering
\includegraphics[width=8cm]{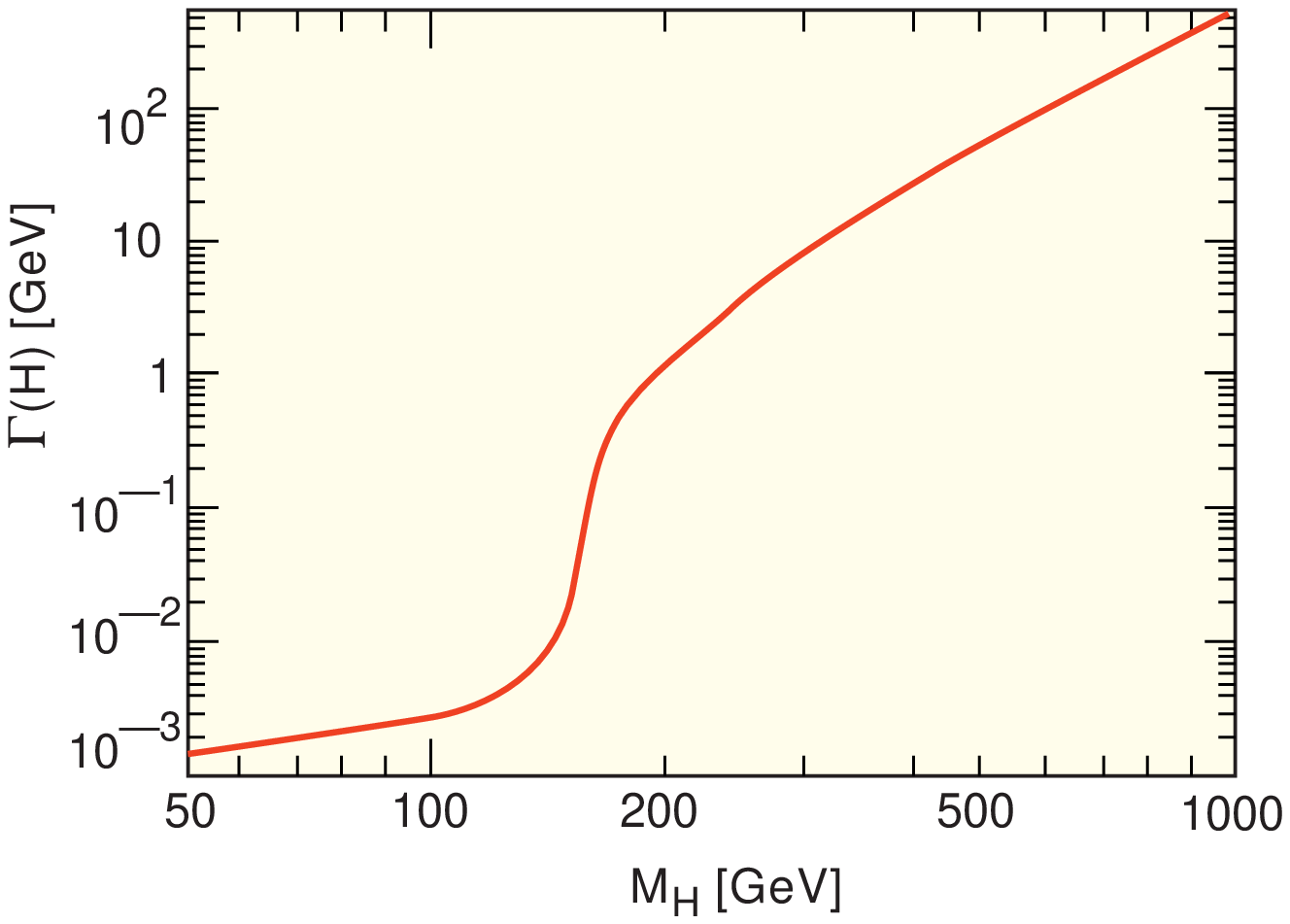}
\end{minipage}
\caption{Branching fractions of the different Higgs decay modes (left)
and total decay width of the Higgs boson (right) as function
of $M_H$ \cite{Denegri}.}
\label{fig:HiggsDecays}
\end{figure}

The couplings of the Higgs boson are always proportional to some
mass scale. The $Hf\bar f$ interaction grows linearly with the
fermion mass, while the $HWW$ and $HZZ$ vertices are proportional to
$M_W^2$ and $M_Z^2$, respectively. Therefore, the most probable
decay mode of the Higgs will be the one into the heaviest possible
final state. This is clearly illustrated in
Fig.~\ref{fig:HiggsDecays}. The $H\to b\bar b$ decay channel is by
far the dominant one below the $W^+W^-$ production threshold. When
$M_H$ is large enough to allow the production of a pair of gauge
bosons, $H\to W^+W^-$ and \ $H\to ZZ$ \ become dominant. For
$M_H>2m_t$, the $H\to t\bar t$ \ decay width is also sizeable,
although smaller than the $WW$ and $ZZ$ ones because of the
different dependence of the corresponding Higgs coupling with the
mass scale (linear instead of quadratic).

The total decay width of the Higgs grows with increasing values of $M_H$.
The effect is very strong above the $W^+W^-$ production threshold.
A heavy Higgs becomes then very broad. At $M_H\sim 600\;\mathrm{GeV}$,
the width is around $100\;\mathrm{GeV}$; while for
$M_H\sim 1\;\mathrm{TeV}$, $\Gamma_H$ is already of the same size
as the Higgs mass itself.

The design of the LHC detectors has taken into account all
these very characteristic properties in order to optimize the
future search for the Higgs boson.


\setcounter{equation}{0}
\section{Flavour \ Dynamics}
\label{sec:flavour}

We have learnt experimentally that there are six different quark
flavours \ $u\,$, $d\,$, $s\,$, $c\,$, $b\,$, $t\,$, three different
charged leptons \ $e\,$, $\mu\,$, $\tau$ \ and their corresponding
neutrinos \ $\nu_e\,$, $\nu_\mu\,$, $\nu_\tau\,$. We can nicely
include all these particles into the SM framework, by organizing
them into three families of quarks and leptons, as indicated in
Eqs.~\eqn{eq:families} and \eqn{eq:structure}. Thus, we have three
nearly identical copies of the same $SU(2)_L\otimes U(1)_Y$
structure, with masses as the only difference.

Let us consider the general case of $N_G$ generations of fermions,
and denote $\nup_j$, $\lp_j$, $\up_j$, $\dop_j$ the members of the
weak family $j$ \ ($j=1,\ldots,N_G$), with definite transformation
properties under the gauge group. Owing to the fermion replication,
a large variety of fermion-scalar couplings are allowed by the gauge
symmetry. The most general Yukawa Lagrangian has the form
\beqn\label{eq:N_Yukawa}
\cL_Y &=&-\,\sum_{jk}\;\left\{
\left(\bar \up_j , \bar \dop_j\right)_L \left[\, c^{(d)}_{jk}\,
\left(\ba \phi^{(+)}\\ \phi^{(0)}\ea\right)\, \dop_{kR} \; +\;
c^{(u)}_{jk}\,
\left(\ba \phi^{(0)*}\\ -\phi^{(-)}\ea\right)\, \up_{kR}\,
\right]
\right.\no\\ && \qquad\!\left. +\;\;
\left(\bar \nup_j , \bar \lp_j\right)_L\, c^{(l)}_{jk}\,
\left(\ba \phi^{(+)}\\ \phi^{(0)}\ea\right)\, \lp_{kR}
\,\right\}
\; +\; \mathrm{h.c.},
\eeqn
where $c^{(d)}_{jk}$, $c^{(u)}_{jk}$ and $c^{(l)}_{jk}$
are arbitrary coupling constants.

After SSB, the Yukawa Lagrangian can be written as
\bel{eq:N_Yuka}
\cL_Y\, =\, - \left(1 + {H\over v}\right)\,\left\{\,
\overline{\bmd}\hskip .6pt'_L \,\bM_d'\,
\bmd\hskip .6pt'_R \; + \;
\overline{\bmu}\hskip .6pt'_L \,\bM_u'\,
\bmu\hskip .6pt'_R
\; + \;
\overline{\bml}\hskip .2pt'_L \,\bM'_l\,
\bml\hskip .2pt'_R \; +\;
\mathrm{h.c.}\right\} .
\ee
Here, $\bmd\hskip .6pt'$, $\bmu\hskip .6pt'$
and $\bml\hskip .2pt'$ denote vectors in
the $N_G$-dimensional flavour
space, and the corresponding mass matrices are given by
\bel{eq:M_c_relation}
(\bM'_d)^{}_{ij}\,\equiv\, c^{(d)}_{ij}\, {v\over\sqrt{2}}\, ,\qquad
(\bM'_u)^{}_{ij}\,\equiv\, c^{(u)}_{ij}\, {v\over\sqrt{2}}\, ,\qquad
(\bM'_l)^{}_{ij}\,\equiv\, c^{(l)}_{ij}\, {v\over\sqrt{2}}\, .
\ee
The diagonalization of these mass matrices determines the mass
eigenstates $d_j$, $u_j$ and $l_j$,
which are linear combinations of the corresponding weak eigenstates
$\dop_j$, $\up_j$ and $\lp_j$, respectively.

The matrix $\bM_d'$ can be decomposed as\footnote{
The condition $\det{\bM'_f}\not=0$ \
($f=d,u,l$)
guarantees that the decomposition
$\bM'_f=\bH_f\bU_f$ is unique:
$\bU_f\equiv\bH_f^{-1}\bM_f'$.
The matrices $\bS_f$
are completely determined (up to phases)
only if all diagonal elements of $\bM_f$
are different.
If there is some degeneracy, the arbitrariness of
$\bS_f$
reflects the freedom to define the physical fields.
If $\det{\bM'_f}=0$,
the matrices $\bU_f$ and
$\bS_f$ are not
uniquely determined, unless their unitarity is explicitly imposed.}
$\bM_d'=\bH^{}_d\,\bU^{}_d=\bS_d^\dagger\, \mathbf{\cM}^{}_d
\,\bS^{}_d\,\bU^{}_d$, where \ $\bH^{}_d\equiv
\sqrt{\bM_d'\bM_d'^{\dagger}}$ is an Hermitian positive-definite
matrix, while $\bU^{}_d$ is unitary. $\bH^{}_d$ can be diagonalized
by a unitary matrix $\bS^{}_d$; the resulting matrix
$\mathbf{\cM}^{}_d$ is diagonal, Hermitian and positive definite.
Similarly, one has \ $\bM_u'= \bH^{}_u\,\bU^{}_u= \bS_u^\dagger\,
\mathbf{\cM}^{}_u\, \bS^{}_u\,\bU^{}_u$ \ and \ $\bM_l'=
\bH^{}_l\,\bU^{}_l= \bS_l^\dagger\, \mathbf{\cM}^{}_l
\,\bS^{}_l\,\bU^{}_l$. In terms of the diagonal mass matrices
\bel{eq:Mdiagonal}
\mathbf{\cM}^{}_d =\mathrm{diag}(m_d,m_s,m_b,\ldots)\, ,\quad
\mathbf{\cM}^{}_u =\mathrm{diag}(m_u,m_c,m_t,\ldots)\, ,\quad
\mathbf{\cM}^{}_l=\mathrm{diag}(m_e,m_\mu,m_\tau,\ldots)\, ,
\ee
the Yukawa Lagrangian takes the simpler form
\bel{eq:N_Yuk_diag}
\cL_Y\, =\, - \left(1 + {H\over v}\right)\,\left\{\,
\overline{\bmd}\,\mathbf{\cM}^{}_d\,\bmd \; + \;
\overline{\bmu}\, \mathbf{\cM}^{}_u\,\bmu \; + \;
\overline{\bml}\,\mathbf{\cM}^{}_l\,\bml \,\right\}\, ,
\ee
where the mass eigenstates are defined by
\beqn\label{eq:S_matrices}
\bmd^{}_L &\!\!\!\!\equiv &\!\!\!\!
\bS^{}_d\, \bmd\hskip .6pt'_L \, ,
\qquad\,\,\,\,\,\,\,\,\,
\bmu^{}_L \equiv \bS^{}_u \,\bmu\hskip .6pt'_L \, ,
\qquad\,\,\,\,\,\,\,\,\,
\bml^{}_L \equiv \bS^{}_l \,\bml\hskip .2pt'_L \, ,
\no\\
\bmd^{}_R &\!\!\!\!\equiv &\!\!\!\!
\bS^{}_d \bU^{}_d\,\bmd\hskip .6pt'_R \, , \qquad
\bmu^{}_R \equiv \bS^{}_u\bU^{}_u\, \bmu\hskip .6pt'_R \, , \qquad
\bml^{}_R \equiv \bS^{}_l\bU^{}_l \, \bml\hskip .2pt'_R \, .
\eeqn
Note, that the Higgs couplings are proportional to the
corresponding fermions masses.

\begin{figure}[tb]\centering
\begin{minipage}[c]{.45\linewidth}\centering
\includegraphics[width=4cm]{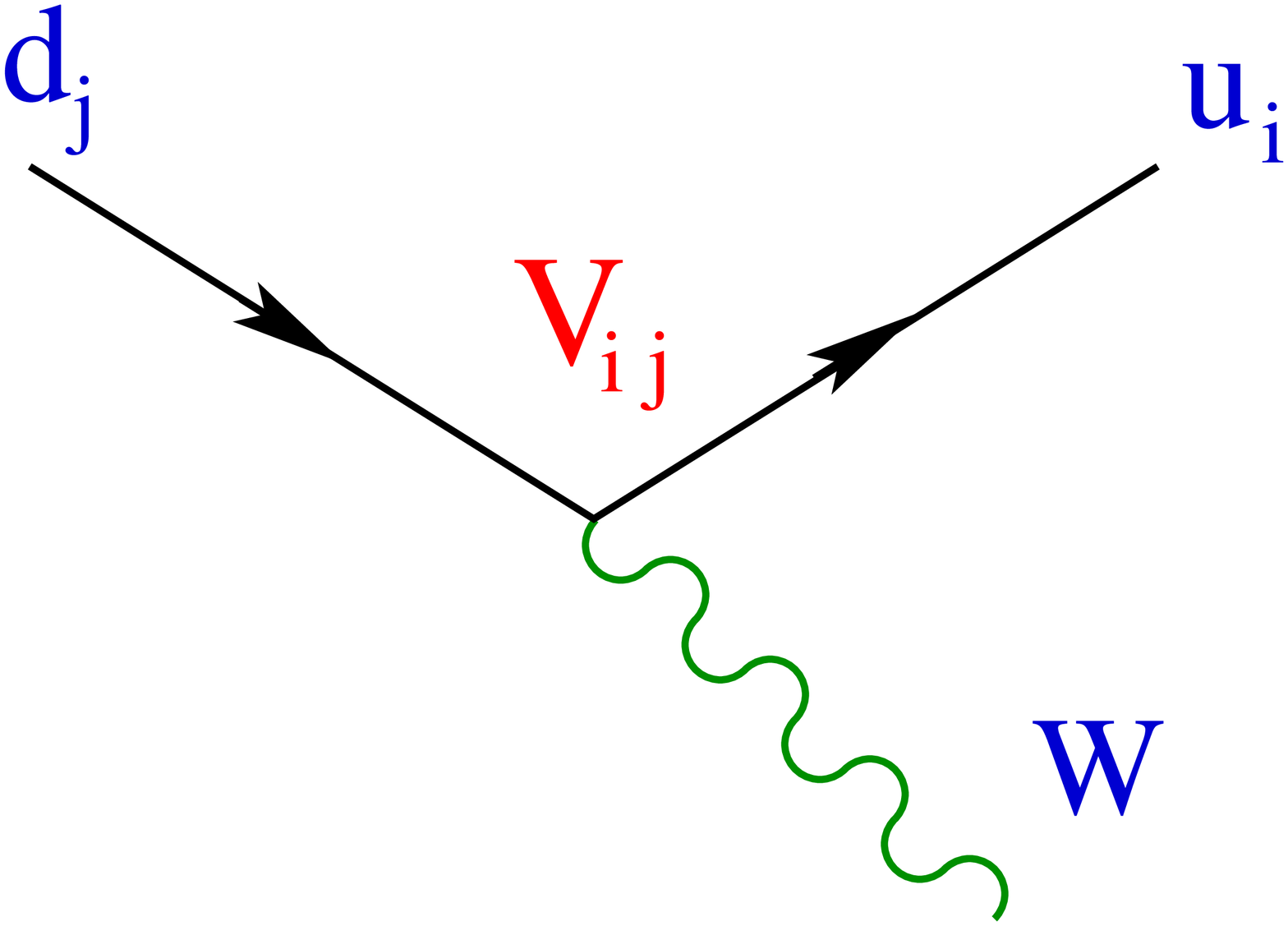}
\end{minipage}
\hskip 1cm
\begin{minipage}[c]{.45\linewidth}\centering
\includegraphics[width=4cm]{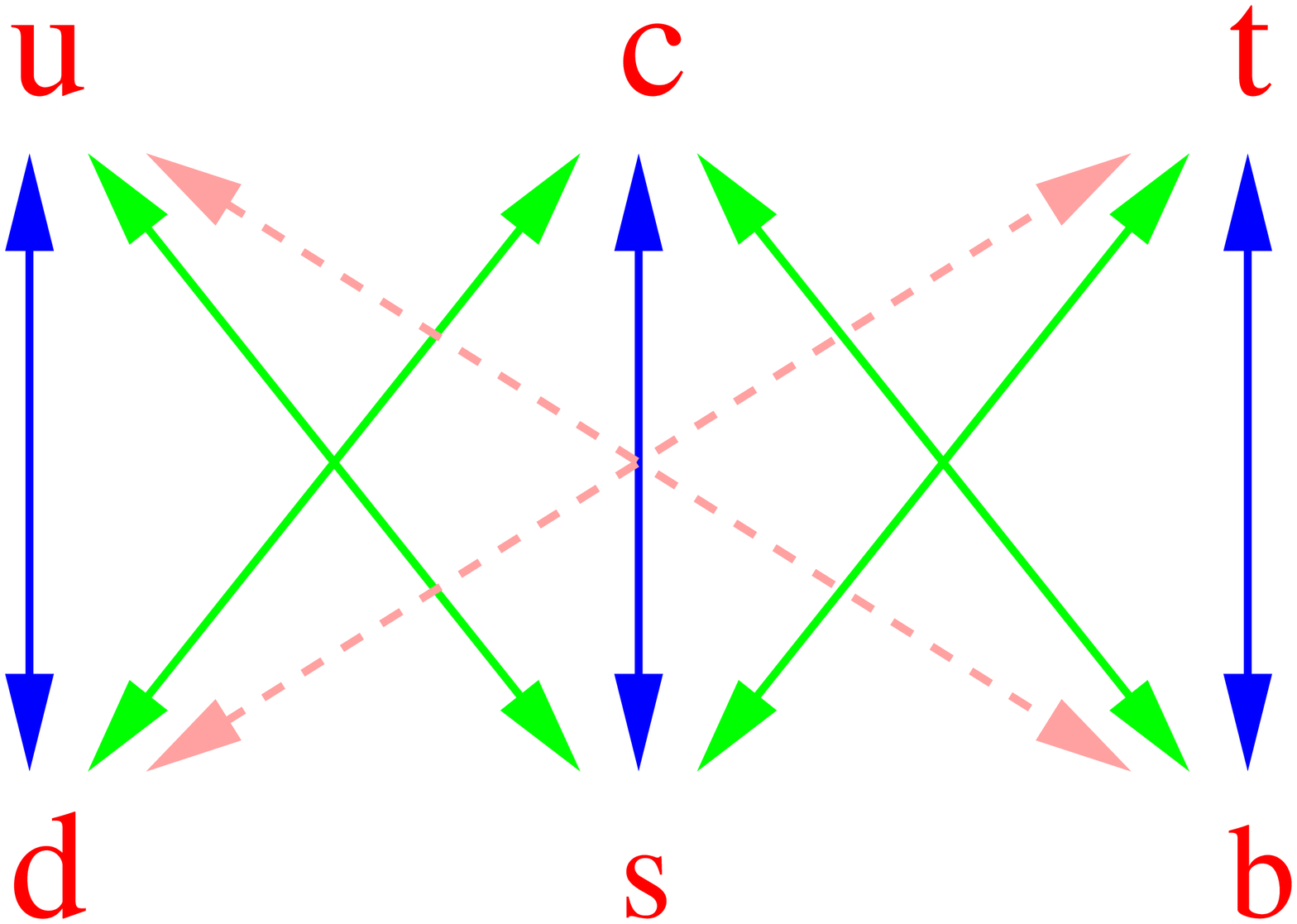}
\end{minipage}
\caption{Flavour-changing transitions through the charged-current
couplings of the $W^\pm$ bosons.}
\label{fig:CKM}
\end{figure}

Since, \ $\overline{\bmf}\hskip .7pt'_L\, \bmf\hskip .7pt'_L =
\overline{\bmf}^{}_L \,\bmf^{}_L$ \ and \
$\overline{\bmf}\hskip .7pt'_R \,\bmf\hskip .7pt'_R =
\overline{\bmf}^{}_R \,\bmf^{}_R$ \
($f=d,u,l$), the form of the neutral-current part of the
$SU(2)_L\otimes U(1)_Y$ Lagrangian does not change when expressed
in terms of mass eigenstates. Therefore, there are no
flavour-changing neutral currents in the SM (GIM
mechanism \cite{GIM:70}). This
is a consequence of treating all equal-charge fermions
on the same footing.

However, $\overline{\bmu}\hskip .7pt'_L \,\bmd\hskip .7pt'_L =
\overline{\bmu}^{}_L \,\bS^{}_u\,\bS_d^\dagger\,\bmd^{}_L\equiv
\overline{\bmu}^{}_L \bV\,\bmd^{}_L$. In general, $\bS^{}_u\not=
\bS^{}_d\,$; thus, if one writes the weak eigenstates in terms of
mass eigenstates, a $N_G\times N_G$ unitary mixing matrix $\bV$,
called the Cabibbo--Kobayashi--Maskawa (CKM) matrix
\cite{cabibbo,KM:73}, appears in the quark charged-current sector:
\bel{eq:cc_mixing}
\cL_{\rms CC}\, = \, - {g\over 2\sqrt{2}}\,\left\{
W^\dagger_\mu\,\left[\,\sum_{ij}\;
\bar u_i\,\gamma^\mu(1-\gamma_5) \,\bV_{\! ij}\, d_j
\; +\;\sum_l\; \bar\nu_l\,\gamma^\mu(1-\gamma_5)\, l
\,\right]\, + \, \mathrm{h.c.}\right\}\, .
\ee
The matrix $\bV$ couples any `up-type' quark with all `down-type'
quarks (Fig.~\ref{fig:CKM}).

If neutrinos are assumed to be massless, we can always redefine the
neutrino flavours, in such a way as to eliminate the analogous
mixing in the lepton sector: $\overline{\mbox{\boldmath
$\nu$}}\hskip .7pt'_L\, \bml\hskip .7pt'_L =
\overline{\mbox{\boldmath $\nu$}}\hskip .7pt_L'\,
\bS^\dagger_l\,\bml^{}_L\equiv \overline{\mbox{\boldmath
$\nu$}}^{}_L\, \bml^{}_L$. Thus, we have lepton-flavour conservation
in the minimal SM without right-handed neutrinos.
If sterile $\nu^{}_R$ fields are included in the model, one would
have an additional Yukawa term in Eq.~\eqn{eq:N_Yukawa}, giving rise
to a neutrino mass matrix \ $(\bM'_\nu)_{ij}\equiv c^{(\nu)}_{ij}\,
{v/\sqrt{2}}\,$. Thus, the model could accommodate non-zero neutrino
masses and lepton-flavour violation through a lepton mixing matrix
$\bV^{}_{\! L}$ analogous to the one present in the quark sector.
Note, however, that the total lepton number \ $L\equiv L_e + L_\mu +
L_\tau$ \ would still be conserved. We know experimentally that
neutrino masses are tiny and there are strong bounds on
lepton-flavour violating decays: \ $\mathrm{Br}(\mu^\pm\to e^\pm e^+
e^-) < 1.0\cdot 10^{-12}$ \cite{BE:88}, $\mathrm{Br}(\mu^\pm\to
e^\pm\gamma) < 1.2\cdot 10^{-11}$ \cite{BR:99},
$\mathrm{Br}(\tau^\pm\to \mu^\pm\gamma) < 4.5\cdot 10^{-8}$
\cite{LFVbelle,LFVbabar} \ldots\ However, we do have a clear
evidence of neutrino oscillation phenomena.

The fermion masses and the quark mixing matrix $\bV$ are all
determined by the Yukawa couplings in Eq.~\eqn{eq:N_Yukawa}.
However, the coefficients $c_{ij}^{(f)}$ are not known; therefore we
have a bunch of arbitrary parameters. A general $N_G\times N_G$
unitary matrix is characterized by $N_G^2$ real parameters: \ $N_G
(N_G-1)/2$ \ moduli and \ $N_G (N_G+1)/2$ \ phases. In the case of
$\,\bV$, many of these parameters are irrelevant, because we can
always choose arbitrary quark phases. Under the phase redefinitions
\ $u_i\to \e^{i\phi_i}\, u_i$ \ and \ $d_j\to\e^{i\theta_j}\, d_j$,
the mixing matrix changes as \ $\bV_{\! ij}\to \bV_{\!
ij}\,\e^{i(\theta_j-\phi_i)}$; thus, $2 N_G-1$ phases are
unobservable. The number of physical free parameters in the
quark-mixing matrix then gets reduced to $(N_G-1)^2$: \
$N_G(N_G-1)/2$ moduli and $(N_G-1)(N_G-2)/2$ phases.

In the simpler case of two generations, $\bV$ is determined by a
single parameter. One then recovers the Cabibbo rotation matrix
\cite{cabibbo}
\bel{eq:cabibbo}
\bV\, = \,
\left(\bat \cos{\theta_C} &\sin{\theta_C} \\[2pt] -\sin{\theta_C}& \cos{\theta_C}\ea
\right)\, .
\ee
With $N_G=3$, the CKM matrix is described by three angles and one
phase. Different (but equivalent) representations can be found in
the literature. The Particle data Group \cite{PDG} advocates the use
of the following one as the `standard' CKM parametrization:
\bel{eq:CKM_pdg}
\bV\, = \, \left[
\begin{array}{ccc}
c_{12}\, c_{13}  & s_{12}\, c_{13} & s_{13}\, \e^{-i\delta_{13}} \\[2pt]
-s_{12}\, c_{23}-c_{12}\, s_{23}\, s_{13}\, \e^{i\delta_{13}} &
c_{12}\, c_{23}- s_{12}\, s_{23}\, s_{13}\, \e^{i\delta_{13}} &
s_{23}\, c_{13}  \\[2pt]
s_{12}\, s_{23}-c_{12}\, c_{23}\, s_{13}\, \e^{i\delta_{13}} &
-c_{12}\, s_{23}- s_{12}\, c_{23}\, s_{13}\, \e^{i\delta_{13}} &
c_{23}\, c_{13}
\ea
\right] .
\ee
Here \ $c_{ij} \equiv \cos{\theta_{ij}}$ \ and \ $s_{ij} \equiv
\sin{\theta_{ij}}\,$, with $i$ and $j$ being `generation' labels
($i,j=1,2,3$). The real angles $\theta_{12}$, $\theta_{23}$ and
$\theta_{13}$ can all be made to lie in the first quadrant, by an
appropriate redefinition of quark field phases; then, \ $c_{ij}\geq
0\,$, $s_{ij}\geq 0$ \ and \ $0\leq \delta_{13}\leq 2\pi\,$.

Notice that $\delta_{13}$ is the only complex phase in the SM
Lagrangian. Therefore, it is the only possible source of $\cCP$-violation
phenomena. In fact, it was for this reason that the third generation
was assumed to exist \cite{KM:73},
before the discovery of the $b$ and the $\tau$.
With two generations, the SM could not explain the observed
$\cCP$ violation in the $K$ system.

\subsection{Quark mixing}

\begin{figure}[tbh]\centering
\includegraphics[width=12cm]{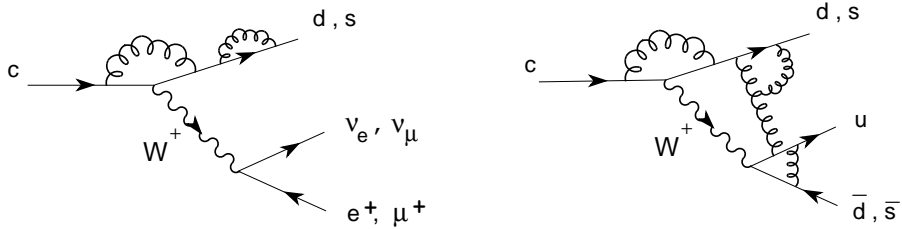}
\vskip -.3cm
\caption{Determinations of $\bV_{\! ij}$ are done
in semileptonic quark decays (left), where
a single quark current is present.
Hadronic decay modes (right) involve two
different quark currents and are more affected
by QCD effects
(gluons can couple everywhere).}
\label{fig:cDecay}
\end{figure}

Our knowledge of the charged-current parameters is unfortunately not
so good as in the neutral-current case. In order to measure the CKM
matrix elements, one needs to study hadronic weak decays of the type
\ $H\to H'\, l^- \bar\nu_l$ \ or \ $H\to H'\, l^+ \nu_l$, which are
associated with the corresponding quark transitions $d_j\to u_i\,
l^-\bar\nu_l$ \  and \ $u_i\to d_j\, l^+\nu_l$
(Fig.~\ref{fig:cDecay}). Since quarks are confined within hadrons,
the decay amplitude
\bel{eq:T_decay}
T[H\to H'\, l^- \bar\nu_l]\;\approx\; {G_F\over\sqrt{2}} \;\bV_{\! ij}\;\,
\langle H'|\, \bar u_i\, \gamma^\mu (1-\gamma_5)\, d_j\, | H\rangle \;\,
\left[\,\bar l \,\gamma_\mu (1-\gamma_5) \,\nu_l\,\right]
\ee
always involves an hadronic matrix element of the weak left current.
The evaluation of this matrix element is a non-perturbative QCD
problem, which introduces unavoidable theoretical uncertainties.

One usually looks for a semileptonic transition where the matrix
element can be fixed at some kinematical point by a symmetry
principle. This has the virtue of reducing the theoretical
uncertainties to the level of symmetry-breaking corrections and
kinematical extrapolations. The standard example is a \ $0^-\to 0^-$
\ decay such as \ $K\to\pi l\nu\,$, $D\to K l\nu$ \ or \ $B\to D
l\nu\,$. Only the vector current can contribute in this case:
\bel{eq:vector_me}
\langle P'(k')| \,\bar u_i \,\gamma^\mu\, d_j\, | P(k)\rangle \; = \; C_{PP'}\,
\left\{\, (k+k')^\mu\, f_+(t)\, +\, (k-k')^\mu\, f_-(t)\,\right\}\, .
\ee
Here, $C_{PP'}$ is a Clebsh--Gordan factor and $t=(k-k')^2\equiv
q^2$. The unknown strong dynamics is fully contained in the form
factors $f_\pm(t)$. In the limit of equal quark masses,
$m_{u_i}-m_{d_j}=0$, the divergence of the vector current is zero;
thus \ $q_\mu\left(\bar u_i \gamma^\mu d_j\right) = 0$, which
implies \ $f_-(t)=0$ \ and, moreover, $f_+(0)=1$ \ to all orders in
the strong coupling because the associated flavour charge is a
conserved quantity.\footnote{
This is completely analogous to the
electromagnetic charge conservation in QED.
The conservation of the
electromagnetic current implies that the proton electromagnetic
form factor does not get any QED or QCD correction at $q^2=0$ and,
therefore,
$Q(p)=2\, Q(u)+Q(d)=|Q(e)|$. A detailed proof can be found in
Ref.~\cite{PI:96}.}
Therefore, one only needs to estimate the corrections induced by the
quark mass differences.

Since
$q^\mu\,\left[\bar l\gamma_\mu (1-\gamma_5)\nu_l\right]\sim m_l$, the contribution
of $f_-(t)$ is kinematically suppressed in the electron and muon modes.
The decay width can then be written as
\bel{eq:decay_width}
\Gamma(P\to P' l \nu)
\; =\; {G_F^2 M_P^5\over 192\pi^3}\; |\bV_{\! ij}|^2\; C_{PP'}^2\;
|f_+(0)|^2\; \cI\; \left(1+\delta_{\rms RC}\right)\, ,
\ee
where $\delta_{\rms RC}$ is an electroweak radiative
correction factor and $\cI$ denotes a phase-space integral,
which in the $m_l=0$ limit takes the form
\bel{eq:ps_integral}
\cI\;\approx\;\int_0^{(M_P-M_{P'})^2} {dt\over M_P^8}\;
\lambda^{3/2}(t,M_P^2,M_{P'}^2)\;
\left| {f_+(t)\over f_+(0)}\right|^2\, .
\ee
The usual procedure to determine $|\bV_{\! ij}|$ involves three steps:
\begin{enumerate}
\item Measure the shape of the $t$ distribution. This fixes 
$|f_+(t)/f_+(0)|$ and therefore determines $\cI$.
\item Measure the total decay width $\Gamma$. Since $G_F$ is already known
from $\mu$ decay, one gets then an
experimental value for the product $|f_+(0)|\, |\bV_{\! ij}|$.
\item Get a theoretical prediction for $f_+(0)$.
\end{enumerate}
It is important to realize that theoretical input is always needed.
Thus, the accuracy of the $|\bV_{\! ij}|$ determination is limited
by our ability to calculate the relevant hadronic input.

\begin{table}[bth]
\caption{Direct determinations of the CKM matrix elements $\bV_{\!
ij}$. For $|\bV_{\! tb}|$, 95\% C.L. limits are given.}
\begin{center}
\renewcommand{\arraystretch}{1.23}
\begin{tabular}{c@{\hspace{.75cm}}cc}    
\hline\hline
 \raisebox{-.75ex}{CKM entry} & \raisebox{-.75ex}{Value} &
 \raisebox{-.75ex}{Source}
\\[7pt] \hline
$|\mathbf{V}_{\! ud}|$ & $0.97377\pm 0.00027$ & Nuclear $\beta$
decay \ \cite{PDG}
\\
& $0.9746\pm 0.0019$ & $n\to p\, e^-\bar\nu_e$ \ \cite{PDG}
\\
& $0.9728\pm 0.0030$ & $\pi^+\to \pi^0\, e^+\nu_e$ \
\cite{PIBETA:04}
\\
& $0.97378\pm 0.00027$ & average
\\ \hline
$|\mathbf{V}_{\! us}|$ & $0.2234\pm 0.0024$ & $K\to\pi l^+\nu_l$ \
\cite{PDG,Kl3,Kl3_th}
\\
& $0.2220\pm 0.0033$ & $\tau$ decays \ \cite{GJPPS:05}
\\
& $0.2226\;{}^{+\;\, 0.0026}_{-\;\, 0.0014}\;\:$ &
$K^+/\pi^+\to\mu^+\nu_\mu$, $\bV_{\! ud}\,$ \
\cite{PDG,MA:04,JOP:06,MILC}
\\
& $0.226\pm 0.005$ & Hyperon decays \ \cite{MP:05,FL:04,CSW:03}
\\
& $0.2230\pm 0.0015$ & average
\\ \hline
$|\mathbf{V}_{\! cd}|$ & $0 .213 \pm 0.022$ & $D \to \pi
l\,\bar\nu_l$ \ \cite{PDG}
\\
& $0.230\pm 0.011$ & $\nu\, d \to c\, X$ \ \cite{PDG}
\\
& $0.227\pm 0.010$ & average
\\ \hline
$|\mathbf{V}_{\! cs}|$ & $0.957\pm 0.095$ & $D \to K l\,\bar\nu_l$ \
\cite{PDG}
\\
& $0.94\;{}^{+\; 0.35}_{-\; 0.29}\;\:$ & $W^+\to c\bar s$ \
\cite{PDG}
\\
& $0.974\pm 0.013$ & $W^+\to \mathrm{had.}\,$, $\bV_{\! uj}\,$,
$\bV_{\! cd}\,$, $\bV_{\! cb}\,$ \ \cite{LEPEWWG,LEPEWWG_SLD:06}
\\ \hline
$|\mathbf{V}_{\! cb}|$ & $0.0392\pm 0.0016$ & $B\to D^*
l\,\bar\nu_l$ \ \cite{PDG,HFAG:07}
\\
& $0.0417\pm 0.0007$ & $b\to c\, l\,\bar\nu_l$ \ \cite{PDG,HFAG:07}
\\
& $0.0413\pm 0.0006$ & average
\\ \hline
$|\mathbf{V}_{\! ub}|$ & $0.0039\pm 0.0006$ &
 $B\to\pi\, l\,\bar\nu_l$ \ \cite{PDG,HFAG:07}     
\\
& $0.0045\pm 0.0003$ & $b\to u\, l\,\bar\nu_l$ \ \cite{PDG,HFAG:07}
\\
& $0.0044\pm 0.0003$ & average
\\ \hline
\raisebox{-.75ex}{ $|\mathbf{V}_{\! tb}|\, / \sqrt{\sum_q
|\mathbf{V}_{\! tq}|^2}$} & \raisebox{-.75ex}{$> 0.78$} &
\raisebox{-.75ex}{$t\to b\, W / q\, W$ \ \cite{CDF:05,D0:06}}
\\[7pt]
$|\mathbf{V}_{\! tb}|$ & $>0.68\quad ; \quad \le 1$ & $p\bar p\to
tb+X$ \ \cite{D0:07}
\\[7pt]  \hline\hline
\end{tabular}
\end{center}
\label{tab:V_CKM}
\end{table}

The conservation of the vector and axial-vector QCD currents in the
massless quark limit allows for accurate determinations of the
light-quark mixings $|\bV_{\! ud}|$ and $|\bV_{\! us}|$. The present
values are shown in Table~\ref{tab:V_CKM}, which takes into account
the recent changes in the $K\to\pi e^+\nu_e$ data \cite{PDG,Kl3} and
the new $|\bV_{\! us}|$ determinations from
Cabibbo suppressed tau decays \cite{GJPPS:05} and from the ratio of
decay amplitudes
$\Gamma(K^+\to\mu^+\bar\nu_\mu)/\Gamma(\pi^+\to\mu^+\bar\nu_\mu)$
\cite{MA:04,JOP:06,MILC}. Since $|\bV_{\! ub}|^2$ is tiny, these two
light quark entries provide a sensible test of the unitarity of the
CKM matrix:
\bel{eq:unitarity_test} |\bV_{\! ud}|^2 + |\bV_{\! us}|^2 + |\bV_{\!
ub}|^2 \, = \, 0.9980\pm 0.0012 \, . \ee
It is important to notice that at the quoted level of uncertainty
radiative corrections play a crucial role.

In the limit of very heavy quark masses, QCD has additional
symmetries \cite{IW:89,GR:90,EH:90,GE:90} which can be used to make
rather precise determinations of $|\bV_{\! cb}|$, either from
exclusive decays such as $B\to D^* l\bar\nu_l$ \cite{NE:91,LU:90} or
from the inclusive analysis of $b\to c\, l\,\bar\nu_l$ transitions.
The control of theoretical uncertainties is much more difficult for
$|\bV_{\! ub}|$, $|\bV_{\! cd}|$ and $|\bV_{\! cs}|$, because the
symmetry arguments associated with the light and heavy quark limits
get corrected by sizeable symmetry-breaking effects.

The most precise determination of $|\bV_{\! cd}|$ is based on
neutrino and antineutrino interactions. The difference of the ratio
of double-muon to single-muon production by neutrino and
antineutrino beams is proportional to the charm cross-section off
valence $d$ quarks and, therefore, to $|\bV_{\! cd}|$. A direct
determination of $|\bV_{\! cs}|$ can be also obtained from
charm-tagged $W$ decays at LEP2. Moreover, the ratio of the total
hadronic decay width of the $W$ to the leptonic one provides the sum
\cite{LEPEWWG,LEPEWWG_SLD:06}
\bel{eq:unitarity_test2}
\sum_{\ba\scriptstyle i\, =\, u,c \\[-6pt]
\scriptstyle j\,  =\,  d, s, b\ea}\; |\bV_{\! ij}|^2\; = \; 1.999\pm
0.025\, . \ee
Although much less precise than Eq.~\eqn{eq:unitarity_test}, this
result test unitarity at the 1.25\% level. From
Eq.~\eqn{eq:unitarity_test2} one can also obtain a tighter
determination of $|\bV_{\! cs}|$, using the experimental knowledge
on the other CKM matrix elements, i.e., \ $|\bV_{\! ud}|^2 +
|\bV_{\! us}|^2 + |\bV_{\! ub}|^2 + |\bV_{\! cd}|^2 + |\bV_{\!
cb}|^2 = 1.0512\pm 0.0058\,$. This gives the most accurate and final
value of $|\bV_{\! cs}|$ quoted in Table~\ref{tab:V_CKM}.

The measured entries of the CKM matrix show a hierarchical pattern, with the
diagonal elements being very close to one, the ones connecting the
two first generations having a size
\bel{eq:lambda} \lambda\approx |\bV_{\!\! us}| = 0.2230\pm 0.0015 \,
, \ee
the mixing between the second and third families being of order
$\lambda^2$, and the mixing between the first and third quark generations
having a much smaller size of about $\lambda^3$.
It is then quite practical to use the
approximate parametrization \cite{WO:83}:

\bel{eq:wolfenstein}
\bV\; =\; \left[ \bath\displaystyle
1- {\lambda^2 \over 2} & \lambda & A\lambda^3 (\rho - i\eta)
\\[8pt]
-\lambda &\displaystyle 1 -{\lambda^2 \over 2} & A\lambda^2
\\[8pt]
A\lambda^3 (1-\rho -i\eta) & -A\lambda^ 2 &  1
\ea\right]\; +\; O\left(\lambda^4 \right) \, ,
\ee
where
\bel{eq:circle} A\approx {|\bV_{\! cb}|\over\lambda^2} = 0.831\pm
0.014 \, , \qquad\qquad \sqrt{\rho^2+\eta^2} \,\approx\,
\left|{\bV_{\! ub}\over \lambda \bV_{\! cb}}\right| \, =\, 0.478\pm
0.033 \, . \ee
Defining to all orders in $\lambda$ \cite{BLO:94}
$s_{12}\equiv\lambda$, $s_{23}\equiv A\lambda^2$ and $s_{13}\,
\e^{-i\delta_{13}}\equiv A\lambda^3 (\rho-i\eta)$,
Eq.~\eqn{eq:wolfenstein} just corresponds to a Taylor expansion of
Eq.~\eqn{eq:CKM_pdg} in powers of $\lambda$.

\subsection{CP Violation}
\label{subsec:CP-Violation}

While parity and charge conjugation are violated by the weak
interactions in a maximal way, the product of the two discrete
transformations is still a good symmetry (left-handed fermions
$\leftrightarrow$ right-handed antifermions). In fact, $\cCP$
appears to be a symmetry of nearly all observed phenomena. However,
a slight violation of the $\cCP$ symmetry at the level of $0.2\%$ is
observed in the neutral kaon system and more sizeable signals of
$\cCP$ violation have been recently established at the B factories.
Moreover, the huge matter--antimatter asymmetry present in our
Universe is a clear manifestation of $\cCP$ violation and its
important role in the primordial baryogenesis.

The $\cCPT$ theorem guarantees that the product of the three
discrete transformations is an exact symmetry of any local and
Lorentz-invariant quantum field theory preserving micro-causality.
Therefore, a violation of $\cCP$ requires a corresponding
violation of time reversal. Since $\cT$ is an antiunitary
transformation, this requires the presence of relative complex
phases between different interfering amplitudes.

The electroweak SM Lagrangian only contains a single complex phase
$\delta_{13}$ ($\eta$). This is the sole possible source of $\cCP$
violation and, therefore, the SM predictions for $\cCP$-violating
phenomena are quite constrained. The CKM mechanism requires several
necessary conditions in order to generate an observable
$\cCP$-violation effect. With only two fermion generations, the
quark mixing mechanism cannot give rise to $\cCP$ violation;
therefore, for $\cCP$ violation to occur in a particular process,
all three generations are required to play an active role. In the
kaon system, for instance, $\cCP$-violation effects can only appear
at the one-loop level, where the top quark is present. In addition,
all CKM matrix elements must be non-zero and the quarks of a given
charge must be non-degenerate in mass. If any of these conditions
were not satisfied, the CKM phase could be rotated away by a
redefinition of the quark fields. $\cCP$-violation effects are then
necessarily proportional to the product of all CKM angles, and
should vanish in the limit where any two (equal-charge) quark masses
are taken to be equal. All these necessary conditions can be
summarized in a very elegant way as a single requirement on the
original quark mass matrices $\bM'_u$ and $\bM'_d$ \cite{JA:85}:
\be
\cCP \:\mbox{\rm violation} \qquad \Longleftrightarrow \qquad
\mbox{\rm Im}\left\{\det\left[\bM_u^\prime \bM^{\prime\dagger}_u\, ,
  \,\bM^{\prime\phantom{\dagger}}_d \bM^{\prime\dagger}_d\right]\right\} \not=0 \, .
\ee

Without performing any detailed calculation, one can make the
following general statements on the implications of the CKM mechanism
of $\cCP$ violation:

\bi
\item[--]
Owing to unitarity, for any choice of $i,j,k,l$ (between 1 and 3),
\beqn\label{eq:J_relation}
\mbox{\rm Im}\left[
\bV^{\phantom{*}}_{ij}\bV^*_{ik}\bV^{\phantom{*}}_{lk}\bV^*_{lj}\right]
\, =\, \cJ \sum_{m,n=1}^3 \epsilon_{ilm}\epsilon_{jkn}\, ,
\qquad\quad\\
\cJ \, =\, c_{12}\, c_{23}\, c_{13}^2\, s_{12}\, s_{23}\, s_{13}\, \sin{\delta_{13}}
\,\approx\, A^2\lambda^6\eta \, < \, 10^{-4}\, .
\eeqn
Any $\cCP$-violation observable involves the product $\cJ$
\cite{JA:85}. Thus, violations of the $\cCP$ symmetry are
necessarily small.
\item[--] In order to have sizeable $\cCP$-violating asymmetries
$\cA\equiv (\Gamma - \overline{\Gamma})/(\Gamma +
\overline{\Gamma})$, one should look for very suppressed decays,
where the decay widths already involve small CKM matrix elements.
\item[--] In the SM, $\cCP$ violation is a low-energy phenomenon,
in the sense that any effect should disappear when the quark mass
difference $m_c-m_u$ becomes negligible.
\item[--] $B$ decays are the optimal place for $\cCP$-violation signals to show up.
They involve small CKM matrix elements and are the lowest-mass
processes where the three quark generations play a direct
(tree-level) role. \ei

The SM mechanism of $\cCP$ violation is based on the unitarity of the
CKM matrix. Testing the constraints implied by unitarity
is then a way to test the source of $\cCP$ violation.
The unitarity tests in Eqs.~\eqn{eq:unitarity_test} and
\eqn{eq:unitarity_test2} involve only the moduli of the CKM parameters,
while $\cCP$ violation has to do with their phases.
More interesting are the off-diagonal unitarity conditions:
\beqn\label{eq:triangles}
\bV^\ast_{\! ud}\bV^{\phantom{*}}_{\! us} \, +\,
\bV^\ast_{\! cd}\bV^{\phantom{*}}_{\! cs} \, +\,
\bV^\ast_{\! td}\bV^{\phantom{*}}_{\! ts} & = & 0 \, ,
\\[5pt]
\bV^\ast_{\! us}\bV^{\phantom{*}}_{\! ub} \, +\,
\bV^\ast_{\! cs}\bV^{\phantom{*}}_{\! cb} \, +\,
\bV^\ast_{\! ts}\bV^{\phantom{*}}_{\! tb} & = & 0 \, ,
\\[5pt]
\bV^\ast_{\! ub}\bV^{\phantom{*}}_{\! ud} \, +\,
\bV^\ast_{\! cb}\bV^{\phantom{*}}_{\! cd} \, +\,
\bV^\ast_{\! tb}\bV^{\phantom{*}}_{\! td} & = & 0 \, .
\eeqn
These relations can be visualized by triangles in a complex
plane which, owing to Eq.~\eqn{eq:J_relation}, have the
same area $|\cJ|/2$.
In the absence of $\cCP$ violation, these triangles would degenerate
into segments along the real axis.

In the first two triangles, one side is much shorter than the other
two (the Cabibbo suppression factors of the three sides are
$\lambda$, $\lambda$ and $\lambda^5$ in the first triangle, and
$\lambda^4$, $\lambda^2$ and $\lambda^2$ in the second one). This is
why $\cCP$ effects are so small for $K$ mesons (first triangle), and
why certain  asymmetries in $B_s$ decays are predicted to be tiny
(second triangle).
The third triangle looks more interesting, since the three sides
have a similar size of about $\lambda^3$. They are small, which
means that the relevant $b$-decay branching ratios are small, but
once enough $B$ mesons have been produced, the $\cCP$-violation
asymmetries are sizeable. The present experimental constraints on
this triangle are shown in Fig.~\ref{fig:UTfit}, where it has been
scaled by dividing its sides by $\bV^\ast_{\!
cb}\bV^{\phantom{*}}_{\! cd}$. This aligns one side of the triangle
along the real axis and makes its length equal to 1; the coordinates
of the 3 vertices are then $(0,0)$, $(1,0)$ and
$(\bar\rho,\bar\eta)\equiv (1-\lambda^2/2) (\rho,\eta)$.

\begin{figure}[bht]\centering
\includegraphics[width=12cm]{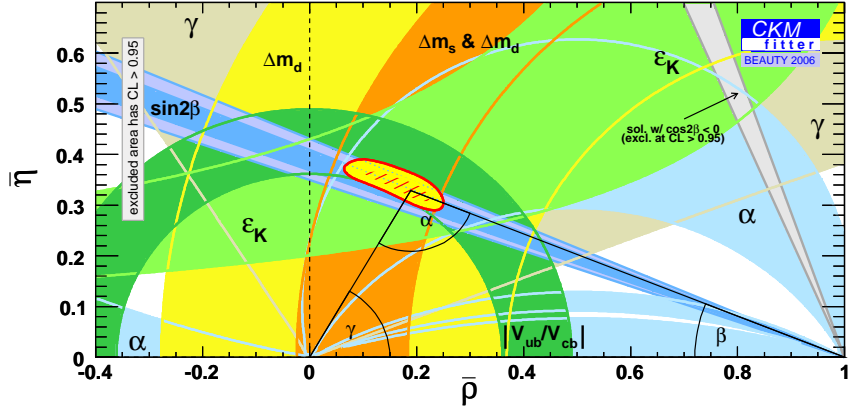}
\caption{Experimental constraints on the SM unitarity triangle
\cite{CKMfitter}.} \label{fig:UTfit}
\end{figure}

\begin{figure}[tbh]\centering
\includegraphics[width=9cm]{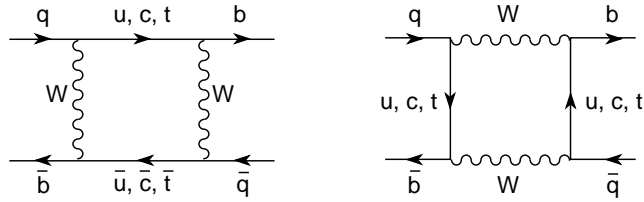}
\caption{$B^0$--$\bar B^0$ mixing diagrams. Owing to the unitarity
of the CKM matrix, the mixing vanishes for equal up-type quark
masses (GIM mechanism). The mixing amplitude is then proportional to
the  mass (squared) splittings between the $u$, $c$ and $t$ quarks,
and is completely dominated by the top contribution.}
\label{fig:Bmixing}
\end{figure}

One side of the unitarity triangle has been already determined in
Eq.~\eqn{eq:circle} from the ratio $|\bV_{\! ub}/\bV_{\! cb}|$. The
other side can be obtained from the measured mixing between the
$B^0_d$ and $\bar B^0_d$ mesons (Fig.~\ref{fig:Bmixing}), $\Delta
M_d = 0.507\pm 0.004\;\mathrm{ps}^{-1}$ \cite{HFAG:07}, which fixes
$|\bV_{\! tb}|$. Additional information has been provided by the
recent observation of $B^0_s$--$\bar B^0_s$ oscillations at CDF,
implying $\Delta M_s = 17.77\pm 0.12\;\mathrm{ps}^{-1}$
\cite{CDF:06}. From the experimental ratio $\Delta M_d/\Delta M_s =
0.0286\pm 0.0003$, one obtains $|\bV_{\! td}|/|\bV_{\! ts}|$. A more
direct constraint on the parameter $\eta$ is given by the observed
$\cCP$ violation in $K^0\to 2\pi$ decays. The measured value of
$|\varepsilon_K| = (2.232\pm 0.007)\cdot 10^{-3}$ \cite{PDG}
determines the parabolic region shown in Fig.~\ref{fig:UTfit}.

$B^0$ decays into $\cCP$ self-conjugate final states provide
independent ways to determine the angles of the unitarity triangle
\cite{CS:80,BS:81}. The $B^0$ (or $\bar B^0$) can decay directly to
the given final state $f$, or do it after the meson has been changed
to its antiparticle via the mixing process. $\cCP$-violating effects
can then result from the interference of these two contributions.
The time-dependent $\cCP$-violating rate asymmetries contain direct
information on the CKM parameters. The gold-plated decay mode is
$B^0_d\to J/\psi K_S$, which gives a clean measurement of
$\beta\equiv -\arg(\mathbf{V}^{\phantom{*}}_{\! cd}\mathbf{V}^*_{\!
cb}/ \mathbf{V}^{\phantom{*}}_{\! td}\mathbf{V}^*_{\! tb})$, without
strong-interaction uncertainties. Including the information obtained
from other $b\to c\bar c s$ decays, one gets \cite{HFAG:07}:
\begin{equation}\label{eq:beta}
\sin{2\beta} = 0.68\pm 0.03\, .
\end{equation}

Many additional tests of the CKM matrix from different $B$ decay
modes are being pursued at the $B$ factories. Determinations of the
other two angles of the unitarity triangle, $\alpha\equiv -\arg(
   \mathbf{V}^{\phantom{*}}_{\! td}\mathbf{V}^*_{\! tb}/
   \mathbf{V}^{\phantom{*}}_{\! ud}\mathbf{V}^*_{\! ub})$
and
$\gamma\equiv -\arg(
   \mathbf{V}^{\phantom{*}}_{\! ud}\mathbf{V}^*_{\! ub}/
   \mathbf{V}^{\phantom{*}}_{\! cd}\mathbf{V}^*_{\! cb})$,
have been already obtained \cite{HFAG:07,BI:07}, and are included in
the global fit shown in Fig.~\ref{fig:UTfit} \cite{CKMfitter,UTfit}.
Complementary and very valuable information could be also obtained
from the kaon decay modes $K^\pm\to\pi^\pm\nu\bar\nu$,
$K_L\to\pi^0\nu\bar\nu$ and $K_L\to\pi^0e^+e^-$ \cite{BSU:04}.

\subsection{Lepton mixing}

%
\begin{figure}[tbh]
\centering
\includegraphics[width=9cm,clip]{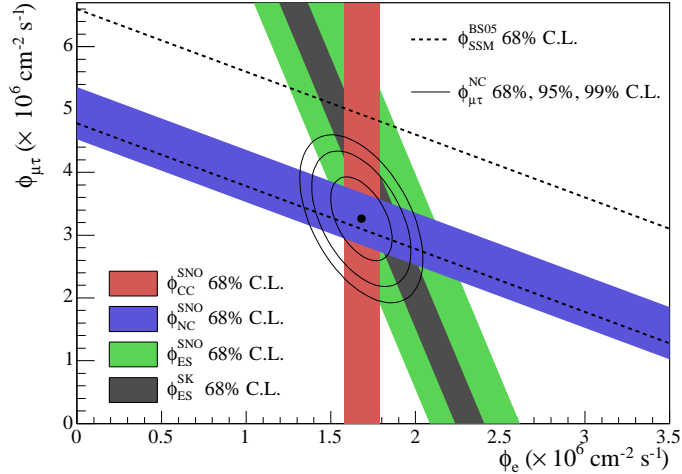}
\caption{Measured fluxes of ${}^8B$ solar neutrinos of $\nu_\mu$ or
$\nu_\tau$ type ($\phi_{\mu,\tau}$) versus the flux of $\nu_e$
($\phi_e$)~\cite{SNO}.} \label{fig:SNO}
\end{figure}

The so-called `solar neutrino problem' has been a long-standing
question, since the very first chlorine experiment at the Homestake
mine \cite{DA:68}. The flux of solar $\nu_e$ neutrinos reaching the
Earth has been measured by several experiments to be significantly
below the standard solar model prediction \cite{BP:04}. More
recently, the Sudbury Neutrino Observatory has provided strong
evidence that neutrinos do change flavour as they propagate from the
core of the Sun \cite{SNO}, independently of solar model flux
predictions. SNO is able to detect neutrinos through three different
reactions: the charged-current process $\nu_e d \to e^-pp$ which is
only sensitive to $\nu_e$, the neutral current transition $\nu_x d
\to \nu_x pn$ which has equal probability for all active neutrino
flavours, and the elastic scattering $\nu_x e^-\to\nu_x e^-$ which
is also sensitive to $\nu_\mu$ and $\nu_\tau$, although the
corresponding cross section is a factor $6.48$ smaller than the
$\nu_e$ one. The measured neutrino fluxes, shown in
Fig.~\ref{fig:SNO}, demonstrate the existence of a non-$\nu_e$
component in the solar neutrino flux at the $5.3\,\sigma$ level. The
SNO results are in good agreement with the Super-Kamiokande solar
measurements \cite{SKsolar} and have been further reinforced with
the more recent KamLAND data, showing that $\bar\nu_e$ from nuclear
reactors disappear over distances of about 180 Km \cite{KamLAND}.

Another evidence of oscillations has been obtained from atmospheric
neutrinos. The known discrepancy between the experimental
observations and the predicted ratio of muon to electron neutrinos
has become much stronger with the high precision and large
statistics of Super-Kamiokande \cite{SKatm}. The atmospheric anomaly
appears to originate in a reduction of the $\nu_\mu$ flux, and the
data strongly favours the $\nu_\mu\to\nu_\tau$ hypothesis. This
result has been confirmed by K2K \cite{K2K} and MINOS \cite{MINOS},
observing the disappearance of accelerator $\nu_\mu$'s at distances
of 250 and 735 Km, respectively. Super-Kamiokande has recently
reported statistical evidence of $\nu_\tau$ appearance at the
$2.4\,\sigma$ level \cite{SKatm}. The direct detection of the
produced $\nu_\tau$ is the main goal of the ongoing CERN to Gran
Sasso neutrino program.

Thus, we have now clear experimental evidence that neutrinos are
massive particles and there is mixing in the lepton sector. Figures
\ref{fig:SolarNu} and \ref{fig:AtmosNu} show the present information
on neutrino oscillations, from solar, atmospheric, accelerator and
reactor neutrino data. A global analysis, combining the full set of
data, leads to the following preferred ranges for the oscillation
parameters \cite{PDG}:
\bel{nu_mix}
 \Delta m^2_{21}\; = \; \left( 8.0\, {}^{+\, 0.4}_{-\, 0.3}\right)
 \cdot 10^{-5}\;\mathrm{eV}^2\; , \qquad
 1.9 \cdot 10^{-3}\; <\; |\Delta m^2_{32}|\; /\; \mathrm{eV}^2 \; <\;
 3.0 \cdot 10^{-3} \; ,
\ee
\bel{nu_mix2}
 \sin^2{(2\theta_{12})}\; =\; 0.86\, {}^{+\, 0.03}_{-\, 0.04}
 \; , \qquad
 \sin^2{(2\theta_{23})}\; >\; 0.92 \; , \qquad
 \sin^2{(2\theta_{13})}\; <\; 0.19 \; ,
\ee
where $\Delta m^2_{ij}\equiv m^2_i - m^2_j$ are the mass squared
differences between the neutrino mass eigenstates $\nu_{i,j}$ and
$\theta_{ij}$ the corresponding mixing angles in the standard
three-flavour parametrization \cite{PDG}. The ranges indicate 90\%
C.L. bounds.
In the limit $\theta_{13}=0$, solar and atmospheric neutrino
oscillations decouple because $\Delta m^2_\odot \ll\Delta
m^2_\mathrm{atm}$. Thus, $\Delta m^2_{21}$, $\theta_{12}$ and
$\theta_{13}$ are constrained by solar data, while atmospheric
experiments constrain $\Delta m^2_{32}$, $\theta_{23}$ and
$\theta_{13}$. The angle $\theta_{13}$ is strongly constrained by
the CHOOZ reactor experiment \cite{CHOOZ}. New planned reactor
experiments, T2K and NO$\nu$A are expected to achieve sensitivities
around $\sin^2{(2\theta_{13})}\sim 0.01$.

\begin{figure}[tbh]
\begin{minipage}[c]{.45\linewidth}\centering\vskip .2cm
\includegraphics[height=6cm,clip]{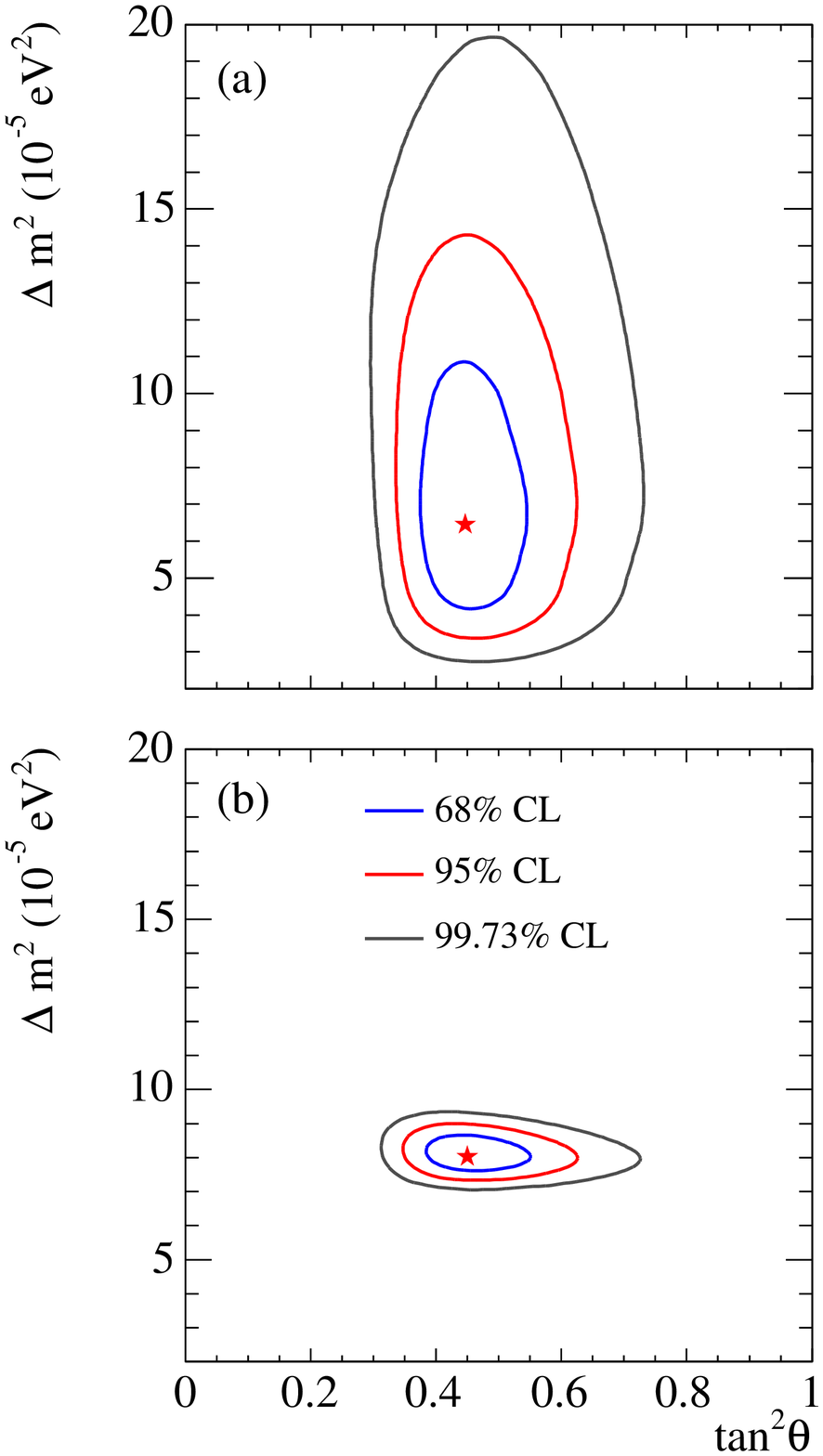}
\caption{Allowed regions for $2\nu$ oscillations for the combination
of solar ($\nu_e$) and KamLAND  ($\bar\nu_e$) data, assuming
$\mathcal{CPT}$ symmetry \cite{SNO}.} \label{fig:SolarNu}
\end{minipage}
\hfill
\begin{minipage}[c]{.45\linewidth}\centering
\includegraphics[height=6.3cm,clip]{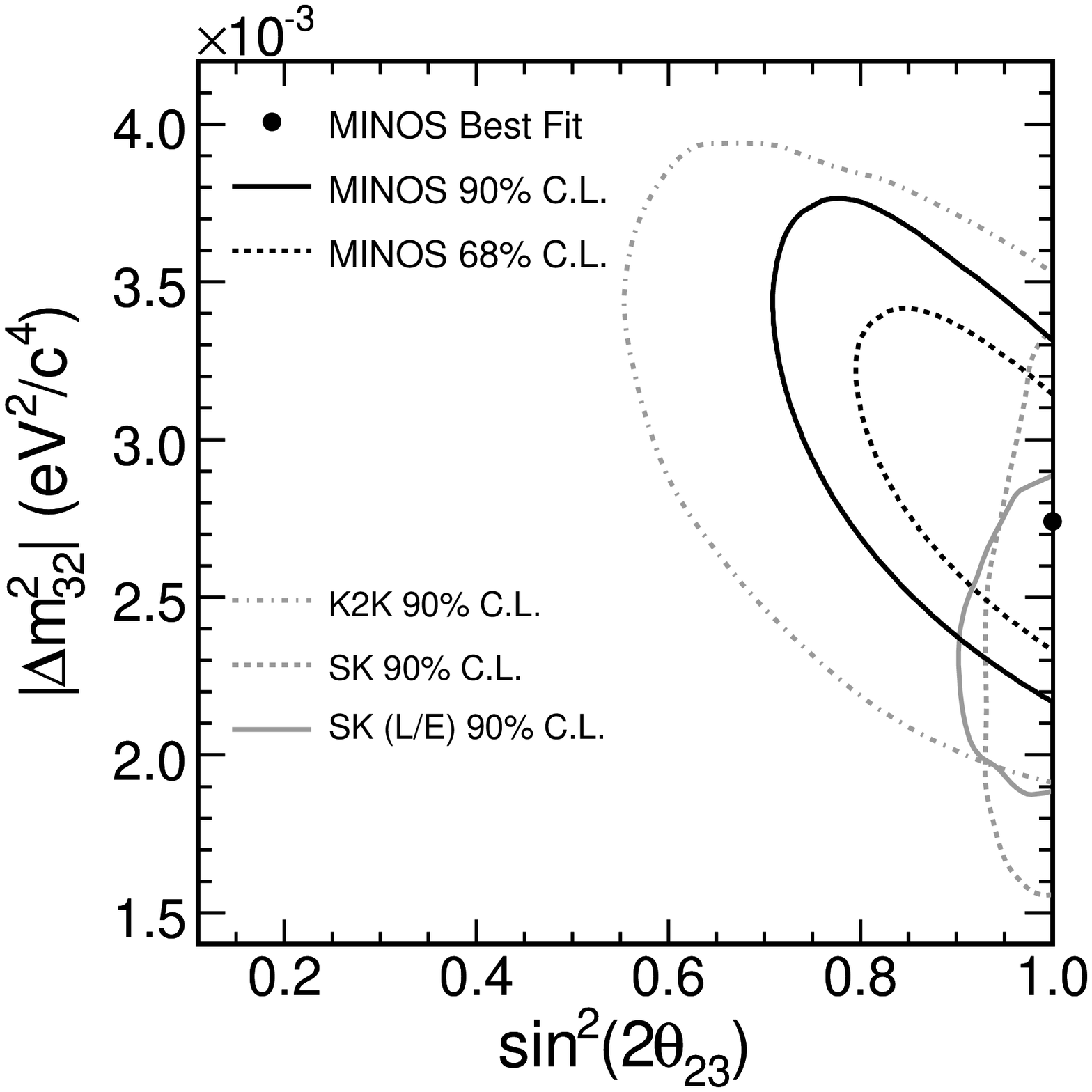}
\vskip -.1cm
 \caption{MINOS allowed regions for $\nu_\mu$ disappearance
 oscillations, compared with K2K and Super-Kamiokande results
\cite{MINOS}.} \label{fig:AtmosNu}
\end{minipage}
\end{figure}

Non-zero neutrino masses constitute a clear indication of new
physics beyond the SM. Right-handed neutrinos are an obvious
possibility to incorporate Dirac neutrino masses. However, the
$\nu_{iR}$ fields would be $SU(3)_C\otimes SU(2)_L\otimes U(1)_Y$
singlets, without any SM interaction. If such objects do exist, it
would seem natural to expect that they are able to communicate with
the rest of the world through some still unknown dynamics. Moreover,
the SM gauge symmetry would allow for a right-handed Majorana
neutrino mass term,
\bel{eq:Majorana} \cL_M = -{1\over 2}\, \overline{\nu_{iR}^c}\,
M_{ij}\, \nu^{}_{jR} \, +\,\mathrm{h.c.} \, , \ee
where $\nu_{iR}^c\equiv\cC\,\bar\nu_{iR}^T$ denotes the
charge-conjugated field. The Majorana mass matrix $M_{ij}$ could
have an arbitrary size, because it is not related to the ordinary
Higgs mechanism. Since both fields $\nu^{}_{iR}$ and
$\overline{\nu_{iR}^c}$ absorb $\nu$ and create $\bar\nu$, the
Majorana mass term mixes neutrinos and anti-neutrinos, violating
lepton number by two units. Clearly, new physics is called for.

Adopting a more general effective field theory language, without any
assumption about the existence of right-handed neutrinos or any
other new particles, one can write the most general $SU(3)_C\otimes
SU(2)_L\otimes U(1)_Y$ invariant Lagrangian, in terms of the known
low-energy fields (left-handed neutrinos only). The SM is the unique
answer with dimension four. The first contributions from new physics
appear through dimension-5 operators, and have also a unique form
which violates lepton number by two units \cite{WE:79}:
\bel{eq:WE} \Delta \cL\; =\; - {c_{ij}\over\Lambda}\; \bar
L_i\,\tilde\phi\, \tilde\phi^t\, L_j^c \; + \; \mathrm{h.c.}\, , \ee
where $L_i$ denotes the $i$-flavoured $SU(2)_L$ lepton doublet,
$\tilde\phi \equiv i\,\tau_2\,\phi^*$ and $L_i^c \equiv \mathcal{C}
\bar L_i^T$. Similar operators with quark fields are forbidden, due
to their different hypercharges, while higher-dimension operators
would be suppressed by higher powers of the new-physics scale
$\Lambda$.
After SSB, $\langle\phi^{(0)}\rangle = v/\sqrt{2}$, $\Delta \cL$
generates a Majorana mass term for the
left-handed neutrinos, with\footnote{
This relation generalizes the well-known see-saw mechanism
($m_{\nu_{L}}\sim m^2/\Lambda$) \cite{GMRS:79,YA:79}.}
$M_{ij} = c_{ij} v^2/\Lambda$. Thus, Majorana neutrino masses should
be expected on general symmetry grounds. Taking $m_\nu\gtrsim
0.05$~eV, as suggested by atmospheric neutrino data, one gets
$\Lambda/c_{ij}\lesssim 10^{15}$~GeV, amazingly close to the
expected scale of Gran Unification.

With non-zero neutrino masses, the leptonic charged-current
interactions involve a flavour mixing matrix $\mathbf{V}_L$. The
data on neutrino oscillations imply that all elements of
$\mathbf{V}_L$ are large, except for $(\mathbf{V}_L)_{e3}< 0.18$;
therefore the mixing among leptons appears to be very different from
the one in the quark sector. The number of relevant phases
characterizing the matrix $\mathbf{V}_L$ depends on the Dirac or
Majorana nature of neutrinos, because if one rotates a Majorana
neutrino by a phase, this phase will appear in its mass term which
will no longer be real. With only three Majorana (Dirac) neutrinos,
the $3\times 3$ matrix $\mathbf{V}_L$ involves six (four)
independent parameters: three mixing angles and three (one) phases.

\begin{table}[t]
\caption{Best published limits (90\% C.L.) on
lepton-flavour-violating decays \cite{PDG,LFVbabar,LFVbelle}.}
\label{table:LFV}\vspace{0.2cm}
\renewcommand{\arraystretch}{1.2} 
\begin{tabular}{@{\hskip .1cm}lll}
\hline\hline
 $\Br (\mu^-\to e^-\gamma) < 1.2\cdot 10^{-11}$ &
 $\Br (\mu^-\to e^-2\gamma) < 7.2\cdot 10^{-11}$ &
 $\Br (\mu^-\to e^-e^-e^+) < 1.0\cdot 10^{-12}\hskip -.2cm $
 \\
 $\Br (\tau^-\to \mu^-\gamma) < 4.5\cdot 10^{-8}$ &
 $\Br (\tau^-\to e^-\gamma) < 1.1\cdot 10^{-7}$ &
 $\Br (\tau^-\to e^-e^-\mu^+) < 1.1\cdot 10^{-7}\hskip -.2cm $
 \\
 $\Br (\tau^-\to e^-K_S) < 5.6\cdot 10^{-8}$ &
 $\Br (\tau^-\to \mu^-K_S) < 4.9\cdot 10^{-8}$ &
 $\Br (\tau^-\to \mu^+\pi^-\pi^-) < 0.7\cdot 10^{-7}\hskip -.2cm $
 \\
 $\Br (\tau^-\to \Lambda\pi^-) < 7.2\cdot 10^{-8}$ &
 $\Br (\tau^-\to e^-\pi^0) < 1.4\cdot 10^{-7}$ &
 $\Br (\tau^-\to e^-\pi^+\pi^-) < 1.2\cdot 10^{-7}\hskip -.2cm $
 \\
 $\Br (\tau^-\to \mu^-\pi^0) < 1.1\cdot 10^{-7}$ &
 $\Br (\tau^-\to \mu^-\eta) < 1.3\cdot 10^{-7}$ &
 $\Br (\tau^-\to \mu^-e^+\mu^-) < 1.3\cdot 10^{-7}\hskip -.1cm $
 \\ \hline\hline
\end{tabular}\end{table}

The smallness of neutrino masses implies a strong suppression of
neutrinoless lepton-flavour-violating processes, which can be
avoided in models with other sources of lepton-flavour violation,
not related to $m_{\nu_i}$. Table~\ref{table:LFV} shows the best
published limits on lepton-flavour-violating decays. The B Factories
are pushing the experimental limits on neutrinoless $\tau$ decays
beyond the $10^{-7}$ level, increasing in a drastic way the
sensitivity to new physics scales. Future experiments could push
further some limits to the $10^{-9}$ level, allowing to explore
interesting and totally unknown phenomena. Complementary information
will be provided by the MEG experiment, which will search for
$\mu^+\to e^+\gamma$ events with a sensitivity of $10^{-13}$
\cite{MEG}. There are also ongoing projects at J-PARC aiming to
study $\mu\to e$ conversions in muonic atoms, at the $10^{-18}$
level.

At present, we still ignore whether neutrinos are Dirac or Majorana
fermions. Another important question to be addressed in the future
concerns the possibility of leptonic CP violation and its relevance
for explaining the baryon asymmetry of our Universe through
leptogenesis.


\setcounter{equation}{0}
\section{Summary}

The SM provides a beautiful theoretical framework which is able to
accommodate all our present knowledge on electroweak and strong
interactions. It is able to explain any single experimental fact
and, in some cases, it has successfully passed very precise
tests at the 0.1\% to 1\% level.
In spite of this impressive phenomenological success, the SM leaves
too many unanswered questions to be considered as a complete
description of the fundamental forces. We do not understand yet
why fermions are replicated in three (and only three)
nearly identical copies. Why the pattern of masses and mixings
is what it is?  Are the masses the only difference among the three
families? What is the origin of the SM flavour structure?
Which dynamics is responsible for the observed $\cCP$ violation?

In the gauge and scalar sectors, the SM Lagrangian contains only
four parameters: $g$, $\gp$, $\mu^2$ and $h$. We can trade them by
$\alpha$, $M_Z$, $G_F$ and $M_H$; this has the advantage of using
the three most precise experimental determinations to fix the
interaction. In any case, one describes a lot of physics with only
four inputs. In the fermionic flavour sector, however, the situation
is very different. With $N_G=3$, we have 13 additional free
parameters in the minimal SM: 9 fermion masses, 3 quark mixing
angles and 1 phase. Taking into account non-zero neutrino masses, we
have three more mass parameters plus the leptonic mixings: three
angles and one phase (three phases) for Dirac (or Majorana)
neutrinos.

Clearly, this is not very satisfactory. The source of this
proliferation of parameters is the set of unknown Yukawa couplings
in Eq.~\eqn{eq:N_Yukawa}. The origin of masses and mixings, together
with the reason for the existing family replication, constitute at
present the main open problem in electroweak physics. The problem of
fermion mass generation is deeply related with the mechanism
responsible for the electroweak SSB. Thus, the origin of these
parameters lies in the most obscure part of the SM Lagrangian: the
scalar sector. The dynamics of flavour appears to be `terra
incognita' which deserves a careful investigation.

The SM incorporates a mechanism to generate $\cCP$ violation,
through the single phase naturally occurring in the CKM matrix.
Although the present laboratory experiments are well described, this
mechanism is unable to explain the matter--antimatter asymmetry of
our Universe. A fundamental explanation of the origin of
$\cCP$-violating phenomena is still lacking.

The first hints of new physics beyond the SM have emerged recently,
with convincing evidence of neutrino oscillations showing that
$\nu_e\to\nu_{\mu,\tau}$ and $\nu_\mu\to\nu_\tau$ transitions do
occur. The existence of lepton-flavour violation opens a very
interesting window to unknown phenomena.

The Higgs particle is the main missing block of the SM framework.
The successful tests of the SM quantum corrections with precision
electroweak data confirm the assumed pattern of SSB, but do not
prove the validity of the minimal Higgs mechanism embedded in the
SM. The present experimental bounds \eqn{eq:MH_limits} put the Higgs
hunting within the reach of the new generation of detectors. The LHC
should find out whether such scalar field indeed exists, either
confirming the SM Higgs mechanism or discovering completely new
phenomena.

Many interesting experimental signals are expected to be seen
in the near future. New experiments will probe
the SM to a much deeper level of sensitivity
and will explore the frontier of its possible extensions.
Large surprises may well be expected, probably
establishing the existence of new physics beyond the SM and
offering clues to the problems of mass generation, fermion
mixing and family replication.


\section*{Acknowledgements}

I want to thank the organizers for the charming atmosphere of this
school and all the students for their many interesting questions and
comments. This work has been supported by the EU MRTN-CT-2006-035482
(FLAVIA{\it net}),  MEC (Spain, FPA2004-00996) and Generalitat
Valenciana (GVACOMP2007-156).

\newpage

\appendix
\renewcommand{\theequation}{A.\arabic{equation}}
\setcounter{equation}{0}
\section{Basic Inputs from Quantum Field Theory}

\subsection{Wave equations}

The classical Hamiltonian of a non-relativistic free particle is
given by $H = \vec{p}^{\: 2}/(2m)$. In quantum mechanics, energy and
momentum correspond to operators acting on the particle wave
function. The substitutions \ $H = i \hbar\,
{\partial\over\partial\, t}$ \ and \ $\vec{p} = -i
\hbar\,\vec{\nabla}$ \ lead then to the Schr\"odinger equation:
\bel{eq:Schrodinger}
i \hbar\, {\partial\over\partial t}\, \psi\left(\vec{x},t\right)
\, =\, -{\hbar^2\over 2 m}\, \vec{\nabla}^2\psi\left(\vec{x},t\right)\, .
\ee
We can write the energy and momentum operators in a relativistic
covariant way as \ $p^\mu = i \,\partial^\mu \equiv i \,
{\partial\over\partial x_\mu}\,$, where we have adopted the usual
natural units convention $\hbar = c = 1$. The relation $E^{\, 2}
=\vec{p}^{\: 2} + m^2$ determines the Klein--Gordon equation for a
relativistic free particle:
\bel{eq:KG} \left( \Box + m^2 \right) \phi(x) = 0 \, ,\qquad\qquad
\qquad\qquad \Box \equiv \partial^\mu \partial_\mu =
{\partial^2\over\partial t^2} - \vec{\nabla}^2\, . \ee

The Klein--Gordon equation is quadratic on the time derivative
because relativity puts the space and time coordinates on an equal
footing. Let us investigate whether an equation linear in
derivatives could exist. Relativistic covariance and dimensional
analysis restrict its possible form to
\bel{eq:Dirac}
\left( i\,\gamma^\mu\partial_\mu - m\right) \psi(x) = 0\, .
\ee
Since the r.h.s. is identically zero, we can fix the coefficient of
the mass term to be $-1$; this just deter\-mines the normalization
of the four coefficients $\gamma^\mu$. Notice that $\gamma^\mu$
should transform as a Lorentz four-vector. The solutions of
Eq.~\eqn{eq:Dirac} should also satisfy the Klein--Gordon relation of
Eq.~\eqn{eq:KG}. Applying an appropriate differential operator to
Eq.~\eqn{eq:Dirac}, one can easily obtain the wanted quadratic
equation:
\bel{trick}
- \left( i\,\gamma^\nu\partial_\nu + m\right)
\left( i\,\gamma^\mu\partial_\mu - m\right) \psi(x) = 0
\;\equiv \;\left( \Box + m^2 \right) \psi(x)\, .
\ee
Terms linear in derivatives cancel identically, while the term with two
derivatives reproduces the operator $\Box \equiv\partial^\mu \partial_\mu$
\ provided the coefficients $\gamma^\mu$ satisfy the
algebraic relation
\bel{eq:GammaAlgebra}
\left\{\gamma^\mu , \gamma^\nu\right\}\equiv
\gamma^\mu\gamma^\nu + \gamma^\nu\gamma^\mu = 2\, g^{\mu\nu}\, ,
\ee
which defines the so-called Dirac algebra. Eq.~\eqn{eq:Dirac} is known
as the Dirac equation.

Obviously the components of the four-vector $\gamma^\mu$ cannot
simply be numbers.
The three $2\times 2$ Pauli matrices satisfy \
$\left\{\sigma^i ,\sigma^j\right\} = 2 \,\delta^{ij}$, which is very
close to the relation \eqn{eq:GammaAlgebra}. The lowest-dimensional
solution to the Dirac algebra is obtained with \ $D=4$ \ matrices.
An explicit representation is given by:
\bel{eq:DiracMatrices}
\gamma^0 = \left(\bat I_2 & 0\cr 0 & -I_2\ea\right)
\, , \qquad\qquad
\gamma^i = \left(\bat 0 & \sigma^i\cr -\sigma^i & 0\ea\right)\, .
\ee
Thus, the wave function $\psi(x)$ is a column vector with four
components in the Dirac space. The presence of the Pauli matrices
strongly suggests that it contains two components of spin
$\frac{1}{2}$. A proper physical analysis of its solutions shows
that the Dirac equation describes simultaneously a fermion of spin
$\frac{1}{2}$ and its own antiparticle \cite{Bjorken}.

It turns useful to define the following combinations of gamma matrices:
\bel{eq:gamma5} \sigma^{\mu\nu} \equiv {i\over 2}\,\left[\gamma^\mu
, \gamma^\nu\right] \, ,\qquad\qquad \gamma_5 \equiv \gamma^5 \equiv
i\,\gamma^0\gamma^1\gamma^2\gamma^3 = -{i\over
4!}\,\epsilon_{\mu\nu\rho\sigma}
\gamma^\mu\gamma^\nu\gamma^\rho\gamma^\sigma\, . \ee
In the explicit representation \eqn{eq:DiracMatrices},
\bel{eq:sigmaForm} \sigma^{ij} = \epsilon^{ijk}\,\left(\bat \sigma^k
& 0\cr 0 & \sigma^k\ea\right) \, , \quad\quad \sigma^{0i} =
i\,\left(\bat 0 & \sigma^i \cr \sigma^i & 0\ea\right) \, ,
\quad\quad \gamma_5 =\left(\bat 0 & I_2 \cr I_2 & 0\ea\right) \, .
\ee
The matrix $\sigma^{ij}$ is then related to the spin operator.
Some important properties are:
\bel{eq:Gproperties} \gamma^0\gamma^\mu\gamma^0 =
{\gamma^\mu}^\dagger \, ,\quad\quad \gamma^0\gamma_5\gamma^0 =
-{\gamma_5}^\dagger = -\gamma_5 \, ,\quad\quad \left\{\gamma_5 ,
\gamma^\mu\right\} = 0 \, ,\quad\quad (\gamma_5)^2 = I_4 \, . \ee
Specially relevant for weak interactions are the chirality projectors \
($P_L+P_R=1$)
\bel{eq:PLPR}
P_L\equiv {1-\gamma_5\over 2}
\, ,\quad\quad
P_R\equiv {1+\gamma_5\over 2}
\, ,\quad\quad
P_R^2 = P_R
\, ,\quad\quad
P_L^2 = P_L
\, ,\quad\quad
P_L P_R = P_R P_L = 0
\, ,
\ee
which allow to decompose the Dirac spinor in its left-handed and right-handed
chirality parts:
\bel{eq:chirality}
\psi(x) = \left[ P_L + P_R \right]\, \psi(x)
\equiv \psi_L(x) + \psi_R(x)\, .
\ee
In the massless limit, the chiralities correspond to the fermion
helicities.

\subsection{Lagrangian formalism}

The Lagrangian formulation of a physical system provides a compact
dynamical description and makes it easier to discuss the underlying
symmetries. Like in classical mechanics, the dynamics is encoded in
the action
\bel{eq:action}
S\, = \int d^{\, 4} x\quad \cL\left[\phi_i(x),\partial_\mu\phi_i(x)\right]\, .
\ee
The integral over the four space-time coordinates preserves
relativistic invariance. The Lagrangian density $\cL$ is a
Lorentz-invariant functional of the fields $\phi_i(x)$ and their
derivatives. The space integral \ $L = \int d^{\, 3} x\;\cL$ \ would
correspond to the usual non-relativistic Lagrangian.

The principle of stationary action requires the variation $\delta S$
of the action to be zero under small fluctuations\ $\delta\phi_i$\
of the fields. Assuming that the variations\ $\delta\phi_i$\ are
differentiable and vanish outside some bounded region of space-time
(which allows an integration by parts), the condition $\delta S = 0$
determines the Euler--Lagrange equations of motion for the fields:
\bel{eq:EulerLagrange}
{\partial\cL\over\partial\phi_i}\, - \,\partial^\mu\!\left(
{\partial\cL\over\partial\left(\partial^\mu\phi_i\right)}\right)
\, =\, 0\, .
\ee

One can easily find appropriate Lagrangians to generate the
Klein--Gordon and Dirac equations. They should be quadratic on the
fields and Lorentz invariant, which determines their possible form
up to irrelevant total derivatives. The Lagrangian
\bel{eq:KGcomplex}
\cL\, =\,\partial^\mu\phi^*\partial_\mu\phi - m^2\, \phi^*\phi
\ee
describes a complex scalar field without interactions. Both the
field $\phi(x)$ and its complex conjugate $\phi^*(x)$ satisfy the
Klein--Gordon equation; thus, $\phi(x)$ describes a particle of mass
$m$ without spin and its antiparticle. Particles which are their own
antiparticles (i.e., with no internal charges) have only one degree
of freedom and are described through a real scalar field. The
appropriate Klein--Gordon Lagrangian is then
\bel{eq:KGreal}
\cL\, =\,\frac{1}{2}\,\partial^\mu\phi\,\partial_\mu\phi -
\frac{1}{2}\, m^2\, \phi^2\, .
\ee

The Dirac equation can be derived from the Lagrangian density
\bel{eq:DiracL}
\cL\, =\, \overline\psi\,\left(
i\,\gamma^\mu\partial_\mu - m\right) \psi\, .
\ee
The adjoint spinor \ $\overline\psi(x) = \psi^\dagger(x)\,\gamma^0$
closes the Dirac indices. The matrix $\gamma^0$ is included
to guarantee the proper behaviour under Lorentz transformations:
$\overline\psi\psi$ is a Lorentz scalar, while
$\overline\psi\gamma^\mu\psi$ transforms as a four-vector
\cite{Bjorken}. Therefore, $\cL$ is Lorentz invariant as it
should.

Using the decomposition \eqn{eq:chirality} of the Dirac field in its
two chiral components, the fermionic Lagrangian adopts the form:
\bel{eq:DiracL_LR}
\cL\, =\, \overline\psi_L\, i\,\gamma^\mu\partial_\mu\psi_L
\, +\,\overline\psi_R\, i\,\gamma^\mu\partial_\mu\psi_R
\, -\, m\,\left(\overline\psi_L\psi_R +\overline\psi_R\psi_L
\right)\, .
\ee
Thus, the two chiralities decouple if the fermion is massless.

\subsection{Symmetries and conservation laws}

Let us assume that the Lagrangian of a physical system is invariant
under some set of continuous transformations
\bel{eq:PhiTransf}
\phi_i(x)\;\to\; \phi'_i(x) = \phi_i(x) + \epsilon\:\delta_\epsilon\phi_i(x)
+ O(\epsilon^2)\, ,
\ee
i.e., \ $\cL\left[\phi_i(x),\partial_\mu\phi_i(x)\right]
=\cL\left[\phi'_i(x),\partial_\mu\phi'_i(x)\right]$. One finds then
that
\bel{eq:VarS}
\delta_\epsilon\cL\; =\; 0\; =\;
\sum_i\left\{\left[
{\partial\cL\over\partial\phi_i}\, - \,\partial^\mu\!\left(
{\partial\cL\over\partial\left(\partial^\mu\phi_i\right)}\right)
\right]\delta_\epsilon\phi_i
\, +\, \partial^\mu\left[
{\partial\cL\over\partial\left(\partial^\mu\phi_i\right)}\,
\delta_\epsilon\phi_i\right]\right\}\, .
\ee
If the fields satisfy the Euler--Lagrange equations of motion
\eqn{eq:EulerLagrange}, the first term is identically zero;
therefore the system has a conserved current:
\bel{eq:Noether} J_\mu \equiv\sum_i
{\partial\cL\over\partial\left(\partial^\mu\phi_i\right)}
\,\delta_\epsilon\phi_i \, , \qquad\qquad\qquad\qquad
\partial^\mu J_\mu = 0\, .
\ee
This allows us to define a conserved charge
\bel{eq:NoetherCharge}
\cQ \equiv \int d^{\, 3} x\; J^0\, .
\ee
The condition \ $\partial^\mu J_\mu = 0$ \ guarantees that \
${d\cQ\over dt} = 0\,$, i.e., that $\cQ$ is a constant of motion.

This result, known as Noether's theorem, can be easily extended to
general transformations involving also the space-time coordinates.
For every continuous symmetry transformation which leaves the
Lagrangian invariant, there is a corresponding divergenceless
Noether's current and, therefore, a conserved charge. The selection
rules observed in Nature, where there exist several conserved
quantities (energy, momentum, angular momentum, electric charge,
etc.), correspond to dynamical symmetries of the Lagrangian.

\subsection{Classical electrodynamics}

The well-known Maxwell equations,
\beqn\label{eq:Maxwell_1}
 \vec{\nabla}\cdot\vec{B} = 0 \, ,\qquad\qquad
&& \qquad\qquad \vec{\nabla}\times\vec{E} +
{\partial\vec{B}\over\partial\, t} = 0\, ,
\\[3pt] \label{eq:Maxwell_2}
 \vec{\nabla}\cdot\vec{E} = \rho\, ,
\qquad\qquad && \qquad\qquad \vec{\nabla}\times\vec{B} -
{\partial\vec{E}\over\partial\, t} = \vec{J}\, ,
\eeqn
summarize a large amount of experimental and theoretical work and
provide a unified description of the electric and magnetic forces.
The first two equations in \eqn{eq:Maxwell_1} are easily solved,
writing the electromagnetic fields in terms of potentials:
\bel{eq:potentials} \vec{E} = -\vec{\nabla} V -
{\partial\vec{A}\over\partial\, t} \, , \qquad\qquad\qquad\qquad
\vec{B} = \vec{\nabla}\times\vec{A}\, . \ee

It is very useful to rewrite these equations in a Lorentz covariant
notation. The charge density $\rho$ and the electromagnetic current
$\vec{J}$ transform as a four-vector \ $J^\mu
\equiv\left(\rho,\vec{J}\,\right)$. The same is true for the
potentials which combine into \ $A^\mu
\equiv\left(V,\vec{A}\right)$. The relations \eqn{eq:potentials}
between the potentials and the fields then take a very simple form,
which defines the field strength tensor:
\bel{eq:Fmunu}
F^{\mu\nu}\equiv\partial^\mu A^\nu-\partial^\nu A^\mu =
\left(\begin{array}{cccc}
0 & -E_1 & -E_2 & -E_3 \\
E_1 & 0 & -B_3 & B_2 \\
E_2 & B_3 & 0 & -B_1 \\
E_3 & -B_2 & B_1 & 0
\ea\right)
\, , \qquad\qquad
\tilde{F}^{\mu\nu}\equiv\frac{1}{2}\, \epsilon^{\mu\nu\rho\sigma}\,
F_{\rho\sigma}\, .
\ee
In terms of the tensor $F^{\mu\nu}$, the covariant form of the
Maxwell equations turns out to be very transparent:
\bel{eq:Maxwell_Cov}
\partial_\mu \tilde{F}^{\mu\nu} = 0
\, , \qquad\qquad\qquad\qquad
\partial_\mu F^{\mu\nu} = J^\nu\, .
\ee
The electromagnetic dynamics is clearly a relativistic phenomenon,
but Lorentz invariance was not very explicit in the original
formulation of Eqs.~\eqn{eq:Maxwell_1} and \eqn{eq:Maxwell_2}. Once
a covariant formulation is adopted, the equations become much
simpler.
The conservation of the electromagnetic current appears now as a natural
compatibility condition:
\bel{eq:CurrentCons}
\partial_\nu J^\nu = \partial_\nu\partial_\mu F^{\mu\nu} = 0\, .
\ee
In terms of potentials, $\partial_\mu \tilde{F}^{\mu\nu}$ is identically zero
while $\partial_\mu F^{\mu\nu} = J^\nu$ adopts the form:
\bel{eq:Maxwell_Cov2}
\Box\, A^\nu -\partial^\nu\left(\partial_\mu A^\mu\right) = J^\nu\, .
\ee

The same dynamics can be described by many different electromagnetic
four-potentials, which give the same field strength tensor
$F^{\mu\nu}$. Thus, the Maxwell equations are invariant under gauge
transformations:
\bel{eq:GaugeTransf}
A^\mu\;\longrightarrow\; A'^\mu = A^\mu + \partial^\mu\Lambda\, .
\ee
Taking the {\it Lorentz gauge} \ $\partial_\mu A^\mu = 0$,
Eq.~\eqn{eq:Maxwell_Cov2} simplifies to
\bel{eq:Maxwell_Cov3}
\Box\, A^\nu = J^\nu\, .
\ee
In the absence of an external current, i.e., with $J^\mu = 0$, the
four components of $A^\mu$ satisfy then a Klein--Gordon equation
with $m= 0$. The photon is therefore a massless particle.

The Lorentz condition \ $\partial_\mu A^\mu = 0$ \ still allows for
a residual gauge invariance under transformations of the type
\eqn{eq:GaugeTransf}, with the restriction \ $\Box\,\Lambda = 0$.
Thus, we can impose a second constraint on the electromagnetic field
$A^\mu$, without changing $F^{\mu\nu}$. Since $A^\mu$ contains
four fields ($\mu = 0, 1, 2, 3$) and there are two arbitrary
constraints, the number of physical degrees of freedom is just two.
Therefore, the photon has two different physical polarizations

\renewcommand{\theequation}{B.\arabic{equation}}
\setcounter{equation}{0}
\section{SU(N) \ Algebra}

$SU(N)$ is the group of\ $N\times N$ unitary matrices,
$U U^\dagger = U^\dagger U =1$, with \ $\det U=1$.
Any $SU(N)$ matrix can be written in the form
\bel{eq:Uexp}
U = \exp{\left\{i\, T^a \theta_a\right\}}
\, , \qquad\qquad\qquad\qquad a=1,2,\ldots,N^2-1\, ,
\ee
with \ $T^a = \lambda^a/2$ \ Hermitian, traceless matrices. Their
commutation relations
\bel{eq:T_com}
[T^a, T^b] \, = \, i\, f^{abc}\, T^c
\ee
define the $SU(N)$ algebra.
The $N\times N$ matrices $\lambda^a/2$ generate the fundamental
representation of the $SU(N)$ algebra.
The basis of generators $\lambda^a/2$ can be chosen so that
the structure constants $f^{abc}$ are real and totally antisymmetric.

For $N=2$,\ $\lambda^a$ are the usual Pauli matrices,
\bel{eq:pauli}
\sigma_1=\left(\bat  0 & 1 \\ 1 & 0 \ea\right)\, , \qquad
  \sigma_2=\left(\bat 0 & -i \\ i & 0 \ea\right)\, , \qquad
  \sigma_3=\left(\bat 1 & 0 \\ 0 & -1 \ea\right)\, ,
\ee
which satisfy the commutation relation
\bel{eq:SU(2)}
\left[ \sigma_i,\sigma_j\right]  =  2\, i \,\epsilon_{ijk}\,\sigma_k \, .
\ee
Other useful properties are: \
$\left\{ \sigma_i,\sigma_j\right\}  =  2\,\delta_{ij}$ \
and \ $\mbox{\rm Tr}\left(\sigma_i\sigma_j\right)  =  2\,\delta_{ij}$.

For $N=3$, the fundamental representation
corresponds to the eight Gell-Mann matrices:
\beqn\label{eq:GM_matrices}
\lambda^1 =\left( \bath 0 & 1 & 0 \\ 1 & 0 & 0 \\ 0 & 0 & 0 \ea \right) ,
\quad\,
\lambda^2 & = &
\left( \bath 0 & -i & 0 \\ i & 0 & 0 \\ 0 & 0 & 0 \ea \right) ,
\quad
\lambda^3 = \left( \bath 1 & 0 & 0 \\ 0 & -1 & 0 \\ 0 & 0 & 0 \ea \right) ,
\quad
\lambda^4 = \left( \bath 0 & 0 & 1 \\ 0 & 0 & 0 \\ 1 & 0 & 0 \ea \right) ,
 \no\\   && \\
\lambda^5 =
\left( \bath 0 & 0 & -i \\ 0 & 0 & 0 \\ i & 0 & 0 \ea \right) \! ,
\;\;
\lambda^6 & = &
\left( \bath 0 & 0 & 0 \\ 0 & 0 & 1 \\ 0 & 1 & 0 \ea \right)\! , \;\;
\lambda^7 = \left( \bath 0 & 0 & 0 \\ 0 & 0 & -i \\ 0 & i & 0 \ea \right)
\! , \;\;
\lambda^8 =
{1\over\sqrt{3}}
\left( \bath 1 & 0 & 0 \\ 0 & 1 & 0 \\ 0 & 0 & -2 \ea \right) \! . \no
\eeqn
They satisfy the anticommutation relation
\bel{eq:anticom}
\left\{\lambda^a,\lambda^b\right\} \, = \,
{4\over N} \,\delta^{ab} \, I_N \,
+ \, 2\, d^{abc} \, \lambda^c \, ,
\ee
where $I_N$ denotes the $N$-dimensional unit matrix and the constants
$d^{abc}$ are totally symmetric in the three indices.

For $SU(3)$, the only non-zero (up to permutations)
$f^{abc}$ and $d^{abc}$ constants are
\beqn\label{eq:constants} &&
{1\over 2}\, f^{123} = f^{147} = -
f^{156} = f^{246} = f^{257} = f^{345} = - f^{367} =
{1\over\sqrt{3}}\, f^{458} = {1\over\sqrt{3}}\, f^{678} =
{1\over2}\; , \qquad
\CR && d^{146} = d^{157} = -d^{247} = d^{256} =
d^{344} = d^{355} = -d^{366} = - d^{377} = {1\over 2}\; ,
\\
&&d^{118} = d^{228} = d^{338} = -2\, d^{448} = -2\, d^{558} = -2\, d^{668} =
-2\, d^{778} = -d^{888} = {1\over \sqrt{3}}\; .
\no
\eeqn

The adjoint representation of the $SU(N)$ group is given by the
$(N^2-1)\!\times\! (N^2-1)$ matrices $(T^a_A)_{bc} \equiv - i
f^{abc}$, which satisfy the commutation relations \eqn{eq:T_com}.
The following equalities
\beqn\label{eq:invariants}
{\rm Tr}\left(\lambda^a\lambda^b\right) =  4 \, T_F \, \delta_{ab}
\, , \qquad\qquad\quad\quad\quad\quad && T_F = {1\over 2} \, ,
\no\\
\left(\lambda^a\lambda^a\right)_{\alpha\beta} = 4\,
C_F \, \delta_{\alpha\beta}\, , \qquad\qquad\quad\quad\quad\quad
&& C_F = {N^2-1\over 2N} \, ,
\\[5pt] \;
{\rm Tr}(T^a_A T^b_A) = f^{acd} f^{bcd} = C_A \,\delta_{ab}
\, ,\qquad\quad\quad\qquad
&& C_A = N \, , \qquad\no
\eeqn
define the $SU(N)$ invariants $T_F$, $C_F$ and $C_A$.
Other useful properties are:
$$
\left(\lambda^a\right)_{\alpha\beta}
\left(\lambda^a\right)_{\gamma\delta}
= 2 \,\delta_{\alpha\delta}\delta_{\beta\gamma}
 -{2\over N} \,\delta_{\alpha\beta}\delta_{\gamma\delta}
\, ,\qquad\qquad
{\rm Tr}\left(\lambda^a\lambda^b\lambda^c\right)
 =  2\, (d^{abc} + i f^{abc})\, ,
$$
\be
{\rm Tr}(T^a_A T^b_A T^c_A)  =  i \, {N\over 2}\, f^{abc}
\, ,\qquad\qquad \sum_b d^{abb} = 0\, ,\qquad\qquad
d^{abc} d^{ebc}  =  \left( N - {4\over N}\right) \delta_{ae} \, ,
\ee
$$
f^{abe} f^{cde} + f^{ace} f^{dbe} + f^{ade} f^{bce} = 0
\, ,\qquad\qquad
f^{abe} d^{cde} + f^{ace} d^{dbe} + f^{ade} d^{bce} = 0 \, .
$$

\renewcommand{\theequation}{C.\arabic{equation}}
\setcounter{equation}{0}
\section{Anomalies}
\label{sec:anomalies}

\begin{figure}[htb]\centering
\includegraphics[width=9cm]{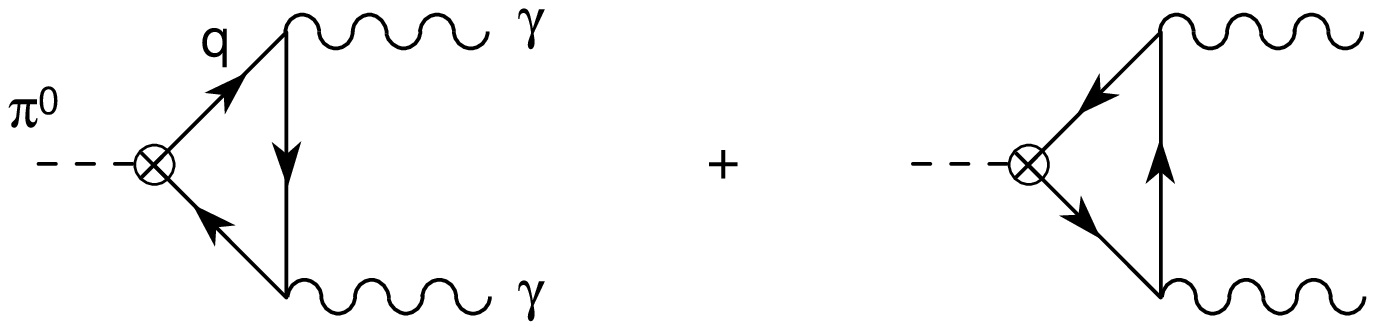}
\caption{Triangular quark loops generating the decay $\pi^0\to\gamma\gamma$.}
\label{fig:triangle}
\end{figure}

Our theoretical framework is based on the local gauge symmetry.
However, so far we have only discussed the symmetries of the
classical Lagrangian. It happens sometimes that a symmetry of $\cL$
gets broken by quantum effects, i.e., it is not a symmetry of the
quantized theory; one says then that there is an `anomaly'.
Anomalies appear in those symmetries involving both axial
($\overline{\psi}\gamma^\mu\gamma_5\psi$) and vector
($\overline{\psi}\gamma^\mu\psi$) currents, and reflect the
impossibility of regularizing the quantum theory (the divergent
loops) in a way which preserves the chiral (left/right) symmetries.

A priori there is nothing wrong with having an anomaly. In fact,
sometimes they are even welcome. A good example is provided by the
decay $\pi^0\to\gamma\gamma$. There is a chiral symmetry of the QCD
Lagrangian which forbids this transition; the $\pi^0$ should then be
a stable particle, in contradiction with the experimental evidence.
Fortunately, there is an anomaly generated by a triangular quark
loop (Fig.~\ref{fig:triangle}) which couples the axial current
$A_\mu^3\equiv (\bar u\gamma_\mu\gamma_5 u - \bar
d\gamma_\mu\gamma_5 d)$ to two electromagnetic currents and breaks
the conservation of the axial current at the quantum level:
\bel{eq:div_A}
\partial^\mu A_\mu^3 \,=\,  {\alpha\over 4\pi}\,
\epsilon^{\alpha\beta\sigma\rho}\,
F_{\alpha\beta}
\, F_{\sigma\rho} \, + \, \cO\left(m_u + m_d\right) .
\ee
Since the $\pi^0$ couples to $A_\mu^3\,$, $\langle
0|A_\mu^3|\pi^0\rangle = 2 i\, f_\pi p_\mu\,$, the
$\pi^0\to\gamma\gamma$ decay does finally occur, with a predicted
rate
\bel{eq:pi_decay}
\Gamma(\pi^0\to\gamma\gamma)\, = \, \left({N_C\over 3}\right)^2
{\alpha^2 m_\pi^3\over 64\pi^3 f_\pi^2}
\, =\, 7.73\, \mbox{\rm eV} ,
\ee
where $N_C=3$ denotes the number of quark colours and the so-called
pion decay constant, $f_\pi=92.4$~MeV, is known from the
$\pi^-\to\mu^-\bar\nu_\mu$ decay rate (assuming isospin symmetry).
The agreement with the measured value, $\Gamma = 7.7\pm 0.6$ eV
\cite{PDG}, is excellent.

Anomalies are, however, very dangerous in the case of local gauge
symmetries, because they destroy the renormalizability of the
Quantum Field Theory. Since the $SU(2)_L\otimes U(1)_Y$ model is
chiral (i.e., it distinguishes left from right), anomalies are
clearly present. The gauge bosons couple to vector and axial-vector
currents; we can then draw triangular diagrams
with three arbitrary gauge bosons ($W^\pm$, $Z$, $\gamma$) in the
external legs. Any such diagram involving one axial and two vector
currents generates a breaking of the gauge symmetry. Thus, our nice
model looks meaningless at the quantum level.

We have still one way out. What matters is not the value of a single
Feynman diagram, but the sum of all possible contributions.
The anomaly generated by the sum of all triangular diagrams
connecting the three gauge bosons $G_a$, $G_b$ and $G_c$ is proportional
to
\bel{eq:an_condition}
\cA\, = \, \mbox{\rm Tr}\left( \{ T^a , T^b \}\, T^c \right)_L -
 \mbox{\rm Tr}\left( \{ T^a , T^b \}\, T^c \right)_R,
\ee
where the traces sum over all possible left- and right-handed
fermions, respectively, running along the internal lines
of the triangle.
The matrices $T^a$ are the generators associated with the corresponding
gauge bosons; in our case, $T^a = \sigma_a/2\, ,\, Y$.

In order to preserve the gauge symmetry, one needs a cancellation of
all anomalous contributions, i.e., $\cA=0$. Since $\mbox{\rm
Tr}(\sigma_k)=0$, we have an automatic cancellation in two
combinations of generators: \ $\mbox{\rm Tr}\left(\{ \sigma_i ,
\sigma_j \}\, \sigma_k \right)=2\,\delta^{ij}\, \mbox{\rm
Tr}(\sigma_k)=0\, $ \ and \ $\mbox{\rm Tr}\left(\{ Y , Y\}\,
\sigma_k \right)\propto\mbox{\rm Tr}(\sigma_k)= 0\,$. However, the
other two combinations, \ $\mbox{\rm Tr}\left(\{ \sigma_i , \sigma_j
\}\, Y \right)$ \ and \ $\mbox{\rm Tr}(Y^3)$ \ turn out to be
proportional to \ $\mbox{\rm Tr}(Q)\,$, i.e., to the sum of fermion
electric charges:
\bel{eq:an_cancellation}
\sum_i Q_i \, = \, Q_e + Q_\nu + N_C \left( Q_u + Q_d \right) \, = \,
-1 + {1\over 3} N_C \, =\, 0\, .
\ee

Equation~\eqn{eq:an_cancellation} conveys a very important message:
the gauge symmetry of the $SU(2)_L\otimes U(1)_Y$ model does not
have any quantum anomaly, provided that $N_C=3$. Fortunately, this
is precisely the right number of colours to understand strong
interactions. Thus, at the quantum level, the electroweak model
seems to know something about QCD. The complete SM gauge theory
based on the group $SU(3)_C\otimes SU(2)_L\otimes U(1)_Y$ is free of
anomalies and, therefore, renormalizable. The anomaly cancellation
involves one complete generation of leptons and quarks: $\nu\, ,\,
e\, ,\, u\, ,\, d$. The SM would not make any sense with only
leptons or quarks.


\end{document}